\documentstyle[sprocl]{article}

\input psfig.sty
\def\nn{\noindent}

\def\Re{{\cal R \mskip-4mu \lower.1ex \hbox{\it e}\,}}
\def\Im{{\cal I \mskip-5mu \lower.1ex \hbox{\it m}\,}}
\def\ie{{\it i.e.}}
\def\eg{{\it e.g.}}

\def\etal{{\it et al.}}
\def\ibid{{\it ibid}.}
\def\sub#1{_{\lower.25ex\hbox{$\scriptstyle#1$}}}

\def\gev{\,{\rm GeV}}

\def\to{\rightarrow}
\def\slash{\not\!}
\def\subw{_{\rm w}}
\def\mh{\ifmmode m\sbl H \else $m\sbl H$\fi}
\def\mch{\ifmmode m_{H^\pm} \else $m_{H^\pm}$\fi}
\def\mt{\ifmmode m_t\else $m_t$\fi}
\def\mc{\ifmmode m_c\else $m_c$\fi}
\def\mz{\ifmmode M_Z\else $M_Z$\fi}
\def\mw{\ifmmode M_W\else $M_W$\fi}
\def\mws{\ifmmode M_W^2 \else $M_W^2$\fi}
\def\mhs{\ifmmode m_H^2 \else $m_H^2$\fi}   
\def\mzs{\ifmmode M_Z^2 \else $M_Z^2$\fi}
\def\mts{\ifmmode m_t^2 \else $m_t^2$\fi}
\def\mcs{\ifmmode m_c^2 \else $m_c^2$\fi}
\def\mchs{\ifmmode m_{H^\pm}^2 \else $m_{H^\pm}^2$\fi}
\def\ztwo{\ifmmode Z_2\else $Z_2$\fi}
\def\zone{\ifmmode Z_1\else $Z_1$\fi}
\def\mtwo{\ifmmode M_2\else $M_2$\fi}
\def\mone{\ifmmode M_1\else $M_1$\fi}
\def\tb{\ifmmode \tan\beta \else $\tan\beta$\fi}
\def\xw{\ifmmode x\subw\else $x\subw$\fi}
\def\ch{\ifmmode H^\pm \else $H^\pm$\fi}
\def\lum{\ifmmode {\cal L}\else ${\cal L}$\fi}
\def\inpb{\ifmmode {\rm pb}^{-1}\else ${\rm pb}^{-1}$\fi}
\def\infb{\ifmmode {\rm fb}^{-1}\else ${\rm fb}^{-1}$\fi}
\def\epem{\ifmmode e^+e^-\else $e^+e^-$\fi}
\def\ppb{\ifmmode \bar pp\else $\bar pp$\fi}
\def\bsg{\ifmmode B\to X_s\gamma\else $B\to X_s\gamma$\fi}
\def\bsll{\ifmmode B\to X_s\ell^+\ell^-\else $B\to X_s\ell^+\ell^-$\fi}
\def\bstt{\ifmmode B\to X_s\tau^+\tau^-\else $B\to X_s\tau^+\tau^-$\fi}

\def\AQ#1{\Pi_{\rm #1}(q^2)}
\def\A0#1{\Pi_{\rm #1}(0)}

\def\AZ#1{\Pi_{\rm #1}(\mz^2)}

\def\AP0#1{\Pi'_{\rm #1}(0)}
\def\APW#1{\Pi'_{\rm #1}(\mw^2)}
\def\APZ#1{\Pi'_{\rm #1}(\mz^2)}

\def\ANEW0#1{\Pi_{\rm #1}^{\rm new}(0)}
\def\ANEWP0#1{{\Pi'_{\rm #1}}^{\rm new}(0)}

\def\B0#1{\Pi'_{\rm #1}(0)}
\def\BW#1{\frac{\Pi_{\rm #1}(\mw^2)-\Pi_{\rm #1}(0)}{\mw^2}}
\def\BZ#1{\frac{\Pi_{\rm #1}(\mz^2)-\Pi_{\rm #1}(0)}{\mz^2}} 
\def\APP0#1{\Pi''_{\rm #1}(0)}
\def\APPP0#1{\Pi'''_{\rm #1}(0)}
\newskip\zatskip \zatskip=0pt plus0pt minus0pt
\def\matth{\mathsurround=0pt}
\def\lsim{\mathrel{\mathpalette\atversim<}}
\def\gsim{\mathrel{\mathpalette\atversim>}}
\def\atversim#1#2{\lower0.7ex\vbox{\baselineskip\zatskip\lineskip\zatskip
  \lineskiplimit 0pt\ialign{$\matth#1\hfil##\hfil$\crcr#2\crcr\sim\crcr}}}

\def\be{\begin{equation}}
\def\ee{\end{equation}}
\def\bea{\begin{eqnarray}}
\def\eea{\end{eqnarray}}

\begin{document}
\rightline{\vbox{\halign{&#\hfil\cr
&SLAC-PUB-7930\cr}}}
\vspace{0.5cm}

\title{THE STANDARD MODEL AND WHY WE BELIEVE IT
\footnote{Lectures given at {\it TASI97: Supersymmetry, Supergravity, and
Supercolliders}, Boulder, CO, June 1997.}
\footnote{Work supported by the Department of 
Energy, Contract DE-AC03-76SF00515}}

\author{ J.L. HEWETT}

\address{Stanford Linear Accelerator Center, Stanford, CA 94309}

\maketitle\abstracts{
The principle components of the Standard Model and the status of their 
experimental verification are reviewed.}


\section{The Standard Model}

The Standard Model (SM), which combines the SU(2)$_L\times$U(1)$_Y$ 
Glashow - Weinberg - Salam theory of electroweak interactions \cite{gws} 
together 
with Quantum Chromodynamics,\cite{qcd} constitutes a remarkable achievement.
The formulation of the theory as a renormalizable quantum theory preserves
its predictive power beyond tree-level computations
and allows for the probing of quantum effects.
An array of experimental results confirm every feature of the theory to
a high degree of precision, at the level of testing higher order perturbation
theory.  In fact, at present there are no compelling pieces of evidence that 
are in conflict with the SM.  In these lectures I will review the components of
the SM and the extent to which they have been tested.  

The strong interactions are described by Quantum Chromodynamics 
\cite{qcd} (QCD), which is a non-abelian gauge theory based on SU(3).
Each quark flavor is a color triplet in the fundamental representation
of SU(3)$_{\rm Color}$ and the SU(3) gauge fields, \ie, the gluons,
lie in the adjoint representation ${\bf \underline 8}$.  All other particles
are color singlets and don't experience strong interactions.  The QCD
Lagrangian may be written as
\be
{\cal L}_{QCD}  =  -{1\over 4}\hat F^a_{\mu\nu}\hat F_a^{\mu\nu}
+\bar\psi_i(i\gamma^\mu\hat D_\mu-m)
\psi_i \,,
\label{qcdla}
\ee
with
\be
\hat F^a_{\mu\nu}=\partial_\mu G^a_\nu-\partial_\nu G^a_\mu+g_sf^{abc}
G_{b\mu}G_{c\nu}
\ee
being the gluon field tensor and the covariant derivative is defined by
$\hat D_\mu=\partial_\mu\delta-ig_s
T_aG^a_\mu$.  Here $g_s$ represents the strong coupling and
the indices are summed over color with $a=1-8$ and $i=1,2,3$.  
$T_a$ and $f_{abc}$ are the SU(3) generators and structure constants,
respectively, which obey the usual commutation relation
\be
[T_a, T_b] = if_{abc} T_c \,.
\ee
The generators are related to the $3\times 3$ Gell-Mann matrices 
\cite{gellmann} by $T_a=\lambda_a/2$.  The Lagrangian is invariant under
infinitesimal local gauge transformations.  Note that in the limit of
equal mass quarks, the QCD Lagrangian possesses a global SU(N)$_f$
flavor symmetry, and in the limit of massless quarks an SU(N)$_L\times$
SU(N)$_R$ chiral symmetry is present.  SU(3) gauge invariance ensures
that the gluons are massless.

The Standard Model of electroweak interactions \cite{gws} is based on the
gauge group SU(2)$_L\times$U(1)$_Y$, where the generators of SU(2)$_L$
correspond to the three components of weak isospin $T_i$ and the U(1)$_Y$
generator to the weak hypercharge $Y$.  These are related to the electric
charge by $Q=T_3+Y/2$.  The Lagrangian describing the electroweak
interactions is
\bea
\label{smlagr}
{\cal L}_{EW} & = & -{1\over 4}F^a_{\mu\nu}F_a^{\mu\nu}-{1\over 4}
G_{\mu\nu}G^{\mu\nu}+i\bar\psi_j\gamma^\mu D_\mu\psi_j\nonumber\\
& & \quad +(D_\mu\Phi)(D^\mu\Phi)^\dagger-\mu^2\Phi^\dagger\Phi
-\lambda(\Phi^\dagger\Phi)^2\\
& & \quad +\lambda_{e_k}\bar L_{L_k}\Phi e_{R_k}
+\lambda_{u_{jk}}\bar Q_{L_j}\tilde\Phi u_{R_k}
+\lambda_{d_{jk}}\bar Q_{L_j}\Phi d_{R_k} + h.c. \,,\nonumber
\eea
with the field strength tensors
\bea
F^a_{\mu\nu} & = & \partial_\mu B^a_\nu-\partial_\nu B^a_\mu+g\epsilon^{abc}
B_{b\mu}B_{c\nu}\,,\nonumber\\
G_{\mu\nu} & = & \partial_\mu C_\nu-\partial_\nu C_\mu\,,
\eea
for the three non-abelian fields of SU(2)$_L$ and the single abelian gauge
field associated with U(1)$_Y$, respectively.  The covariant derivative is
\be
D_\mu=\partial_\mu-igT_aB^a_\mu-ig'{Y\over 2}C_\mu\,,
\ee
with $g\,, g'$ being the SU(2)$_L$, U(1)$_Y$ coupling strength, respectively.
The SU(2) generators obey the usual relation $[T_a,T_b]=i\epsilon_{abc}T_c$
and are related to the Pauli spin matrices by $T_a=\tau_a/2$.  This Lagrangian
is invariant under the infinitesimal local gauge transformations for SU(2)$_L$
and U(1)$_Y$ independently.  Being in the adjoint representation, the SU(2)$_L$ 
massless gauge fields form a weak isospin triplet with the charged fields
being defined by $W_\mu^\pm=(B_1\mp iB_2)_\mu/\sqrt 2$.  The neutral component
of $B_a$ mixes with the abelian gauge field to form the physical states
\bea
Z_\mu & = & B^3_\mu\cos\theta_w+C_\mu\sin\theta_w \,,\nonumber\\
A_\mu & = & C_\mu\cos\theta_w-B^3_\mu\sin\theta_w\,,
\eea
where $\tan\theta_w=g'/g$ is the weak mixing angle.

To generate the left-handed structure of the weak charged current interactions,
the SU(2) symmetry is applied to left-handed fermion fields only.
The fermion fields are thus given by
\be
\psi_L :\quad L_{L_k}={1\over 2}(1-\gamma_5)\left( \begin{array}{c}
\nu_k \\
e_k 
\end{array}\right)\,,\quad\quad Q_{L_k}={1\over 2}(1-\gamma_5)\left( 
\begin{array}{c}
u_k \\
d_k 
\end{array}\right)\,,
\ee
for the SU(2)$_L$ left-handed doublets and
\be
\psi_R :\quad e_{R_k}={1\over 2}(1+\gamma_5)e\,, \quad
u_{R_k}={1\over 2}(1+\gamma_5)u\,, \quad
d_{R_k}={1\over 2}(1+\gamma_5)d\,, \nonumber
\ee
for the right-handed singlets, with $k=1-3$ being a generation index.
The usual convention is that right-handed neutrinos are not introduced.
With quarks having three colors, the quantum numbers for this set of fermions
ensures the cancellation of divergent chiral anomaly diagrams.  The $\lambda_i$
in Eq. (\ref{smlagr}) above are the Yukawa couplings of the quarks and
leptons.

Masses for the non-abelian gauge fields and fermions are generated by the
Higgs mechanism \cite{phiggs} via spontaneous symmetry breaking (SSB) which 
preserves \cite{renorm} the renormalizability of the gauge theory.  The Higgs 
fields are complex scalar iso-doublets $(\phi^+,\, \phi^0)$ with electroweak 
interactions described by the second line in Eq. (\ref{smlagr}).  For the
choice $\mu^2<0$, the ground state of the theory is obtained when the
neutral member of the Higgs doublet acquires a vacuum expectation value (vev),
\be
\langle\Phi\rangle = \left(\begin{array}{c}
                             0 \\
                             {v\over\sqrt 2}
                     \end{array}\right) \,,
\ee
where $v$ is given by $v^2=-\mu^2/\lambda$.  This non-vanishing vev selects
a preferred direction in SU(2)$_L\times$U(1)$_Y$ space and spontaneously breaks
the theory, leaving the U(1)$_{\rm em}$ subgroup intact.  U(1)$_{\rm em}$ would
be broken as well if the $\phi^+$ field were also allowed to obtain a vev.
$\phi^0$ is then redefined by $\phi^0=(H+v)/\sqrt 2$, such that the physical
field $H$ has a vanishing vev and positive mass squared.  The remaining degrees
of freedom, \ie, the Goldstone bosons, are gauged away from the scalar sector,
but essentially reappear in the gauge sector, providing the longitudinal
modes for the $W$ and $Z$ bosons.  An examination of
the $v^2$ terms in the kinetic piece of the scalar Lagrangian reveals the
mass terms for the physical gauge bosons,
\be
M_W={1\over 2}gv\,,\quad\quad M_Z={v\over 2}\sqrt{g^2+g'^2}\,,\quad\quad
M_A=0\,.
\label{massterms}
\ee
In order to obtain Quantum Electrodynamics (QED) the massless $A_\mu$ is
identified with the photon and $e\equiv g\sin\theta_w$.  The QED and weak
charged and neutral currents become
\bea
J_\mu^{em} & = & \bar\psi\gamma_\mu Q\psi\,,\nonumber\\
J_\mu^{CC} & = & \bar\psi\gamma_\mu T_L^\pm\psi\,,\\
J_\mu^{NC} & = & \bar\psi(T_{3L}-x_wQ)\psi\,,\nonumber
\label{smcurs}
\eea
with the interaction terms being
\be
{\cal L} = eJ_\mu^{em}A^\mu+{g\over\sqrt 2}J_\mu^{CC}W^\mu+{g\over\cos\theta_w}
J_\mu^{NC}Z^\mu\,.
\ee
Here, $x_w\equiv\sin^2\theta_w$ and 
the $T_L^\pm\equiv (T_1\pm iT_2)_L/\sqrt 2=\tau^\pm/2$ operations act on
the left-handed isodoublets $\psi_L$, and vanish on $\psi_R$.
Comparison with the Fermi theory of weak interactions yields the relation
\be
{G_F\over\sqrt 2} = {g^2\over 8M_W^2}\,,
\ee
where $G_F$ is the Fermi constant and is well determined from $\mu$ decay.
Eq. (\ref{massterms}) relates the gauge bosons masses by $M_Z=M_W/\cos\theta_w$.
A commonly used parameter, defined by
\be
\rho={M_W^2\over M_Z^2\cos^2\theta_w}\,,
\ee
measures the ratio of the charged to neutral current strengths.  It is
unity at tree-level in the SM with one Higgs doublet.  For a more general set of
Higgs representations with weak isospin $T_i$ and vev's $v_i$ it is given by
\be
\rho={\sum_i[T_i(T_i+1)-T^2_{i3}]v_i^2\over\sum_i2T^2_{i3}v_i^2}\,.
\ee
After SSB, the last line of Eq. (\ref{smlagr}) gives the physical 
Higgs - fermion interactions and generates masses for the fermions,
$m_f=\lambda_f v/\sqrt 2$.  Diagonalization of the masses in the quark
sector introduces the weak mixing, or Cabibbo-Kobayashi-Maskawa (CKM)
matrix,\cite{ckm} which then appears in the hadronic weak charged current.

Having briefly introduced the essential ingredients of the SM, we now pause
to examine its general features and success as a theory.  Three principle
assumptions went into the building of the theory:
\begin{itemize}
\item The gauge group is SU(3)$_C\times$SU(2)$_L\times$U(1)$_Y$
\item There is one Higgs doublet
\item The fermion representations are left-handed weak isodoublets and 
right-handed singlets
\end{itemize}
In addition to these assumptions the theory contains 21 {\it a priori} free 
parameters:
\begin{itemize}
\item 3 coupling constants
\item 12 fermion masses
\item 4 fermion mixing parameters
\item 1 Higgs mass
\item 1 independent gauge boson mass
\end{itemize}
These parameters are inserted into the framework of the SM by hand.  The 
missing ingredients of the model are, of course, the Higgs boson which has yet 
to be discovered, and the $\tau$-neutrino for which there is only indirect 
evidence at present.  However, since experiment is now sensitive to loop-level
effects, indirect constraints on the Higgs boson mass have been obtained.
These bounds will be discussed at length below.
The successes of the SM as a theory can be listed as:
\begin{itemize}
\item Renormalizability
\item Unitarity
\item Unification of the electromagnetic and weak forces
\item Prediction of a specific relationship between $W$ and $Z$ boson masses
\item The weak charged and neutral current structure agrees with experiment
\item All aspects have impressive agreement with all experimental data
\end{itemize}
Despite these successes there remain a number of important questions which the 
SM does not address. These include:
\begin{itemize}
\item The fermion masses and mixings and the nature of CP-violation
\item Neutrino masses and oscillations
\item The number of generations
\item Parity violation in the weak interactions
\item Suppression of strong CP phase
\item What is the electroweak symmetry breaking mechanism (is there really
a light Higgs doublet, or do the gauge bosons become strongly interacting
at the TeV scale) and how is the hierarchy maintained
\item Unification with the strong and gravitational forces
\item Charge quantization, \ie, why does $Q_e=-Q_p$
\item Origin of dark matter
\item Baryogensis
\item Cosmological constant
\item Why is spacetime 4 dimensional?
\end{itemize}
This list of unanswered questions provides the principle motivation for
consideration of physics beyond the SM.  Numerous theories are studied in the
hope that they will address at least one of these issues.  However, no
single theory has been invented that successfully addresses all of these
questions simultaneously.  Unfortunately, since all data agree with the SM,
there is not one shred of experimental evidence to provide guidance for
extending the theory, or to indicate that physics beyond the SM exists.

For the remainder of these lectures, I will discuss the theoretical predictions
of the SM at the quantum level, and the experimental techniques and 
accuracy by which the elements of the SM have been determined.  Here,
it is interesting to provide a short historical perspective.  In the mid
1980's, when I was a student at TASI, the elements of the SM which still
awaited experimental confirmation were:
\begin{itemize}
\item Verify the multiplet structure of fermions (find top!)
\item Demonstrate asymptotic freedom over a wide range of $Q^2$
\item Confirm universality
\item Extract $\sin^2\theta_w$ from numerous experiments and compare
\item Measure the $W$ and $Z$ boson properties
\item Prove it's a gauge theory (measure self-interactions)
\item Find the Higgs
\end{itemize}
Except for discovering the Higgs, this list of measurements has now been 
completed at
various levels of sensitivity.  For example, $\sin^2\theta_w$ is now determined
to the level of $0.1\%$ (!), whereas we are just now entering the era
where the gauge self-interactions are being probed at an interesting level.
It is clear that we have made substantial progress verifying the SM in the last
decade, however, it is time to move on (experimentally and theoretically) to 
what lies beyond.

\section{Tests of QCD}

The basic building blocks of the strong interactions have been presented in 
the previous section and here we focus on the experimental verification of
the various components of the QCD Lagrangian.  Due to the principle of
asymptotic freedom, which we discuss below, the renormalized strong
coupling is small only at high energies.  Hence perturbation theory is
only valid and precision experimental tests can only be performed in
this domain.  Although much progress has recently been made \cite{lattice}
in quantifying QCD predictions in the nonperturbative region via lattice
gauge theory or for soft hadronic processes, we will concentrate solely
on the perturbative regime.

\begin{figure}[htbp]
\centerline{
\psfig{figure=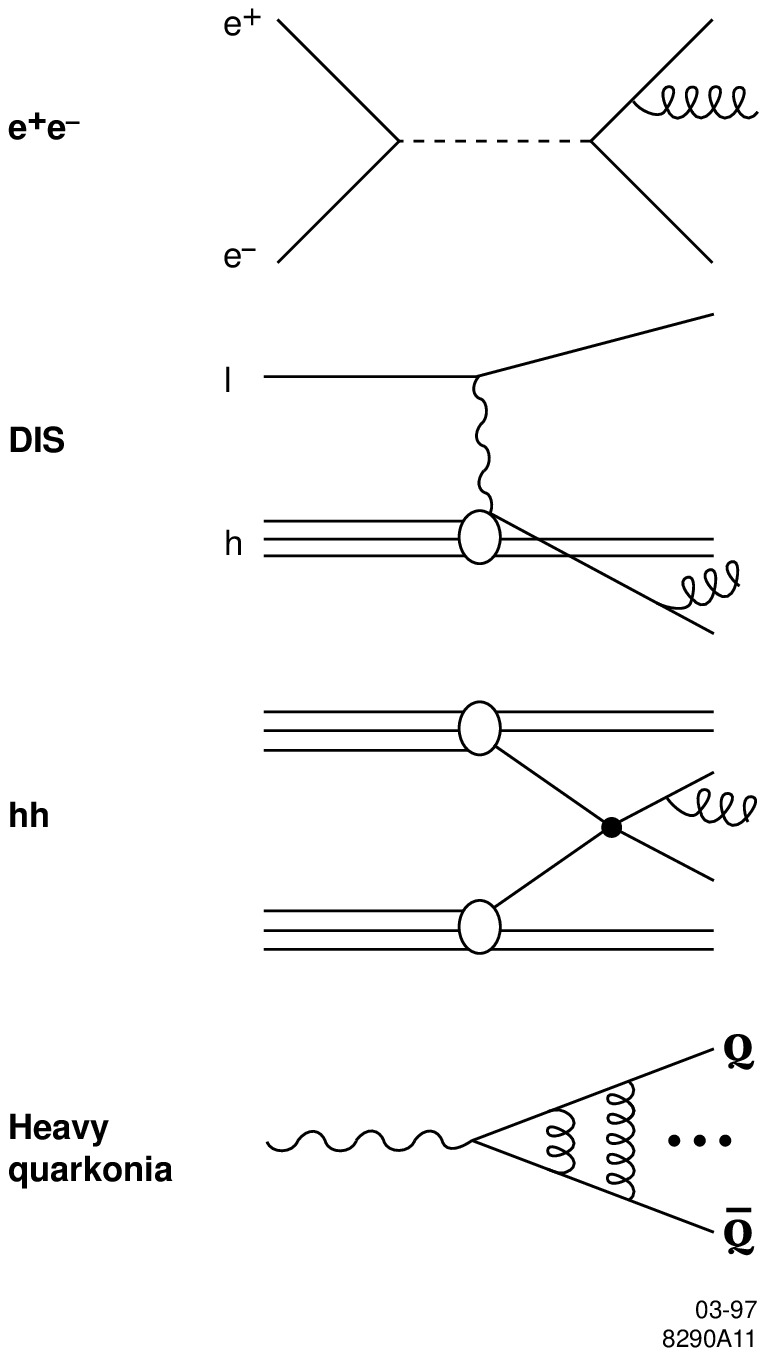,height=16cm,width=12cm,angle=0}}
\vspace*{-0.25cm}
\caption{Diagrams representing the basic processes of hadronic production in 
\epem\ annihilation, deep inelastic lepton-hadron scattering, hadronic 
collisions, and heavy quarkonia production and decay.}
\label{qcdprod}
\end{figure}

The basic perturbative processes responsible for the production of 
hadronic final states in a variety of interactions are schematically 
represented in Fig. \ref{qcdprod}.  In contrast to the other reactions,
we see that in \epem\ annihilation all hadronic activity is confined, by
construction, to the final state; there are no beam remnants to consider,
the hadronic center of mass frame coincides with the lab frame, there are
no parton density uncertainties, and there are fewer Feynman diagrams to
compute at a given order in perturbation theory.  Electron-positron 
colliders thus provide a clean and less complicated laboratory for precise 
QCD studies, both from the experimental and theoretical points of 
view, and our review is biased in this direction.

\vspace*{-0.5cm}
\nn

We now discuss the principle features of QCD and the experimental
evidence which supports QCD as being the underlying theory of the
strong interactions.\cite{phil}  We first list these features:
\begin{itemize}
\item Quarks exist as spin $1/2$ color triplets
\item Gluons exist as spin $1$ color octets 
\item The $q\bar qg$ coupling exists
\item The non-abelian triple and quartic gluon couplings exist
\item Asymptotic Freedom
\item Flavor independence of $\alpha_s$,
\end{itemize}
and then demonstrate their validity by concentrating on $n$-jet events in
\epem\ annihilation.

\subsection{2-jet Events} 
\begin{figure}[htbp]
\centerline{
\psfig{figure=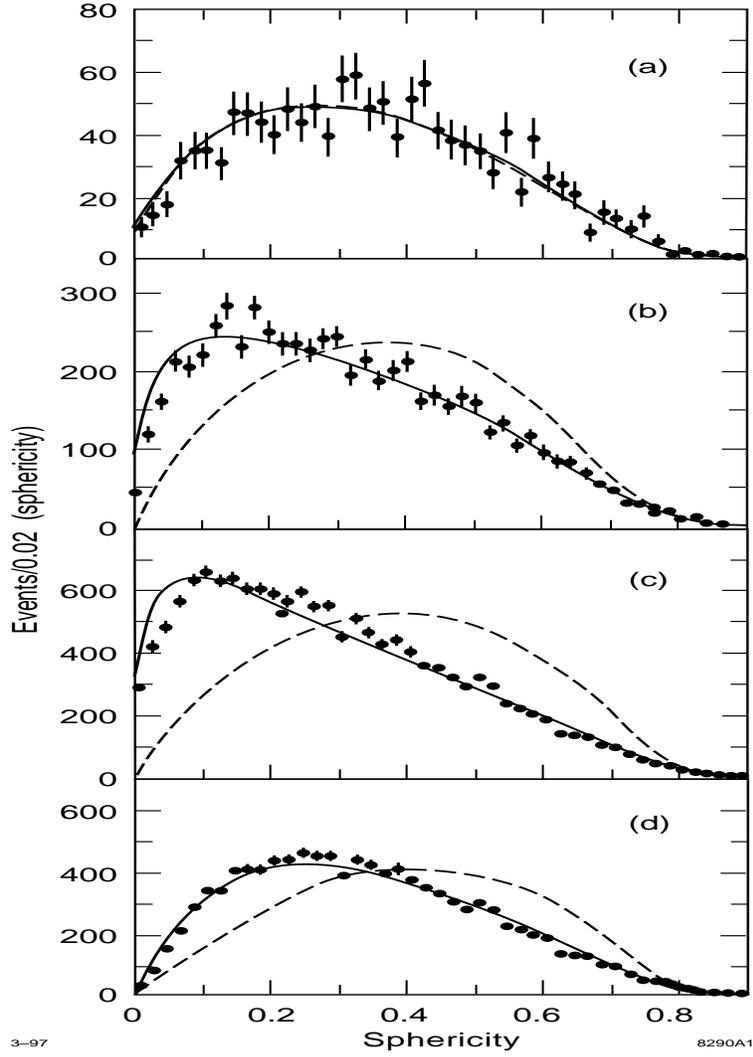,height=14cm,width=10cm,angle=0}}
\vspace*{0.5cm}
\caption{Sphericity distributions from Mark I~\protect\cite{markone} measured 
at $\sqrt s=$ (a) 3.0 GeV, (b) 6.2 GeV, and (c) 7.4 GeV. (d) Distribution
at 7.4 GeV for a subset of events with particles having scaled momentum
$2p/Q < 0.4$.  The dashed curves represent the expectations for a phase
space model of hadron production, while the solid curves correspond to
parton model jet production.}
\label{mark1}
\end{figure}

The first direct evidence for the existence of quarks came from 
deep inelastic electron-nucleon scattering at SLAC in the 1960s,
which established that the electrons were scattering off of point-like
particles in the nucleus.  Jet production was first observed in \epem\
annihilation via $\epem\to q\bar q$ by the Mark I experiment at SPEAR.
At higher energies, the quark-parton model suggests that the transverse
momenta of quark fragmentation products remains small since it arises 
mainly from soft processes, whereas the longitudinal momenta increases
with quark energy, leading to jet production.  In order to describe the 
degree of jet-like behavior, event shape variables, which characterize the
spatial distribution of particles in hadronic events, were introduced.
One such variable, which describes the extent of isotropy in the particle
flow, is sphericity, 
\be
S={3\over 2}{ {\rm Min}\left[ \sum_i (p_{Ti})^2\right]\over\sum_i p_i^2}\,,
\ee
where the sum is carried out over all the final state particles in an event,
and the subscript $T$ denotes the transverse momentum component.  The
axis which minimizes the sum in the numerator is known as the sphericity
axis.  Sphericity lies in the range $0\leq S\leq 1$, where $0$ represents
perfectly collimated back-to-back jets, and $S=1$ corresponds to completely
isotropic or spherical events.  The sphericity distributions from Mark I
\cite{markone} are presented in Fig. \ref{mark1}.  
A marked shift towards lower sphericity
with increasing center of mass energy is clearly present, indicating the
onset of jet production.  These observations have since been corroborated at
higher energy, such as on the $Z$-pole where the jets are almost perfectly
collimated.

\vspace*{-0.5cm}
\nn

Although sphericity is an experimentally useful variable, it is not
directly calculable in perturbation theory, essentially due to its quadratic 
form.  At higher order in perturbation theory a quark can split into a 
collinear quark-gluon pair which not only results in an divergence in 
the massless limit, but also gives substantially different contributions
to the quadratic function than that of the original quark.  Monte Carlo
simulations of the fragmentation process must then be performed, introducing
additional uncertainties.  One means around this is to integrate the cross
section over finite ranges of energy and angles.  Sterman and Weinberg
\cite{sterman} computed the cross section for the case where a fraction of
$1-\epsilon$ or more of the total energy is emitted within two oppositely
directed cones of half-angle $\delta$, making an angle $\theta$ with respect
to the beam axis.  In this case the collinear singularities approximately
cancel to a given order in perturbation theory as long as $\epsilon,\delta
\ll 1$.  Another way to avoid the problems associated with collinear 
splittings is to introduce jet variables which are based on linear sums of 
particle momenta.  Three such collinear and infrared safe shape variables that 
are commonly used are thrust, acoplanarity, and spherocity, defined by
\bea
T & = & {\rm Max}\, \left[ {\sum_i|{\bf p}_i\cdot {\bf n}|\over \sum_i
|{\bf p}_i|}\right]\,,\nonumber\\
A & = & 4\, {\rm Min}\, \left[ {\sum_i|{\bf p}_i\cdot {\bf n}'|\over \sum_i
|{\bf p}_i|}\right]^2\,,\\
S' & = & {16\over\pi^2}\, {\rm Min}\, {\left[ \sum_i E_i|\sin\theta_i|
\right]^2\over E^2_{tot}}\,.\nonumber
\eea
Here ${\bf n}({\bf n}')$ is a unit vector chosen to maximize (minimize) the 
numerator and define the thrust axis (event plane), and $\theta_i$ is the
angle between ${\bf p}_i$ and ${\bf n}$.  These variables have
the ranges, $1/2\le T\le 1$, $0\le A, S'\le 1$.  The thrust distribution
becomes narrower and more peaked towards unity as the total energy increases, 
while the acoplanarity defines the flatness of an event.

In \epem\ collisions the spin of the quark was determined by examining
the angular distributions of the sphericity and thrust axes.  A fit
to the functional forms
(where $P$ is the degree of transverse polarization that built up at
SPEAR due to a synchrotron radiation effect)
\bea
{dN\over d\cos\theta} & \propto & 1+\alpha\cos^2\theta+P^2\alpha\sin^2\theta
\cos 2\phi \,,\nonumber\\
{dN\over d\cos\theta} & \propto & 1+a_{S,T}\cos^2\theta_{S,T}\,,
\eea
yielded $\alpha=0.78\pm 0.12$ at 7.4 GeV from Mark I \cite{markfit} and
$a_S=1.03\pm 0.07$ and $a_T=1.01\pm 0.06$ at 35 GeV from TASSO.\cite{tasso}
These fits are all close to unity, which is what is expected \cite{collphys}
for the pair production of fermions.

\begin{figure}[htbp]
\centerline{
\psfig{figure=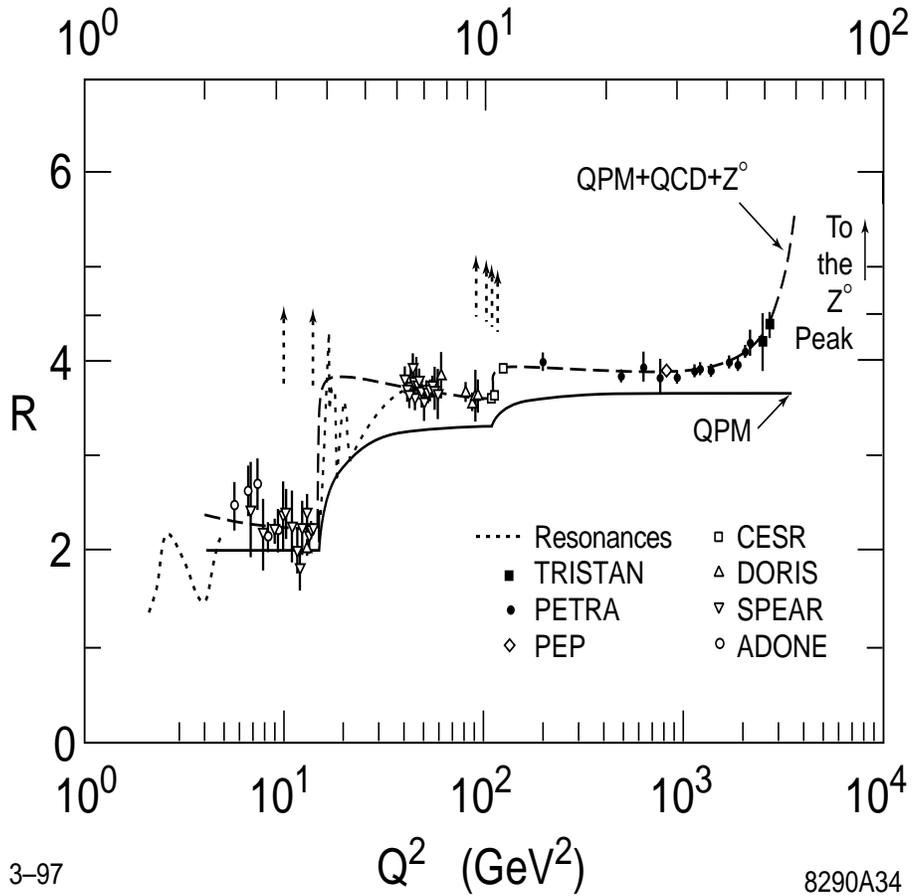,height=12cm,width=12cm,angle=0}}
\vspace*{0.5cm}
\caption{Summary of measurements of the $R$-ratio as a function of
center of mass energy.  The expectations of the quark parton model alone
and including QCD and electroweak radiative corrections are given by the 
solid and dashed curves, respectively.  The dashed arrows represent onset
of the charmonium and upsilon resonances.}
\label{rratio}
\end{figure}

That fact that quarks are color triplets was established experimentally
in \epem\ annihilation by the $R$-ratio, \ie, the ratio of the hadronic
cross section to the point cross section,
\be
R^{(0)}={\sigma(\epem\to{\rm hadrons})\over \sigma_{QED}(\epem\to\mu^+\mu^-)}
= N_c\sum_f Q_f^2\,,
\label{rrat}
\ee
where we have given the leading order parton model expectation at values of 
$s$ not close to quark thresholds, with $Q_f$ being the charge 
of the quark flavor $f$.  A summary of measurements of the $R$-ratio
below the $Z$ boson resonance from Ref. \cite{rmeas} is shown in Fig.
\ref{rratio}.  We see that the leading order quark parton model prediction is
consistent with the data only if each quark flavor has 3 colors.
The excess in the $R$-ratio over the
these parton model predictions provides support for QCD radiative corrections;
this additional contribution arises from gluon emission.  The higher-order
corrections have been calculated to 3-loops \cite{levan} and can be expressed as
\be
R=R^{(0)}\left[ 1+{\alpha_s\over\pi}+C_2\left({\alpha_s\over\pi}\right)^2
+C_3\left({\alpha_s\over\pi}\right)^3 +\cdots\right]\,,
\ee
with $C_2=1.411$ and $C_3=-12.8$.  From looking at the figure, we also see
that the relative size of this excess tends to decrease, roughly 
logarithmically, with $Q^2$; this is a consequence of asymptotic freedom, 
which will be discussed further below.  Production cross sections in \epem\
annihilation are often quoted in units of $R$, where $R$ represents the
point cross section for $\mu$ pair production in QED, \ie\ the denominator 
in Eq. (\ref{rrat}) above.  $1\, R=4\pi\alpha^2/3s=87 {\rm fb}/s 
({\rm TeV}^2)$.

Of course, the fact that quarks are color triplets is also required by the
quark parton model in order not to violate the Pauli exclusion principle
when forming the spin-$3/2$ baryon states such as the $\Delta^{++}$
and the $\Omega^-$.

\vspace*{-0.5cm}
\nn

\subsection{Three-jet events} 
The existence of the gluon and its coupling to quarks was established
by the observation of 3-jet events at $\sqrt s\approx 30 $ GeV at PETRA 
\cite{qqglue}.  These events were interpreted in terms of the process
$\epem\to q\bar qg$ and provided direct evidence for the $q\bar qg$
coupling.  Simply by event counting, it was determined that \cite{saulan}
\be
{ {\rm Number~(3-jet~events)}\over {\rm Number~(2-jet~events)}}
\approx 0.15 
\ee
at this center of mass energy.  At lowest order in perturbative QCD,
this ratio is just the probability of gluon emission and thus provided
a first direct measurement of the strong coupling constant $\alpha_s$.

\begin{figure}[htbp]
\centerline{
\psfig{figure=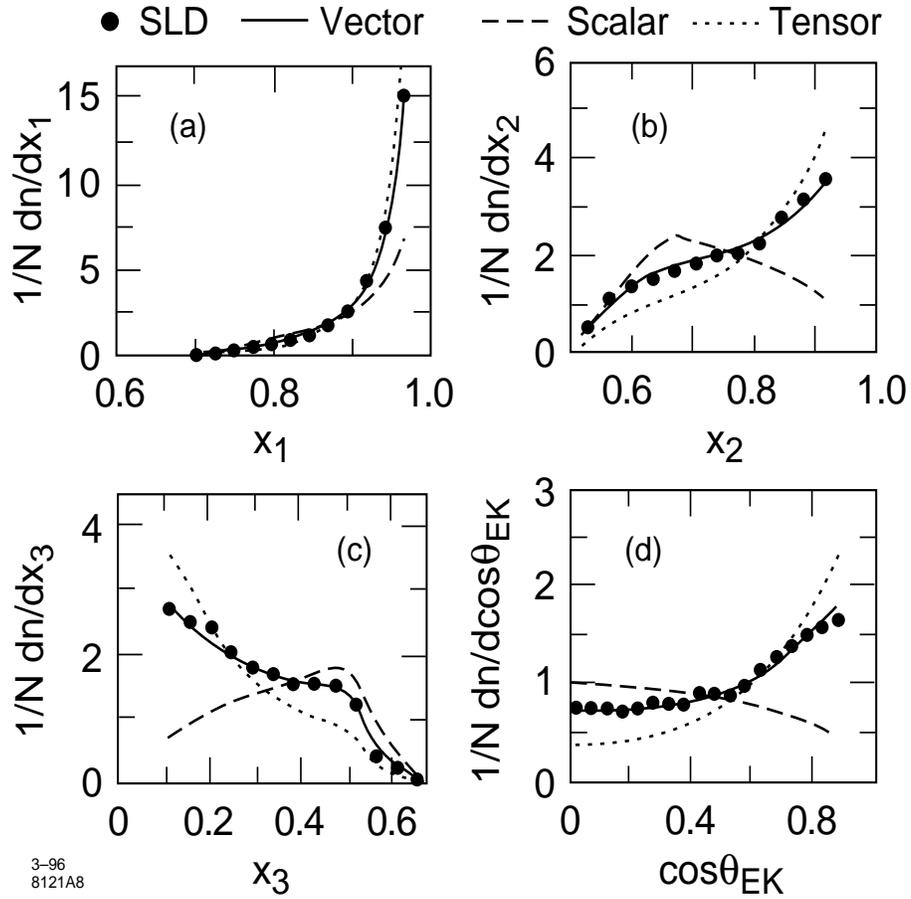,height=12cm,width=12cm,angle=0}}
\vspace*{0.5cm}
\caption{Scaled jet energy and Ellis-Karliner angle distributions in 
3-jet events as measured by SLD.  Also shown are the predictions from
leading-order vector, scalar, and tensor gluon models. 
From~\protect\cite{sldtgr}.}
\label{gluspin}
\end{figure}

The spin of the gluon can be determined by analyzing the jet energy and
angular distributions of three-jet events.  It is common practice to
label the three jet energies in terms of the ordering $E_1>E_2>E_3$ and 
to define the scaled jet energies $x_i\equiv 2E_i/Q$ such that the
relation $x_1+x_2+x_3=2$ holds.  Performing a Lorentz boost into the
rest frame of jets 2 and 3, the Ellis-Karliner angle \cite{marek} can
be defined as the angle between jets 1 and 2 in this frame.  At lowest
order with massless partons, we have
\be
\cos\theta_{EK}={x_2-x_3\over x_1} \,.
\ee
This angle is particularly sensitive to the spin of the gluon.  
Figure \ref{gluspin} displays the $x_i$ and Ellis-Karliner angle
distributions for 3-jet events as determined \cite{sldtgr} by SLD at 
$\sqrt s=M_Z$.  In this figure, the data is compared with expectations
for the distributions from leading-order vector, scalar, and tensor gluon 
models; it is clear that the data prefer the vector case.

\vspace*{-0.5cm}
\nn

We now return to the subject of collinear singularities in 3-jet events,
which we briefly discussed above.  The leading order normalized differential 
cross section for $\epem\to q\bar qg$ in the massless limit is
\be
{1\over \sigma_{\rm total}}{d\sigma\over dx_1dx_2}={2\alpha_s\over 3\pi}
{x_1^2+x_2^2\over (1-x_1)(1-x_2)} \,.
\ee
Here, the indices 1 and 2 now label the scaled jet energies of the quark
and anti-quark.  This expression is clearly singular when $x_i\to 1$;
this behavior is exhibited in two distinct physical situations.  (i) 
Infrared divergences occur when the gluon is soft, \ie, when both
$x_1,x_2\sim 1$.  (ii)  Collinear singularities occur when the gluon
is emitted parallel to either of the quark jets, \ie, either $x_1$ or
$x_2\sim 1$.  These singularities are cancelled by terms higher order in 
perturbation theory.  However, in practice a real-life detector cannot 
resolve 3-jet events if the gluon is very soft or collinear.  The
procedure introduced by Sterman-Weinberg \cite{sterman} is 
sometimes employed.  The definition of the 3-jet fraction is then
\bea
(1-f) & = & {\sigma_{\rm 3-jet}\over \sigma_{\rm total}}\nonumber\\
& = & \int_{\epsilon,\delta}
{1\over\sigma_{\rm total}}{d^2\sigma_{\rm 3-jet}\over dx_1dx_2} \\
& = & {4\alpha_s\over 3\pi}
[4\ln{\delta}\ln{2\epsilon}+3\ln{\delta}+{\pi^2
\over 3}-{7\over 4}]\,,\nonumber
\eea
where $\epsilon$ and $\delta$ are as defined in the previous section, and $f$
represents the fraction of total cross section in which all but $\epsilon$
of the total energy is deposited into two cones of opening angle $\delta$.

An alternative approach in defining the 3-jet cross section which has
become more popular is to introduce the variable $y_{\rm cut}$.  At
the parton level for $\epem\to q(p_1)+\bar q(p_2)+g(p_3)$, one demands
that the invariant mass of any pair of final state momenta be greater
than some minimum value, \ie, $(p_i+p_j)^2/s\ge y_{\rm cut}$. With
this requirement, the integration over the infrared and collinear
divergences is rendered finite, since all jet (\ie, parton) energies
are now required to exceed $y_{\rm cut}\sqrt s$ and the opening angle
between two partons satisfies $x_ix_j(1-\cos\theta_{ij})\ge 2y_{\rm cut}$.
In this case one obtains
\be
{\sigma_{\rm 3-jet}\over\sigma_{\rm total}}=\int_{2y_{\rm cut}}^{1-y_{\rm cut}}
dx_1\int_{1+y_{\rm cut}-x_1}^{1-y_{\rm cut}}dx_2\, {1\over\sigma_{\rm total}}
{d^2\sigma_{\rm 3-jet}\over dx_1dx_2} \,.
\ee

Another property of QCD that can be verified in 3-jet events is that the
strong coupling between quarks and gluons be independent of quark flavor.
This is required by gauge invariance and renormalizability.  While
theoretical and experimental uncertainties, such as those discussed above, 
limit the absolute precision to which $\alpha_s$ can be determined in
such events, the ratio of the couplings for different quark flavors
can provide an accurate test of flavor independence as most of the
uncertainties cancel in the ratio.  The first such comparisons \cite{asflav}
were performed at PETRA with the limited precision of $\delta\alpha_s^c/
\alpha_s^{\rm all}=0.41$ and $\delta\alpha_s^b/\alpha_s^{\rm all}=0.57$;
these measurements were hampered by low statistics and poor heavy quark
tagging capabilities.  Much better accuracy has recently been obtained
at the $Z$-boson resonance due to the large available data sample and
the use of micro-vertex detectors for improved heavy quark tagging.
For example, the SLD Collaboration finds \cite{assld}
\bea
\alpha_s^{uds}/\alpha_s^{\rm all} & = & 0.987\pm 0.010(stat)^{+0.012}_{-0.010}
(sys)^{+0.009}_{-0.008}(theory)\,,\nonumber\\
\alpha_s^{c}/\alpha_s^{\rm all} & = & 1.023\pm 0.034(stat)^{+0.032}_{-0.036}
(sys)^{+0.018}_{-0.014}(theory)\,,\\
\alpha_s^{b}/\alpha_s^{\rm all} & = & 0.993\pm 0.016(stat)^{+0.020}_{-0.023}
(sys)^{+0.019}_{-0.027}(theory)\,,\nonumber
\eea
which is consistent with the strong coupling being independent of quark
flavor.

\subsection{Multi-Jet Events}

\begin{figure}[htbp]
\centerline{
\psfig{figure=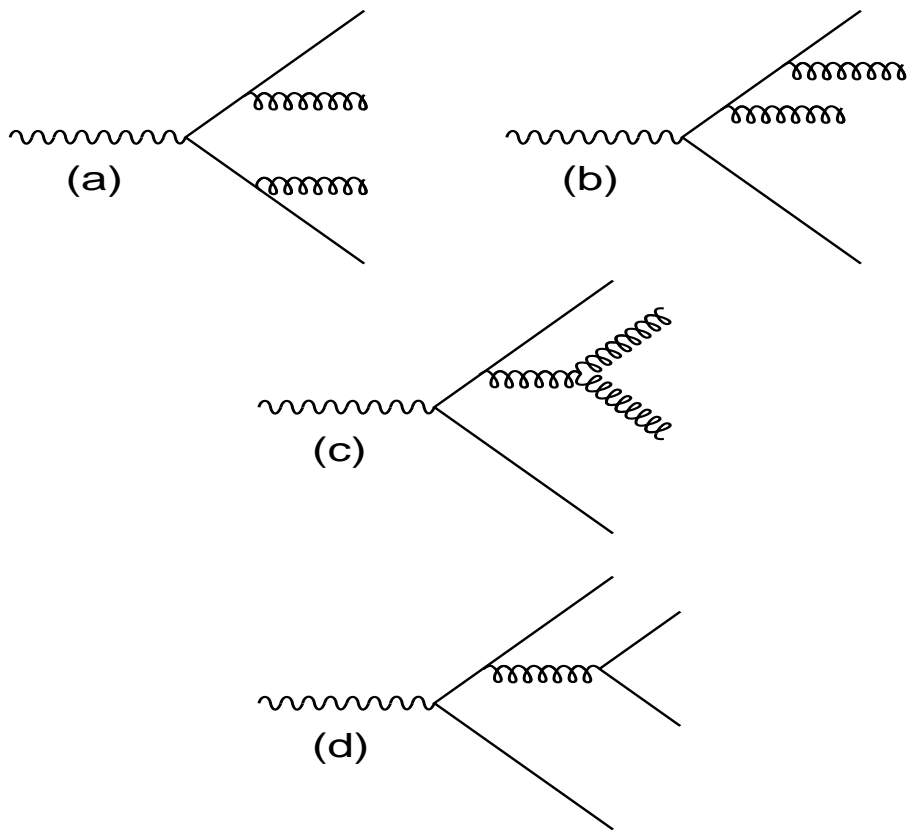,height=12cm,width=12cm,angle=0}}
\vspace*{0.5cm}
\caption{Set of tree-level diagrams responsible for 4-jet production in
\epem\ collisions.}
\label{4jet}
\end{figure}

The non-abelian triple gluon vertex and the Casimir classification of
the tree-level QCD couplings can be probed in 4-jet production in \epem\
annihilation.  The tree-level Feynman diagrams responsible for 4-jet 
production are displayed in Fig. \ref{4jet}.  Figures \ref{4jet}(a) and (b)
correspond to double bremsstrahlung diagrams, and have the characteristic
that the primary jets originating from the $q$ and $\bar q$ are the most
energetic.  Figure \ref{4jet}(d) illustrates the gluon splitting into
a quark pair; here the radiated $q\bar q$ tend to be produced along the
axis normal to the primary $q\bar q$ plane.  Figure \ref{4jet}(c)
represents the non-abelian triple gluon vertex.  In this case the gg pair
tend to be produced in the plane of the primary $q\bar q$ pair.  This
diagram has no analog in QED.

\vspace*{-0.5cm}
\nn

\begin{figure}[t]
\centerline{
\psfig{figure=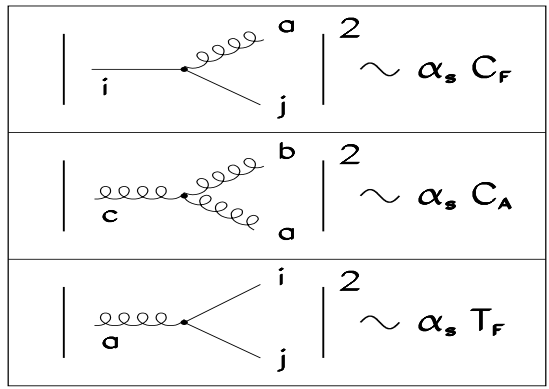,height=10cm,width=10cm,angle=0}}
\vspace*{-1.5cm}
\caption{Casimir classification of the tree-level couplings present in
QCD.}
\label{casi}
\end{figure}

In any general gauge group obeying a Lie algebra the couplings of the fermions
to the gauge fields and of the gauge self-interactions in the non-abelian
case are determined by the coupling constant and the Casimir operators of
the gauge group.  The Casimir operators are invariant operators which commute
with the generators of the gauge group.
Determination of the eigenvalues of these operators probes
the underlying structure of the theory in a gauge invariant manner.
The relations
\bea
\sum_a(T^aT^{\dagger a})_{ij} & = & 
\delta_{ij} C_F \,,\nonumber\\
\sum_{a,b}(f^{abc}f^{*abd}) & = & \delta^{cd}C_A \,,\\
\sum_{ij}T^a_{ij}T^{\dagger b}_{ji} & = & 
\delta^{ab}T_F\,, \nonumber
\eea
define the eigenvalues $C_F\,, C_A\,, T_F$, which are the color factors
of QCD.  The Casimir factors for various gauge groups, including SU(3)$_C$, are 
listed in Table \ref{castab}.  The tree-level couplings present in QCD can 
be classified in terms of the eigenvalues of the Casimir operators as 
illustrated in Fig. \ref{casi}.  Note that the direction of momentum flow is 
relevant in these diagrams.

\begin{table}
\centering
\begin{tabular}{|c|c|c|c|} \hline\hline
Group & $C_A$ & $C_F$ & $T_F$ \\ \hline
U(1) & 0 & 1 & 1\\
U(1)$_3$ & 0 & 1 & 3\\
SU(N) & $N$ & $(N^2-1)/2N$ & $1/2$\\
SU(3) & 3 & $4/3$ & $1/2$ \\ \hline
\end{tabular}
\caption{Casimir factors for some common gauge groups.  U(1)$_3$ represents
the abelian gluon model.}
\label{castab}
\end{table}

\nn

\begin{figure}[t]
\centerline{
\psfig{figure=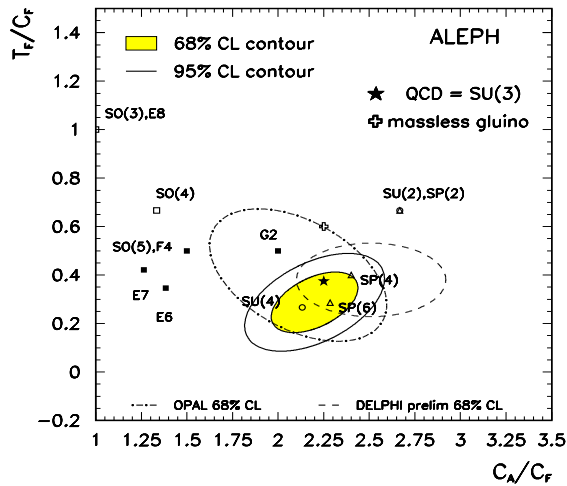,height=11cm,width=12cm,angle=0}}
\vspace*{-1.5cm}
\caption{Measurement of the QCD color factors from Ref.~\protect\cite{casdet}.}
\label{casimeas}
\end{figure}

How well can these color factors be probed in 4-jet events?  To
answer this, one must first compute the 4-jet production cross section
to a consistent order in perturbation theory.  To lowest order, this
entails computing the tree-level diagrams in Fig. \ref{4jet} as well
as the 2- and 3-jet amplitudes to a consistent order in perturbation
theory.  The terms which correspond to 4-jet production can then be
identified in a gauge consistent manner, and yield the expression
\bea
{1\over\sigma_0}d\sigma^{(4)} & = & \left( {\alpha_s C_F\over\pi}\right)^2
\left[ F_A +\left(1-{C_A\over 2C_F}\right)F_B+{C_A\over C_F} F_C
\right.\nonumber\\
& & \quad \left. + {T_F\over C_F}N_fF_D+\left( 1-{C_A\over 2C_F}\right)
F_E\right] \,,
\eea
where $F_{A...E}$ are kinematical functions.  We note that the complete 
next-to-leading (NLO) 4-jet production cross section has
recently been computed \cite{lance} and that it is in excellent agreement
with experiment.  We see that the kinematical
distributions of the above cross section depend on the ratios $C_A/C_F$ and
$T_F/C_F$, as well as the number of flavors $N_f$, 
which in principle can then be determined by experiment.
In fact, numerous 4-jet event shape observables have been proposed which
are sensitive to these ratios of color factors.  A determination \cite{casdet}
of these ratios by ALEPH is presented in Fig. \ref{casimeas}; also shown
in the figure are the expectations from several potential gauge groups.
We see that only the groups SU(3), SU(4), SP(4), and SP(6) are
consistent with the data.  However, the latter three groups, SU(4), SP(4), and 
SP(6), do not contain 3 degrees of freedom for the quark color representation, 
leaving SU(3) as the only viable gauge theory for QCD.  We also note
that the observation of a nonzero value of $C_A/C_F$ provides evidence
for the existence of the gluon three-point function.  In order that
the theory of QCD be gauge-invariant and self-consistent, this implies that
the quartic gluon self-interaction must also exist.

\nn

However, it should come as no surprise that direct experimental verification 
of the quartic gluon
vertex requires 5-jet production in \epem\ annihilation!  Performing
a computation of the 5-jet cross section similar to that above for 4jets,
which is gauge invariant and consistent in perturbation theory, schematically 
yields
\be
{1\over\sigma_0}d\sigma^{(5)} = {1\over\sigma_0}d\sigma^{2q3g}+
{1\over\sigma_0}d\sigma^{4q1g}\,.
\ee
The first term contributes approximately $85\%$ of the total cross section
and can be written as
\be
{1\over\sigma_0}d\sigma^{2q3g}=\left( {\alpha_sC_F\over\pi}\right)^3
\left[ G_A +{C_A\over C_F}G_B+\left( {C_A\over C_F}\right)^2G_C
\right] \,,
\ee
where $G_{A,B,C}$ are again kinematical functions.  It is the last term in this
expression which corresponds to the quartic gluon coupling.  In principle
a set of 5-jet event shape observables can be analogously defined, however,
the data is not yet sufficient to show statistically significant evidence
for the quartic gluon term.\cite{casdet}

\subsection{Renormalization of the Strong Coupling and Asymptotic Freedom}

When evaluating processes beyond tree-level it is necessary to define a 
renormalized coupling constant.  Loop-level diagrams contain divergent
integrals over the loop momenta which must be regularized by some
procedure that re-expresses the divergences in a well defined manner.  The
regularized divergences are then removed by absorbing them into the
definitions of physical quantities via renormalization.  This procedure is
performed by a specified prescription, of which many are available, and
introduces a new scale, $\mu$.  All renormalized quantities depend on this
scale, however different prescriptions must lead to the same observable
amplitudes.

As an example, consider a single particle irreducible Green's function $\Gamma$,
which cannot be disconnected by a cut on any single internal line.  The
introduction of an ultra-violet cut-off, $\Lambda$, in the loop momentum
integrals controls the divergences and yields the unrenormalized Green's
function $\Gamma_U(p_i,g_0,\Lambda)$, where $p_i$ represents the momenta of
the external particles and $g_0$ is the bare coupling.  In a renormalizable
theory it is possible to define a renormalized Green's function
\be
\Gamma_R(p_i,g,\mu)=Z_\Gamma(g_0,\Lambda/\mu)\Gamma_U(p_i,g_0,\Lambda)\,,
\ee
such that $\Gamma_R$ is then finite in the limit $\Lambda\to\infty$, but
depends on the prescription parameter $\mu$ and the renormalized coupling $g$.
The function $Z_\Gamma$ is a product of renormalization factors, with one 
factor for each external particle.  Now, the unrenormalized Green's function
is independent of $\mu$ and thus
\be
{d\Gamma_U\over d\mu}=0\,,
\ee
giving
\be
\left( \mu{\partial\over\partial\mu}+\beta{\partial\over\partial g}+
\gamma \right) \Gamma_R(p_i,g,\mu)=0\,,
\label{rge}
\ee
with $\Lambda$ being held constant and then taking the subsequent limit
$\Lambda\to\infty$.
Here the beta-function $\beta(g)$ and the anomalous dimension $\gamma$ are
defined by
\be
\beta=\mu{\partial g\over\partial\mu}\,, \quad\quad\quad \gamma={\mu\over
Z_\Gamma}{\partial Z_\Gamma\over\partial\mu} \,.
\ee
The beta-function is universal, but the anomalous dimension clearly depends
on the Green's function.  In the case where there is a single large momentum
scale, $Q$, all momenta $p_i$ can be expressed as a fixed fraction $x_i$ of $Q$.
Then introducing a momentum dependent, or `running'
coupling via the integral equation
\be
t\equiv \int_{g(0)}^{g(t)} {dg'\over\beta(g')}\,,
\ee
where $t=(1/2)\ln(Q^2/\mu^2)$, yields the general solution
\be
\Gamma(t,g(0),x_i)=\Gamma(0,g(t),x_i)\exp\int_{g(0)}^{g(t)}dg'{\gamma(g')\over
\beta(g')}
\ee
to the Renormalization Group Equation (RGE) of Eq. (\ref{rge}).
This explicitly demonstrates that the entire dependence of $\Gamma$ on the
scale $Q$ arises through the running coupling $g(t)$.

The renormalization scale dependence of the QCD effective coupling $\alpha_s=
g^2/4\pi$ is determined by the beta-function.  Expanding this function in a
power series in $\alpha_s$ gives
\be
\beta=\mu{\partial\alpha_s\over\partial\mu}=
-{\beta_0\over 2\pi}\alpha_s^2-{\beta_1\over 4\pi^2}\alpha_s^3
-{\beta_2\over 64\pi^3}\alpha_s^4-\cdots
\label{betaeq}
\ee
with
\bea
\beta_0 & = & 11-{2\over 3}N_f \,,\nonumber\\
\beta_1 & = & 51-{19\over 3}N_f \,,\\
\beta_2 & = & 2857-{5033\over 9}N_f+{325\over 27}N_f^2 \,,\nonumber
\label{betas}
\eea
where $N_f$ is the number of flavors with mass less than the scale $\mu$.
The solution to the differential equation (\ref{betaeq}) introduces a
constant of integration; this constant is
scheme (or prescription) dependent and must be determined from experiment.
The conventional prescription choice \cite{pdg98} for QCD is the modified
minimal subtraction scheme, $\overline{\rm MS}$, (which is discussed in
the next section) and results in (to third order)
\bea
\alpha_s(\mu) & = & {4\pi\over\beta_0\ln(\mu^2/\Lambda^2)}\left[ 1-
{2\beta_1\over\beta_0^2}{\ln[\ln(\mu^2/\Lambda^2)]\over\ln(\mu^2/\Lambda^2)}
+{4\beta_1^2\over\beta_0^4\ln^2(\mu^2/\Lambda^2)}\right. \nonumber\\
& & \quad \left. \times\left( \left[\ln[\ln(\mu^2/\Lambda^2)]-{1\over 2}
\right]^2+{\beta_2\beta_0\over 8\beta_1^2}-{5\over 4}\right)\right] \,.
\label{asyf}
\eea
This equation demonstrates the principle of asymptotic freedom, that is that
the running coupling $\to 0$ as $\mu\to\infty$.  This property allows for
the RGE-improved perturbative calculations at large values of $\mu$.  We
note that the expression for $\beta_2$ above (39) is scheme
dependent and assumes $\overline{\rm MS}$.

We stress again that all physical quantities or observables are independent
of the renormalization scheme.  However, our calculations are truncated
at some order in perturbation theory, and this termination of the perturbation
series introduces a residual renormalization scheme dependence.  The magnitude
of this leftover scheme dependence can be sizeable since the expansion
parameter, $\alpha_s$, is large.  This introduces a source of uncertainty
in comparing QCD predictions to data, which can only be reduced by higher
order computations.

\begin{figure}[htbp]
\centerline{
\psfig{figure=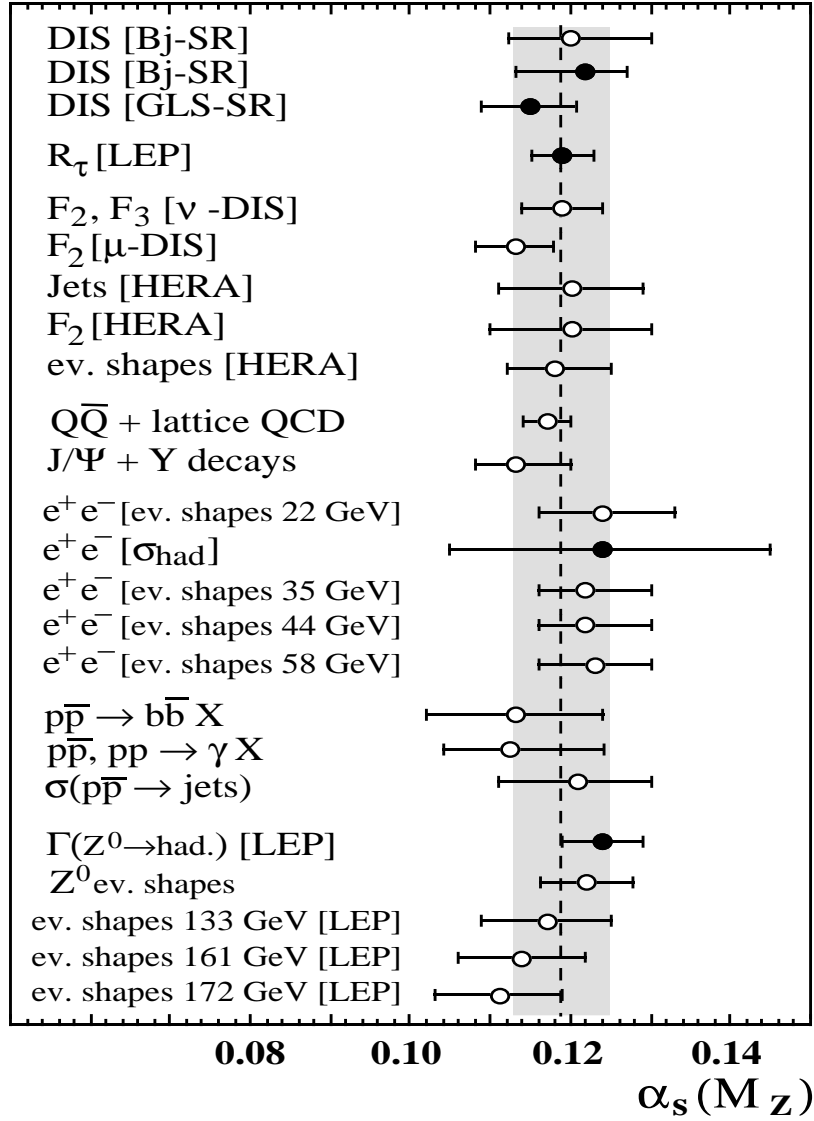,height=15cm,width=11cm,angle=0}}
\vspace*{0.5cm}
\caption{Summary of the various determinations of $\alpha_s$, evaluated
at the scale $\mu=M_Z$.  From Bethke~\protect\cite{phil}.}
\label{alphas}
\end{figure}

\begin{figure}[htbp]
\centerline{
\psfig{figure=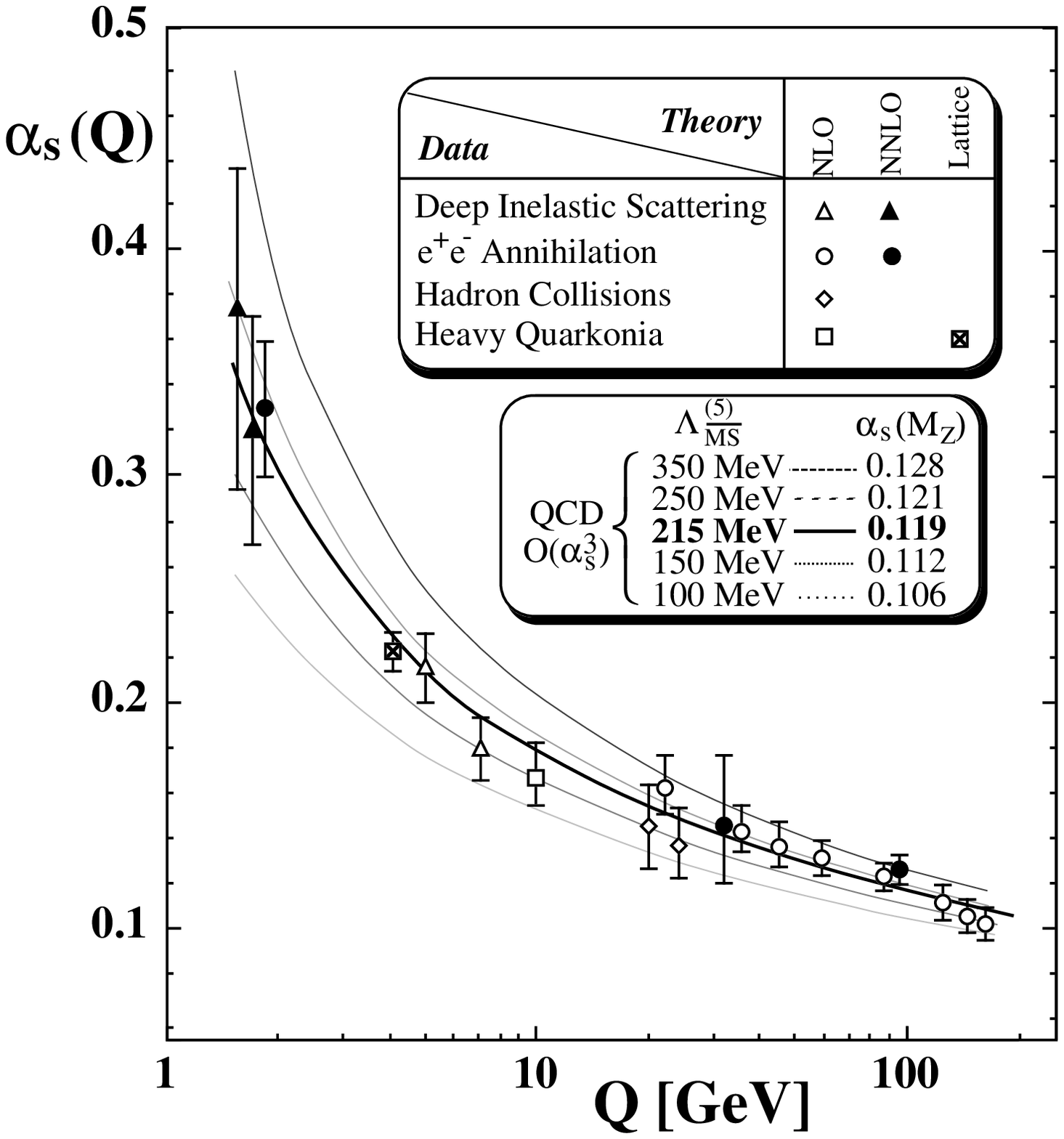,height=12cm,width=12cm,angle=0}}
\vspace*{0.5cm}
\caption{Demonstration of the property of asymptotic freedom via comparison
of data from various processes as labeled with the theoretical predictions
for several values of $\Lambda^{(5)}_{\overline{\rm MS}}$ as labeled.
The best fit is indicated by the thick solid line.  
From Bethke~\protect\cite{phil}.}
\label{asympf}
\end{figure}

A quantitative test of QCD and the property of asymptotic freedom is given
by the measurement of $\alpha_s$ in a variety of processes at numerous
scales $Q$.  Determinations of $\alpha_s$ have been obtained from tests of
sum rules in low-energy Deep Inelastic Scattering (DIS), hadronic $\tau$
decays, neutrino DIS, high-energy DIS at HERA, lattice QCD calculations,
heavy quarkonium decays, hadronic cross sections and event shape observables
in \epem\ annihilation, and jet, prompt photon, and $b\bar b$ production in
hadronic collisions.  The corresponding scales for these processes range
from $1\lsim Q\lsim 500 $ GeV and the theoretical computation for each
process has been performed at least to order NLO, and
in some cases to (next-)next-to-leading order (NNLO), in perturbation theory.
A review of the procedure by which $\alpha_s$ is extracted in each case,
as well as a discussion of the associated theoretical errors is given in
the Particle Data Book.\cite{pdg98}  An up-to-date summary of this 
information from the 1998 summer conferences \cite{qcdvan} is reproduced
here in Table \ref{altab}.  We that in many cases, the dominant source of
error is theoretical.  This is related to the residual scheme dependence
discussed above, and can be resolved only by performing higher order
calculations.  Due to the large data sample available at the
$Z$ boson resonance, it has become conventional to use the scale $Q=M_Z$ as
the standard candle by which to compare the various measurements.  This
is shown graphically in Fig. \ref{alphas}.  It is apparent that the 
measurements are all consistent within the uncertainties.  The present world 
average value of $\alpha_s(M_Z)$ is
\be
\alpha_s(M_Z) = 0.1190\pm 0.0058\,,
\ee
which implies
\bea
\Lambda^{(5)}_{\overline{\rm MS}} & = & 220^{+78}_{-63} \, {\rm MeV} \,,
\nonumber\\
\Lambda^{(4)}_{\overline{\rm MS}} & = & 305^{+94}_{-79} \, {\rm MeV} \,.
\eea
Finally, demonstration that the property of asymptotic freedom is experimentally
verified is given in Fig. \ref{asympf} from Bethke.\cite{phil}

\begin{table}
\centering
\begin{tabular}{|c|c|c|c|c|c|c|} \hline\hline
Process & $Q$ & $\alpha_s(Q)$ & $\alpha_s(M_Z)$ & 
\multicolumn{2}{c}{$\Delta\alpha_s(M_Z)$} & Theory \\ 
 & GeV & & & expt. & theor. & \\ \hline
DIS [pol. strct. fun.] & $0.7-8$ & & $0.120^{+0.010}_{-0.008}$ &
$~^{+0.004}_{-0.005}$ & $~^{+0.009}_{-0.006}$ & NLO \\
DIS [Bj-SR] & $1.58$ & $0.375^{+0.062}_{-0.081}$ & $0.121^{+0.005}_{-0.009}$ &
-- & -- & NNLO \\
DIS [GLS-SR] & $1.73$ & $0.295^{+0.092}_{-0.073}$ & $0.114^{+0.010}_{-0.012}$ &
$~^{+0.005}_{-0.006}$ & $~^{+0.009}_{-0.010}$ & NNLO \\
$\tau$ Decays & $1.78$ & $0.339\pm 0.021$ & $0.120\pm 0.003$ &
$0.001$ & $0.003$ & NNLO \\
DIS [$\nu$; $F_2$ and $F_3$] & $5.0$ & $0.215\pm 0.016$ & $0.119\pm 0.005$ &
$0.002$ & $0.004$ & NLO \\
DIS [$\mu$; $F_2$] & $7.1$ & $0.180\pm 0.014$ & $0.113\pm 0.005$ &
$0.003$ & $0.004$ & NLO \\
DIS [HERA; $F_2$] & $2-10$ & & $0.120\pm 0.010$ & 
$0.005$ & $0.009$ & NLO \\
DIS [HERA; jets] & $10-100$ & & $0.118\pm 0.008$ &
$0.003$ & $0.008$ & NLO \\
DIS [HERA; ev. shps] & $7-100$ & & $0.118^{+0.007}_{-0.006}$ &
$0.001$ & $~^{+0.007}_{-0.006}$ & NLO \\
$Q\bar Q$ states & $4.1$ & $0.223\pm 0.009$ & $0.117\pm 0.003$ &
$0.000$ & $0.003$ & LGT \\
$\Upsilon$ Decays & $4.13$ & $0.220\pm 0.027$ & $0.119\pm 0.008$ &
$0.001$ & $0.008$ & NLO \\
\epem\ [$\sigma_{\rm had}$] & $10.52$ & $0.20\pm 0.06$ & 
$0.130^{+0.021}_{-0.029}$ & $~^{+0.021}_{-0.029}$ & -- & NNLO \\
\epem\ [ev. shapes] & $22.0$ & $0.161^{+0.016}_{-0.011}$ & 
$0.124^{+0.009}_{-0.006}$ & $0.005$ & $~^{+0.008}_{-0.003}$ & resum \\
\epem\ [$\sigma_{\rm had}$] & $34.0$ & $0.146^{+0.031}_{-0.026}$ & 
$0.123^{+0.021}_{-0.019}$ & $~^{+0.021}_{-0.019}$ & -- & NLO \\
\epem\ [ev. shapes] & $35.0$ & $0.145^{+0.012}_{-0.007}$ & 
$0.123^{+0.008}_{-0.006}$ & $0.002$ & $~^{+0.008}_{-0.005}$ & resum \\
\epem\ [ev. shapes] & $44.0$ & $0.139^{+0.010}_{-0.007}$ & 
$0.123^{+0.008}_{-0.006}$ & $0.003$ & $~^{+0.007}_{-0.005}$ & resum \\
\epem\ [ev. shapes] & $58.0$ & $0.132\pm 0.008$ & 
$0.123\pm 0.007$ & $0.003$ & $0.007$ & resum \\
$p\bar p\to b\bar bX$ & $20.0$ & $0.145^{+0.018}_{-0.019}$ & 
$0.113\pm 0.011$ & $~^{+0.007}_{-0.006}$ & $~^{+0.008}_{-0.009}$ & NLO \\
$p\bar p,\, pp\to\gamma X$ & $24.2$ & $0.137^{+0.017}_{-0.014}$ & 
$0.111^{+0.012}_{-0.008}$ & $0.006$ & $~^{+0.010}_{-0.005}$ & NLO \\
$\sigma(p\bar p\to {\rm jets})$ & $30-500$ & & 
$0.121\pm 0.009$ & $0.001$ & $0.009$ & NLO \\
\epem\ [$\Gamma(Z\to {\rm had})$] & $91.2$ & $0.122\pm 0.004$ & 
$0.122\pm 0.004$ & $0.004$ & $0.003$ & NNLO \\
\epem\ [ev. shapes] & $91.2$ & $0.122\pm 0.006$ & 
$0.122\pm 0.006$ & $0.001$ & $0.006$ & resum \\
\epem\ [ev. shapes] & $133.0$ & $0.111\pm 0.008$ & 
$0.117\pm 0.008$ & $0.004$ & $0.007$ & resum \\
\epem\ [ev. shapes] & $161.0$ & $0.105\pm 0.007$ & 
$0.114\pm 0.008$ & $0.004$ & $0.007$ & resum \\
\epem\ [ev. shapes] & $172.0$ & $0.102\pm 0.007$ & 
$0.111\pm 0.008$ & $0.004$ & $0.007$ & resum \\
\epem\ [ev. shapes] & $183.0$ & $0.109\pm 0.005$ & 
$0.121\pm 0.006$ & $0.002$ & $0.006$ & resum \\\hline
\end{tabular}
\caption{World summary of measurements of $\alpha_s$.  Abbreviations:
DIS$=$ deep inelastic scattering; GLS-SR$=$Gross-Llewellyn-Smith sum rules;
Bj-SR$=$ Bjorken sum rules; (N)NLO$=$(next)next-to-leading order;
LGT$=$ lattice gauge theory; resum$=$ resummed NLO. 
From~\protect\cite{qcdvan}.}
\label{altab}
\end{table}

\vspace*{-0.5cm}
\nn

\vspace*{-0.5cm}
\nn

\subsection{Inclusive Jet Cross Sections in Hadronic Collisions}

\begin{figure}[t]
\centerline{
\psfig{figure=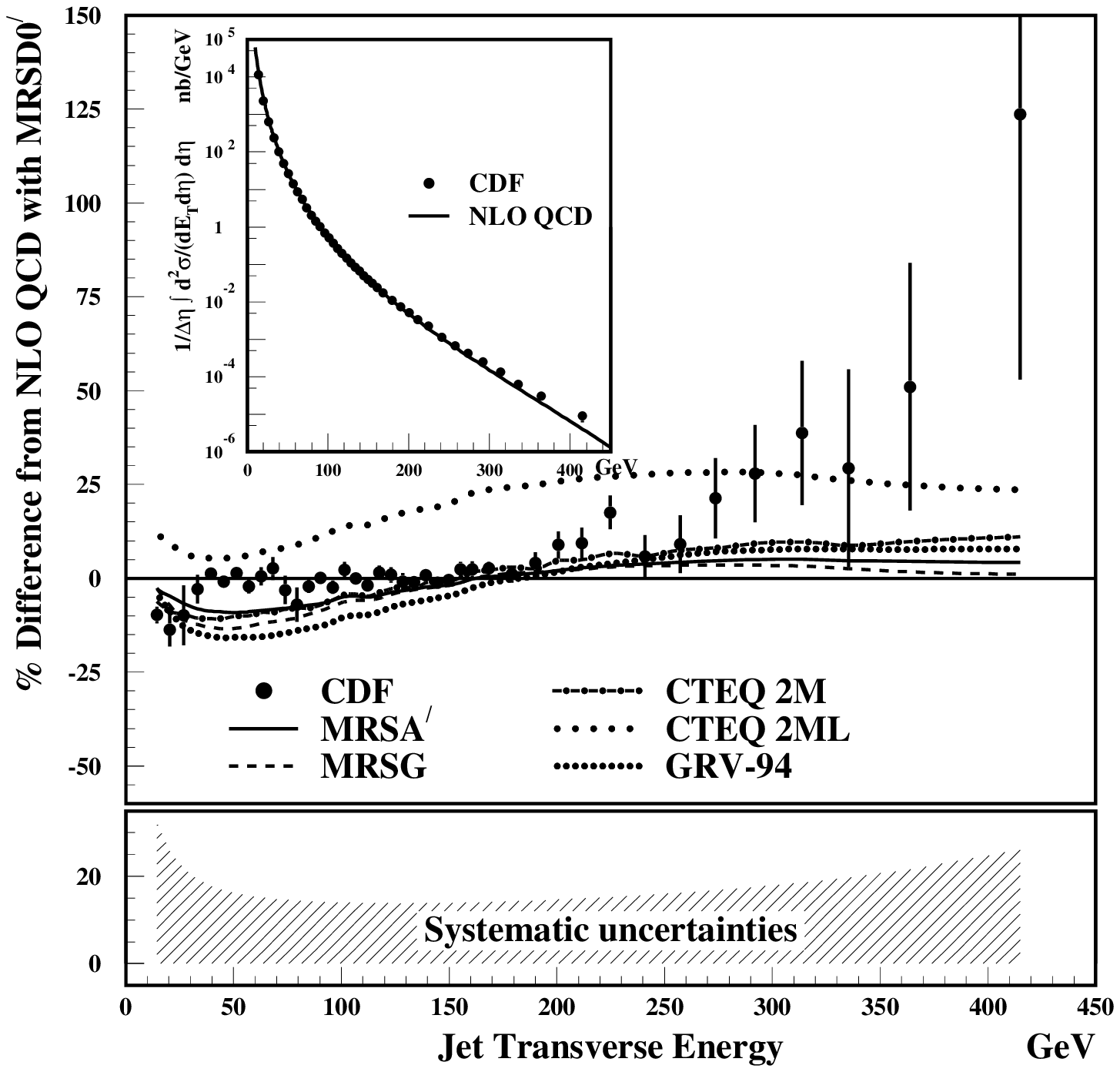,height=12cm,width=12cm,angle=0}}
\vspace*{-3.5cm}
\caption{Percentage difference between the CDF inclusive jet cross section
and the predictions of NLO QCD for various parton densities as indicated.
The error bars represent the uncorrelated uncertainties, while the shaded
region at the bottom indicates the quadratic sum of the correlated
systematic errors.  The differential cross section is shown in the insert.
From~\protect\cite{inclujet}.}
\label{hadrons}
\end{figure}

Hadronic collisions have also provided several valuable tests of QCD, but
a detailed review \cite{hadrev} will not be given here.  Perhaps the most
impressive confirmation of QCD is given by the spectacular agreement of the
inclusive jet cross section with the NLO predictions of QCD.  Numerous
parton-level subprocesses, involving all the interactions present in the
QCD Lagrangian, contribute to this cross section.  At leading order, this
cross section is described by all $2\to 2$ QCD reactions initiated with
$gg\,, g^(\bar q^)\,, q^(\bar q^)$ scattering, while higher order QCD 
subprocesses give more elaborate configurations of partons.   
The radiative corrections
have been computed \cite{nlojet} to NLO and the $O(\alpha_s^2)$ parton cross 
section is convoluted with the initial parton distributions.  The inclusive
differential cross section for jet production measured by CDF \cite{inclujet}
for jet transverse energies from 15 to 440 GeV in the central pseudorapidity
region $0.1\le |\eta| \le 0.7$ is displayed in the insert of Fig. \ref{hadrons}.
The pseudorapidity of a final state particle or jet is defined as 
$\eta\equiv -\ln\tan(\theta/2)$ with $\cos\theta =p_z/p$.
We see that the agreement between theory and experiment is very good,
both in the shape and the normalization of the distribution, spanning over
many orders of magnitude of the falling cross section.  This agreement
establishes the QCD Lagrangian of Eq. (\ref{qcdla}) as the correct theory of
the strong interactions at perturbative energies.  It also illustrates the
existence of all the initial scattering states, directly showing that
the gluon gauge symmetry is non-abelian, and demonstrates the importance of 
the $Q^2$ evolution of the parton density functions.  The remainder of the
figure shows the percentage difference between the CDF measurement and the
expectations of NLO QCD, for various parameterizations of the parton densities,
as a function of the jet transverse energy.  The error bars represent
uncertainties uncorrelated from point to point.  The apparent disagreement
at high values of the jet transverse energy has not been explicitly corroborated
by measurements from D0,\cite{d0jet} and can be easily explained by 
modifying the gluon distribution inside the proton.\cite{wuki}

\nn

\section{QED}

Quantum electrodynamics is a renormalizable local gauge theory.  The QED
Lagrangian is invariant under the local gauge transformation
\bea
\psi(x) & \to & \exp[ieQ\Lambda(x)]\psi(x)\,,\nonumber\\
A_\mu(x) & \to & A_\mu(x)+\partial\Lambda(x)/\partial x_\mu \,,
\eea
with $\Lambda(x)$ being arbitrary and the phase factor originating from the
symmetry group U(1) with generator $Q$.  The photon must remain massless to
preserve exact gauge invariance.  The electromagnetic current (12)
is conserved, \ie, $\partial_\mu J^\mu_{em}=0$, implying conservation of
electric charge.  The success of QED as the first and simplest
gauge theory made 
it a starting point in forming gauge theories for the strong and weak
interactions.

\subsection{Renormalization of the QED Coupling}

As discussed in the introduction, 
in addition to the boson and fermion masses and CKM mixings, the SM has
three {\it a priori} free parameters which are inserted into the theory and
must be used as input into all
calculations.  Due to their precise experimental determination two of these
are taken to be $\alpha_{QED}\equiv e^2/4\pi$ and $G_F$, which play 
fundamental roles in many tests of the SM.  The fact that QED is renormalizable
and hence $\alpha$ can be treated as a `running' coupling is essential
in analyzing SM radiative corrections.  
It is instructive to briefly review this concept.

\begin{figure}[htbp]
\centerline{
\psfig{figure=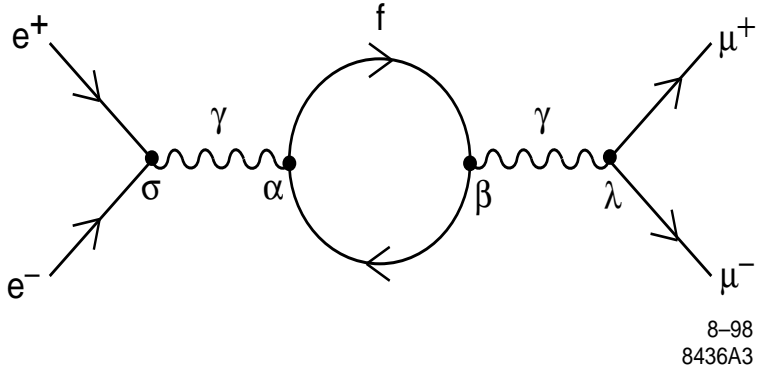,height=5cm,width=10cm,angle=0}}
\vspace*{-0.5cm}
\caption{Feymann diagram for the photon self-energy.}
\label{gamself}
\end{figure}

\begin{figure}[htbp]
\centerline{
\psfig{figure=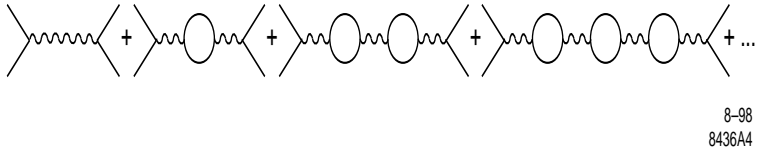,height=2cm,width=10cm,angle=0}}
\vspace*{-0.5cm}
\caption{Bubble string contributing to the photon self-energy.}
\label{bubble}
\end{figure}

Consider the process $\epem\to\gamma\to\mu^+\mu^-$ at $s=q^2\gg m_e^2,
m_\mu^2$.  At lowest order the amplitude can be written schematically as
\be
A_0 \sim e^2J^e_\sigma\cdot {-ig^{\sigma\lambda}\over q^2}\cdot J^\mu_\lambda 
\,,
\ee
where $J^{e,\mu}$ are the conserved electron and muon electromagnetic currents.
At the next order in perturbation theory there is a correction due to the 
1-loop vacuum polarization (or photon self-energy) diagram of 
Fig. \ref{gamself}.  This yields the 1-loop amplitude
\be
A_1 \sim e^2J_\sigma^e{-ig^{\sigma\alpha}\over q^2} \Pi_{\alpha\beta}(q^2)
{-ig^{\beta\lambda}\over q^2} J_\lambda^\mu\,,
\ee
where
\be
\Pi_{\alpha\beta}(q^2)=-e^2Q^2_f\int {d^nk\over(2\pi)^n}(\mu^2)^{2-n/2}
{Tr[\gamma_\alpha(\slash k+m_f)\gamma_\beta(\slash k - \slash q+m_f)]
\over (k^2-m^2_f)\left( (k-q)^2-m_f^2\right) } 
\ee
using dimensional regularization with $k$ representing the loop momenta of
the fermion.  Here the log divergent integral over $k$ has been regulated
by performing the integration in $n$ dimensions and the factor $\mu^(2-n/2)$
has been introduced to keep $e$ dimensionless in $n$ dimensions.
Due to electromagnetic gauge invariance we can write this as
\be
\Pi_{\alpha\beta}(q^2)=-ie^2(q_\alpha q_\beta-g_{\alpha\beta}q^2)\Pi(q^2)
\ee
so that the scattering amplitude becomes
\be
A = A_0+A_1\sim e^2[1+e^2\Pi(q^2)]J^e_\sigma {-ig^{\sigma\lambda}\over q^2}
J^\mu_\lambda \,.
\ee
The bubble string of a series of vacuum polarization contributions depicted in 
Fig. \ref{bubble} forms a geometric series and can be resummed so that
\be
A = {e^2\over 1-e^2\Pi(q^2)} J^e_\sigma {-ig^{\sigma\lambda}\over q^2}
J^\mu_\lambda \,,
\ee
and the factor $e^2/[1-e^2\Pi(q^2)]$ can be thought of as an effective
charge.  However, $\Pi(q^2)$ is divergent and must be renormalized
before using this definition.  One explicitly obtains
\bea
\Pi(q^2)& = & {Q^2_f\over 12\pi^2}\left\{ {1\over 2-n/2}+\ln 4\pi-\gamma_E
\right.\nonumber\\
& & \quad\quad\quad \left. +6\int^1_0 dx~x(1-x)\ln\left[ 
{\mu^2\over m_f^2-q^2x(1-x)}\right]\right\}\,.
\label{pisq}
\eea
Here the first term clearly diverges as $n\to 4$ and $\gamma_E$ denotes the
Euler-Mascheroni constant.

\nn

\nn

There are several renormalization schemes that can be employed.  The two
most popular are the on-shell scheme (OS) (upon which we concentrate here)
and the $\overline{\rm MS}$ scheme.  In the OS scheme $e^2$ is defined
to be the quantity determined by low-energy experiments such as Thompson
scattering and the Josephson Junction \cite{pdg98} or Quantum Hall 
Effect.\cite{Hall}  In this case $\Pi(q^2)$ is renormalized such that
$\Pi(q^2=0)=0$, giving
\be
\Pi_{\rm OS}(q^2)={Q^2_f\over 2\pi^2}\int^1_0 dx~x(1-x)\ln \left[
{m_f^2\over m_f^2-q^2x(1-x)}\right] \,.
\ee
Note that if the fermion is heavy, $4m_f^2\gg q^2$, then $\Pi_{\rm OS}$
exhibits decoupling, \ie, contributions to the integral are highly
suppressed by powers of $q^2/m_f^2$.  In this case we obtain
\be
\Pi_{\rm OS}(q^2\ll 4m_f^2)\simeq {-Q_f^2\over 60\pi^2}\left[ {q^2\over
m_f^2} +{3\over 28}{q^4\over m_f^4} \right]\,.
\label{smallq}
\ee
In the opposite limit of $q^2\gg 4m_f^2$ one must be careful if $q^2$ is
time-like as $\Pi_{\rm OS}$ develops an imaginary, or absorptive, part in
this case.  However, the real part can be written as
\be
\Re\Pi_{\rm OS}(q^2\gg 4m_f^2)\simeq {Q^2_f\over 12\pi^2}\left[
\ln{q^2\over m_f^2}-{5\over 3}-6{m_f^2\over q^2} +\cdots\right] \,.
\ee
Since $e^2$ is real, convention dictates that only $\Re\Pi(q^2)$ is used to 
define the renormalized coupling.  For example, taking $M_Z^2=q^2\gg 4m_f^2$,
the fermion contribution is
\be
\alpha_{\rm OS}(M_Z^2)= {\alpha(0)\over 1-\Delta\alpha_{\rm OS}(M_Z^2) }
\ee
with
\be
\Delta\alpha_{\rm OS}(M_Z^2)\simeq{\alpha\over 3\pi}Q_f^2\left[\ln{M_Z^2\over 
m_f^2}-{5\over 3}-6{m_f^2\over M_Z^2}\right]\,.
\ee
Similarly, in keeping with the philosophy of the OS scheme and performing
perturbative expansions in terms of physical observables (such as $\alpha$),
the renormalized masses of particles in this
scheme correspond to the positions of propagator poles and are the
actual physical masses of the particles.

In contrast, in the $\overline{\rm MS}$ scheme one subtracts the
$1/(2-n/2)+\ln 4\pi-\gamma_E$ piece in $\Pi(q^2)$ leaving a logarithmically
dependent quantity, as can be seen from Eq. (\ref{pisq}).  This is then 
absorbed into the definition of the running
coupling.  In this case, $\alpha_{\overline{\rm MS}}(\mu^2)$ and the 
corresponding running masses $m_f(\mu^2)$ are not directly related to 
physical quantities but are more easily dealt with in perturbation theory.

\begin{figure}[htbp]
\centerline{
\psfig{figure=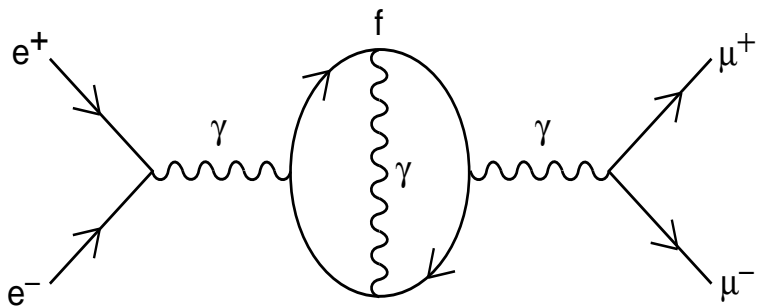,height=5cm,width=10cm,angle=0}}
\vspace*{-0.5cm}
\caption{Higher order QED corrections to the photon vacuum
polarization.}
\label{twoloops}
\end{figure}

Before proceeding further we note that the
summing of the geometric bubble series above leads to a summation
of all of the large logarithms of the form $(\alpha\ln q^2)^n$ that appear in
$\Delta\alpha(M_Z^2)$.  This is known as the leading-log approximation (LLA).
To go beyond this approximation two-loop (and higher order) graphs, such
as that depicted in Fig. \ref{twoloops}, need to be considered.  
These higher order
diagrams lead to sub-leading terms of order $\alpha(\alpha\ln q^2)^n$
in $\Delta\alpha(M_Z^2)$.  The summation of these terms provides the
next-to-leading logarithmic (NLL) estimate of $\Delta\alpha(M_Z^2)$.

\nn

Given the results above, it would appear that a calculation of $\Delta\alpha
(M_Z^2)$ in the OS scheme would be rather straightforward.  It is
instructive, however, to decompose $\Delta\alpha$ in terms of the various
loop contributions
given by $\Delta\alpha_{lept}\,, \Delta\alpha_{had}^{(5)}$, and
$\Delta\alpha_{top}$.  For leptons, which only have electroweak interactions,
it is straightforward to obtain
\be
\Delta\alpha_{lept}(M_Z^2)={\alpha\over 3\pi}\sum_\ell
\left[\ln{M_Z^2\over m_\ell^2}-{5\over 3}-6{m_\ell^2\over M_Z^2}
+O\left({m_\ell^4\over M_Z^4}\right)\right] +O(\alpha^2)+O(\alpha^3)\,,
\ee
where both the $O(\alpha^2)$ and $O(\alpha^3)$ terms have been recently
computed,\cite{KS} resulting in
\be
\Delta\alpha_{lept}(M_Z^2)=314.97686\times 10^{-4}.
\ee
For the top-quark contribution, the QCD corrections are expected to be 
relatively small
and calculable in perturbation theory due to the large top-quark mass.  
Rescaling the result in Eq. (\ref{smallq}) for the $4m_f^2\gg q^2$ limit,
taking $Q_t=2/3$, and noting
that the top-quark is a color triplet, we obtain 
\be
\Delta\alpha_{top}(M_Z^2)={-4\alpha M_Z^2\over 45\pi m_t^2}\left[
a+b{3\over 28}{M_Z^2\over m_t^2}+\cdots\right]\,,
\ee
with the coefficients $a, b$ representing a power series in the QCD coupling.
These coefficients are now evaluated through $O(\alpha_s^2)$,\cite{CKS} so that
\be
\Delta\alpha_{top}(M_Z^2)\simeq (-0.71\pm 0.05)\times 10^{-4}
\ee
for $m_t=173.8\pm 5.0$ GeV.

\begin{figure}[htbp]
\centerline{
\psfig{figure=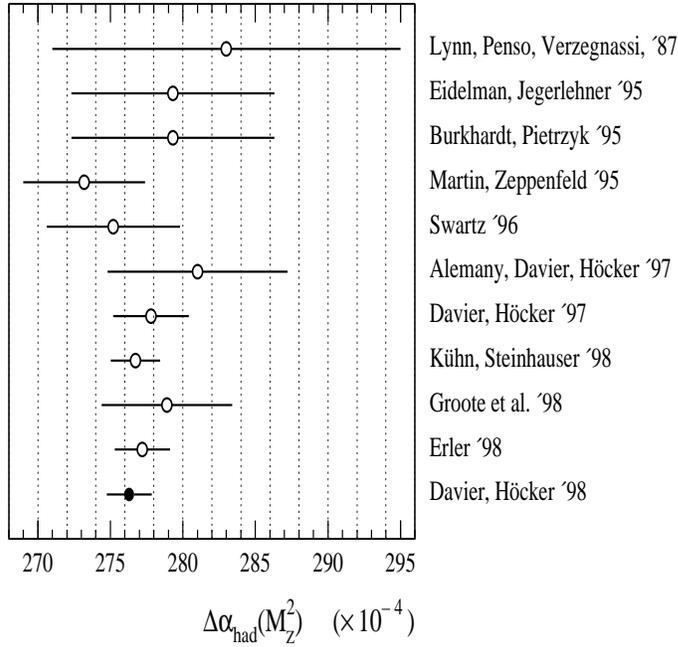,height=9cm,width=12cm,angle=0}}
\vspace*{0.5cm}
\caption{Summary of recent analyses for $\Delta\alpha_{had}^{(5)}$.
From~\protect\cite{DH}.}
\label{alphahad}
\end{figure}

In computing the corrections for the light quark (u, d, s, c, b) 
contributions, $\Delta\alpha_{had}^{(5)}$, the problem arises that
perturbation theory is no
longer reliable since scales of order 1 GeV or less are involved.  To
handle this situation, dispersion relations are employed to obtain
\be
\Delta\alpha_{had}^{(5)}={-\alpha M_Z^2\over 3\pi}\Re\int^\infty_{4m_\pi^2}
ds{R(s)\over s(s-M_Z^2)-i\epsilon} \,,
\ee
where $R(s)$ represents the ratio of hadronic to $\mu$ pair cross sections
in \epem\ annihilation, as discussed in section 2.1.  As we will
see in the next section, a similar situation occurs for the 
hadronic contribution to the anomalous $(g-2)$ of the electron or muon.
In this case one finds
\be
a_\mu^{had}={\alpha^2\over 3\pi^2}\Re\int^\infty_{4m_\pi^2} ds
{K(s)\over s}R(s)\,,
\ee
where $K(s)$ is a known QED determined weight function.
Over the last three years, a large effort has been performed
on modelling $R(s)$ using both low-energy and $\tau$ decay data as well as
sophisticated QCD calculations.  A summary, from \cite{DH}, of the results from
the various analyses is presented in Fig. \ref{alphahad}.
The most recent analyses by Davier and
H\" ocker \cite{DH} and by K\" uhn and Steinhauser \cite{KS} give
essentially identical results:
\bea
\Delta\alpha_{top}+\Delta\alpha_{had}^{(5)}& = & 
(276.3\pm 1.6)\times 10^{-4}\quad \mbox{[DH]}\,,\\
& = & (277.4\pm 1.7)\times 10^{-4}\quad \mbox{[KS]}\,.\nonumber
\eea
When combined with $\Delta\alpha_{lept}(M_Z^2)$ these yield the predictions
\be
\alpha^{-1}(M_Z)=\left\{ \begin{array}{c}
128.933\pm 0.021\quad \mbox{[DH]}\,,\\
128.928\pm 0.023\quad \mbox{[KS]} \,. \end{array}
                 \right.
\ee
It is important to remember that new physics may also
contribute to $\Delta\alpha$ even though new massive states apparently
decouple reasonably rapidly.  This possibility is usually neglected due to the 
assumption that the electroweak SM and QCD correctly describe the data below 
the $Z$ pole.  However, one can imagine some new scenarios which would have
an effect, \eg\ a new $Z'$ boson could modify $R(s)$ somewhat in this 
energy region and lead to an apparent shift in $\Delta\alpha$.  

Better measurements of $R(s)$ in the $\sqrt s\sim 1$ GeV region and improved 
theoretical analyses may lead
to a further reduction in the uncertainties in $\alpha(M_Z^2)$.

\nn

\subsection{$g-2$ of the electron and muon}

The previous discussion of QED and the renormalization of $\alpha$ naturally 
leads to the related topic of the anomalous magnetic moment of leptons,
$a_{e,\mu}\equiv {1\over 2}(g-2)_{e,\mu}$\,, which are the most precisely
calculated quantities in QED.\cite{AIS}  They can be expressed in terms
of the various contributions
\be
a_\ell^{Th}=a_\ell^{QED}+a_\ell^{had}+a_\ell^{EW}\,.
\ee
The first term is given by the usual QED expansion,
\be
a_\ell^{QED}=\sum_n c_n^\ell\left({\alpha\over\pi}\right)^n\,,
\ee
which is known to $n=4$ for electrons and $n=5$ for muons.  Results for
$a_\ell^{had}$ are obtained from vacuum polarization contributions
and light-by-light
scattering involving hadrons, while the perturbative SM electroweak
contributions to $a_\ell^{EW}$ are computed to two-loops.\cite{twl}
For electrons, these two contributions are quite small,
\be
a_e^{had}=1.635\times 10^{-12}\,,\quad\quad\quad a_e^{EW}=0.030\times
10^{-12}\,.
\ee
Using the value of $\alpha$ from Ref. \cite{Hall}, the theoretical expectation
for $a_e$ is
\be
a_e^{Th}=1159652156.4(1.2)(22.9)\times 10^{-12}, 
\ee
where the first (last) error arises from 
uncertainties from higher orders (in $\alpha$ itself).
The latest experimental result \cite{aeexp} is given by
\be
a_e^{exp}=1159652188.2(3.0)\times 10^{-12}\,,
\ee
and agrees within $1\sigma$ of the theoretical calculation.
Since the electroweak contributions are so small it is doubtful that new physics
can make a significant impact here.

\begin{figure}[t]
\centerline{
\psfig{figure=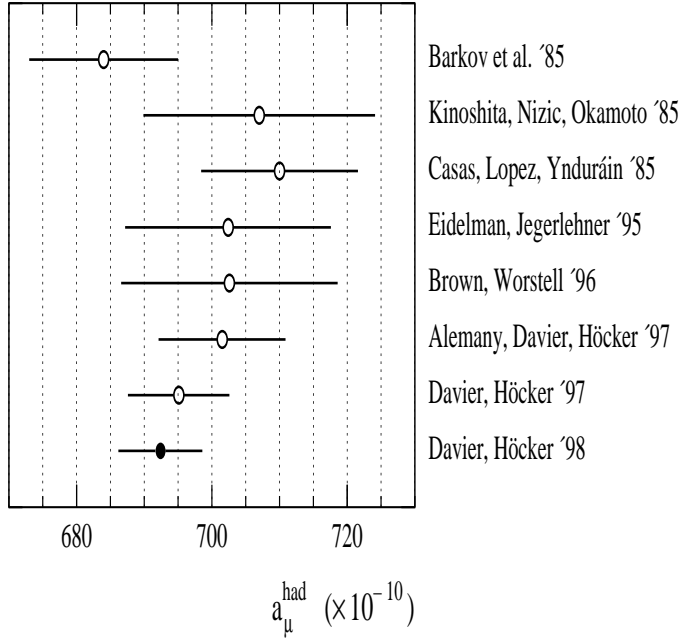,height=9cm,width=12cm,angle=0}}
\vspace*{0.5cm}
\caption{Comparison of the evaluations of $a_\mu^{had}$.  
From~\protect\cite{DH}.}
\label{amuhad}
\end{figure}

For muons, using the results of \cite{DH,twl} one obtains
\bea
a_\mu^{QED} & = & 116584705 (2)\times 10^{-11}\,,\nonumber\\
a_\mu^{had} & = & 6739 (67)\times 10^{-11} \,,\\
a_\mu^{EW} & = & 151 (4) \times 10^{-11} \,,\nonumber
\eea
or
\be
a_\mu^{Th}  =  11659159.6 (6.7)\times 10^{-10} \,.
\ee
Note that the uncertainties in $a_\mu^{had}$ are much larger than those for
the QED and EW contributions.  As discussed in the previous section, these
arise from the non-perturbative nature of QCD in the $\sim 1$ GeV region,
where dispersion relations must be employed in the evaluation of these
contributions.  A compilation, from \cite{DH}, of recent analyses of the 
hadronic contributions to $a_\mu$ is shown in Fig. \ref{amuhad}, where we
see that the calculations are becoming more precise.

On the experimental side, the Particle Data Group \cite{pdg98} gives the world
average result $a_\mu^{PDG}=11659230(84.0)\times
10^{-10}$ and a new result from BNL E821 \cite{Tim} yields $11659250(153.0)
\times 10^{-10}$ to give $a_\mu^{exp}=1165923.5(7.4)\times 10^{-10}$.  The
difference is then found to be
\be
\Delta a_\mu=a_\mu^{exp}-a_\mu^{Th}=(75.4\pm 74.3)\times 10^{-10}\,,
\ee
which is approximately $1\sigma$ high, leaving room for potential new physics 
contributions at the same level (or larger) than the SM.\cite{AIS,jacques}
Future running of the E821 experiment may lower the error by a further
factor of $\sim 20$, which would render new electroweak contributions visible
if they are at least half as large as those corresponding to the SM. 

\nn

\section{Electroweak Interactions}

As discussed in the introduction, the Standard Model of electroweak
interactions has been probed at the quantum level.  Presently, precision
electroweak measurements at the $Z$-pole have tested the SM 
at the $\sim 0.1\%$ level, while low-energy measurements are at the $\sim 1\%$
level.  In this section we first review the main aspects of the computation of
raditative corrections in the SM, and then examine the comparison between the 
predictions of the perturbatively corrected theory and the data.

\subsection{$\mu$ Decay}

The constant $G_F$ is defined by the $\mu$ lifetime, $\tau_\mu\equiv
\Gamma_\mu^{-1}$, as calculated within the local Fermi theory which is finite
to first order in $G_F$ and to all orders in $\alpha$.\cite{bs}  This implies
that $\Gamma_\mu$ may be written as $\Gamma_\mu=\Gamma_0(1+\Delta)$ where 
$\Delta$ is a power series in $\alpha$, 
\be
\Delta=\sum_n\alpha^n\Delta_n\,, 
\ee
and $\Gamma_0$ is the usual tree-level expression
$\Gamma_0=G_F^2m_\mu^5/192\pi^3$.  The lowest order term $\Delta_0$
arises due to finite $x=m_e^2/m_\mu^2$ phase space corrections (assuming 
massless neutrinos) and $W$-boson mass corrections, and is given by
\be
\Delta_0=-8x-12x^2\ln x+8x^3-x^4+{3m_\mu^2\over 5M_W^2} +O\left( {m_\mu^4
\over M_W^4}\right)\,.
\ee
Similarly,\cite{ksb} the higher order terms originate from the $n^{th}$-loop
pure QED corrections, which are found to be
\be
\Delta_1={\alpha(m_\mu)\over 2\pi}\left[ {25\over 4}-\pi^2\right] 
+O(\alpha(m_\mu)x\ln x) \,,
\ee
and the recently computed \cite{rs}
\be
\Delta_2= (6.701\pm 0.002)\left( {\alpha(m_\mu)\over\pi}\right)^2 \,,
\ee
where the $\overline{\rm MS}$ scheme is employed \cite{rs} in evaluating
$\alpha(m_\mu)$,
\be
\alpha(m_\mu)={\alpha\over 1+\mbox{$\alpha\over 3\pi$}\ln x}
-{\alpha^3\over 4\pi^2}\ln x \,.
\ee
This result contains all
corrections of $O(\alpha^2)$, $O(\alpha^3\ln x)$ and $O(\alpha^n\ln^{n-1}x)$
for $n\geq 2$.  Using the PDG \cite{pdg98} values for $m_\mu$, $m_e$, and
$\tau_\mu^{-1}=(2.19703\pm 0.00004)\mu s$, one obtains
\be
G_F=(1.16637\pm 0.00001)\times 10^{-5}\gev^2\,.
\ee
This has shifted from the previous value of $G_F=(1.16639\pm 0.00002)\times
10^{-5}\gev^2$ computed before the above $\Delta_2$ corrections were known.
With this recent higher order result, the theory error is now of $O(10^{-8})$.
Hence the main uncertainty in $G_F$ now arises solely from the measurement of
$\tau_\mu$.  A factor of 10 improvement in the determination of
$\tau_\mu$ may be achieved in proposed experiments at BNL, PSI, and RIKEN.

\subsection{Lorentz Structure of the Weak Interactions}

The Lorentz structure of the charged current weak interactions may be cleanly
tested in the absence of hadronic interference by the leptonic decays 
$\ell^-\to \nu_\ell\ell'^-\bar\nu_{\ell'}$.  For this purpose it is
sufficient to consider these reactions at the Born level. Since the
momentum transfer carried by the $W$-boson is very small compared to
$M_W$ in this case, the vector boson propagator can be reduced to a contact 
interaction.  Assuming $V-A$ interactions this gives the usual effective
Hamiltonian
\be
{\cal H}_{eff}={G_F\over\sqrt 2}\bar\nu_\ell\gamma^\alpha(1-\gamma_5)\ell
\bar \ell'\gamma_\alpha(1-\gamma_5)\nu_{\ell'}\,.
\ee
The couplings can be determined by the final state lepton energy spectrum,
which is easily calculated for $V-A$ interactions with massless 
neutrinos to be
\be
{1\over\Gamma_0}{d\Gamma_0\over dx}=2x^2(3-2x)\,,
\ee
with $0\le x=2p_{\ell'}\cdot p_\ell/m^2_\ell\le 1$.  Deviations from this 
behavior would indicate new physics arising from new gauge or scalar boson 
exchange or from modifications in the SM $W$-boson couplings.  Giving up the 
assumption of $V-A$ interactions yields the most general Lorentz invariant 
effective Hamiltonian (assuming non-derivative couplings)
\be
{\cal H}_{eff}={G_F\over\sqrt 2}\sum_{\gamma,\epsilon,\omega}g^
\gamma_{\epsilon\omega}\langle \bar\ell'_\epsilon|\Gamma^\gamma|\nu_{\ell'}
\rangle\langle \bar\nu_\ell|\Gamma_\gamma|\ell_\omega\rangle\,,
\ee
where $\Gamma^\gamma$ represents either a scalar ($I$), vector 
($\gamma_\alpha$), or tensor ($\sigma_{\alpha\beta}/\sqrt 2$) interaction and 
$\epsilon,\omega=L,R$ denote the chiralities of the two charged leptons.  Tensor
interactions exist only if the charged leptons have opposite chiralities.  Here
we have assumed that neutrinos are massless.  This
leads to 10 complex coupling constants, $g^\gamma_{\epsilon,\omega}$, for
which the SM predicts $g^V_{LL}=1$ with all others vanishing.  
The inclusion of these additional interactions modifies the decay lepton energy
spectrum, which then takes the general form 
\bea
{d\Gamma\over x^2dxd\cos\theta} & = & (3-2x)+({4\over 3}\rho-1)(4x-3)
+12\eta({m_{\ell'}\over m_\ell}x)(1-x)\nonumber\\
&   & \quad\quad -[(2x-1)+({4\over 3}\delta -1)(4x-3)]\xi\cos\theta \,,
\eea
with $\theta$ being the angle between the direction of the momentum of the
final state lepton and the spin vector of the decaying lepton.  
We remind the reader that the
spin vector is given by $\vec s=(\vec p/m,\, E\vec p/m|\vec p|)$.  Note that
only four combinations of the coupling constants, $\rho\,, \eta\,, \xi$, and 
$\delta$, determine the shape of the lepton energy spectra.  These four
parameters are known as the Michel parameters,\cite{mpara} which are defined by
\bea
\rho & = & {3\over 4}|g^V_{LL}|^2+{3\over 4}|g^V_{RR}|^2+{3\over 16}|g^S_{LL}|^2
+{3\over 16}|g^S_{LR}|^2+{3\over 16}|g^S_{RL}|^2+{3\over 16}|g^S_{RR}|^2
\nonumber\\
& & \quad +{3\over 4}|g^T_{LR}|^2+{3\over 4}|g^T_{RL}|^2-{3\over 4}\Re
(g^S_{LR}g^{T*}_{LR})-{3\over 4}\Re (g^S_{RL}g^{T*}_{RL})\,,\nonumber\\
\eta & = & {1\over 2}\Re\left[ g^V_{LL}g^{S*}_{RR}+g^V_{RR}g^{S*}_{LL}
+g^V_{RL}(g^{S*}_{LR}+6g^{T*}_{LR})+g^V_{LR}(g^{S*}_{RL}+6g^{T*}_{RL})\right]
\,,\nonumber\\
\xi & = & |g^V_{LL}|^2+3|g^V_{LR}|^2-3|g^V_{RL}|^2-|g^V_{RR}|^2
+5|g^T_{LR}|^2-5|g^T_{RL}|^2\\
& & \quad {1\over 4}|g^S_{LL}|^2-{1\over 4}|g^S_{LR}|^2+{1\over 4}|g^S_{RL}|^2
-{1\over 4}|g^S_{RR}|^2+4\Re(g^S_{LR}g^{T*}_{LR})-4\Re(g^S_{RL}g^{T*}_{RL})
\,,\nonumber\\
\xi\delta & = & {3\over 4}|g^V_{LL}|^2-{3\over 4}|g^V_{RR}|^2
+{3\over 16}|g^S_{LL}|^2-{3\over 16}|g^S_{LR}|^2+{3\over 16}|g^V_{RL}|^2
-{3\over 16}|g^V_{RR}|^2\nonumber\\
& & \quad -{3\over 4}|g^T_{LR}|^2+{3\over 4}|g^T_{RL}|^2+{3\over 4}\Re
(g^S_{LR}g^{T*}_{RL})-{3\over 4}\Re(g^S_{RL}g^{T*}_{RL}) \,.\nonumber
\eea
The SM and
experimentally determined \cite{pdg98} values of these parameters are given
in Table \ref{michelp}.  We see that both the $V-A$ structure of the weak 
charged current and lepton universality are confirmed at this level of
sensitivity.

\begin{table}
\centering
\begin{tabular}{|c|c|c|c|c|} \hline\hline
 & $V-A$ & $\mu\to e$ & $\tau\to e$ & $\tau\to\mu$ \\ \hline
$\rho$ & ${3\over 4}$ & $0.7518\pm 0.0026$ & $0.745\pm 0.012$  & 
$0.741\pm 0.030$  \\
$\eta$ & $0$ & $-0.007\pm 0.013$  & $0.01\pm 0.07$ & $-0.10\pm 0.18$  \\
$\xi\delta$ & ${3\over 4}$ & $0.7506\pm 0.0074$ & $0.733\pm 0.033$  &
$0.78\pm 0.05$ \\
$\xi$ & $1$ & $1.0027\pm 0.0085$ & $0.98\pm 0.05$  & $1.07\pm 0.08$ \\ \hline
\end{tabular}
\caption{The SM predictions and world average measured
values~\protect\cite{pdg98} for the Michel parameters.}
\label{michelp}
\end{table}

\subsection{Radiative Corrections I: $M_W$, $\sin^2\theta_w$, $\Delta r$}

As discussed in the introduction,
the SM contains a number of natural tree-level relationships between the
lowest order parameters
\begin{enumerate}
\item {$\sin^2\theta_0 =e^2_0/g_0^2$,}
\item{$\sin^2\theta_0=1-(M_W^0/M_Z^0)^2$, assuming only
Higgs Doublets,}
\item{$G_F/\sqrt 2=g_0^2/8(M_W^0)^2=e_0^2/8s_0^2(M_W^0)^2$.}
\end{enumerate}
When radiative corrections are included it is impossible to simultaneously 
maintain all of these relationships; the choice of which to keep and
which to surrender then defines the renormalization scheme.  The OS scheme in 
its absolute form selects $\alpha\,, M_W\,, M_Z$ (together with all the fermion
as well as Higgs masses) to be the input parameters.  However, since $G_F$ is
much more precisely determined, it is usually traded for $M_W$ as one of
the input parameters, with the relationship $x=\sin^2\theta_w=1-M_W^2/M_Z^2$ 
(\ie, the second relation above) being maintained to
all orders.  In order to use $G_F$ as an input parameter, the
third relation above must be employed, together with the definition of
$G_F$ from $\mu$-decay.
From this discussion we recall that $G_F$ was defined within the local
Fermi theory including QED corrections up to $O(\alpha^2)$.  In the full
electroweak theory however, $\mu$ decay proceeds through $W$ exchange,
and the higher order corrections involve many more diagrams than the
simple QED vertex and bremsstrahlung diagrams.

\nn
\begin{figure}[htbp]
\centerline{
\psfig{figure=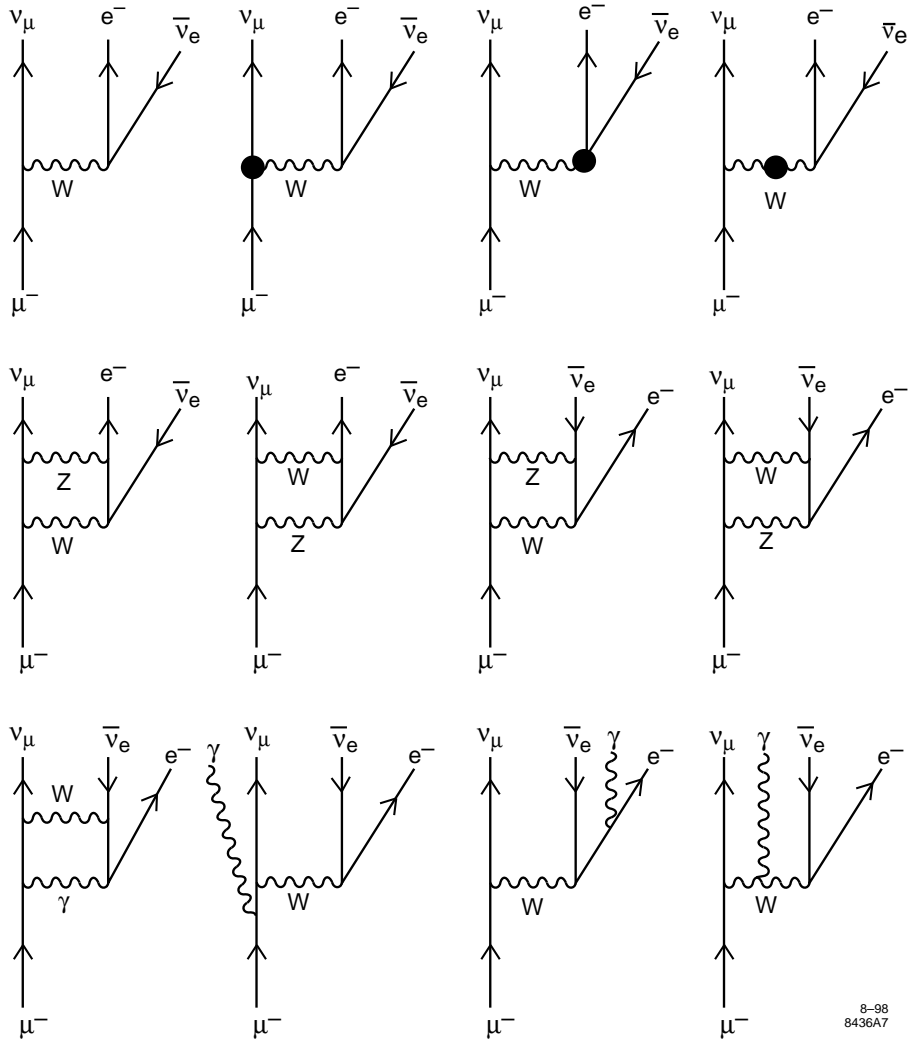,height=14cm,width=12cm,angle=0}}
\vspace*{0.5cm}
\caption{Complete set of Feymann diagrams to order $\alpha^3$ for $\mu$ 
decay.}
\label{mudk}
\end{figure}

Figure \ref{mudk} displays all of the one-loop SM diagrams 
which mediate $\mu$-decay.
These can be separated into several classes beyond those associated with the
pure QED contributions in the local theory limit.  As can be seen from the
figure, these involve corrections to the $\bar\nu_eeW$ and $\nu_\mu\mu W$
vertices, $W$-boson self-energies, box diagrams involving $Z$ exchange, and
$\gamma$ emission off the exchanged $W$-boson.  Since the definition of
$G_F$ given in Section 4.1 
only included the local QED terms, it is clear that the third
relation above is modified and becomes
\be
{G_F\over\sqrt 2}={e^2\over 8xM_W^2}(1+\Delta r) \,,
\ee
where $e^2=e^2(0)$, $M_W$ is the on-shell value, and $\Delta r$ represents
the non-QED corrections to $\mu$ decay.  Employing $x\equiv 1-M_W^2/M_Z^2$,
gives
\be
x(1-x)={A\over M_Z^2}(1+\Delta r)\,,
\ee
with $A\equiv \pi\alpha/\sqrt 2G_F=(37.2805 \gev)^2$.

In the $\overline{\rm MS}$ scheme, the relationship between $M_W$ and $M_Z$ is 
surrendered while $\hat e^2=\hat g^2\sin^2\hat\theta$ is maintained, such
that $M_W^2=\hat c^2\hat\rho M_Z^2$ holds for the physical $M_{W,Z}$, where 
$\hat c^2=1-\hat s^2$.  In this case the value of $\hat x\equiv\hat s^2(M_Z^2)$
is given by
\be
\hat x(1-\hat x)={\hat A\over \hat\rho M_Z^2}(1+\Delta\hat r)\,,
\ee
where $\hat A=\pi\hat\alpha(M_Z^2)/\sqrt 2G_F$ with $\hat\alpha(M_Z^2)$
being the $\overline{\rm MS}$ running $\alpha$ defined at the $Z$ scale,
and $\Delta\hat r$ evaluates the diagrams in Fig. \ref{mudk} in the
$\overline{\rm MS}$ scheme.  Both of the OS and $\overline{\rm MS}$ schemes do 
equally well at describing radiative corrections in the SM provided the known 
two-loop and higher order corrections are included.\cite{dgs}
For example, the OS and $\overline{\rm MS}$ 
predictions for $M_W$ differ by only $2-3$ MeV which is more than an order
of magnitude smaller than the anticipated sensitivity of LEPII and the
Tevatron.

\begin{figure}[htbp]
\centerline{
\psfig{figure=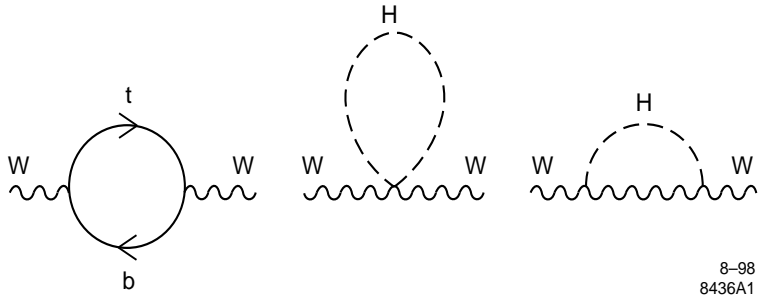,height=4cm,width=10cm,angle=0}}
\vspace*{0.5cm}
\caption{One-loop heavy fermion and Higgs-boson corrections to the $W$
self-energy.}
\label{wloops}
\end{figure}

In order to consider these higher order corrections it is instructive to
first contemplate what lies within the self-energy blob in the $W$-boson
propagator (shown in Fig. \ref{mudk}), 
evaluated at $q^2=0$, \ie, $\Sigma^W(0)$.  In the SM the
dominant contribution arises from the top-bottom quark and Higgs loops depicted
in Fig. \ref{wloops}.  These are of order $m_t^2/M_W^2$ and $\ln m_h/M_W$,
respectively, whereas light fermion contributions are small being of order
$m_f^2/M_W^2$.  The pure gauge boson loops are also suppressed.
This implies that new physics contributions to $\Delta r$ can occur if the
new interactions yield sizeable contributions to the gauge boson
self-energies.  (New interactions \cite{lyb} could occur in the box graphs as 
well, with \eg, $Z\to Z'$ or a SM box being replaced by a supersymmetric
box.)  To isolate the large terms in $\Delta r$ we note from Fig. \ref{mudk} 
that we can schematically write
\be
{G_F\over\sqrt 2} = {e_0^2\over 8s_0^2(M_W^0)^2}\left[ 1+{\Sigma^W(0)\over
M_W^2} + \mbox{box and vertex graphs}\right]\,,
\ee
where the $1/M_W^2$ factor represents an extra $W$ propagator suppression
in the $q^2=0$ limit.  At the one-loop level one must make the replacements
\cite{hol}
\bea
e_0^2 & \to & (e+\delta e)^2 = e^2(1+{2\delta e\over e}) \,,\nonumber\\
(M_W^0)^2 & \to & M_W^2(1+{\delta M_W^2\over M_W^2}) \,,\\
s_0^2 & \to & 1-{M_W^2+\delta M_W^2\over M_Z^2+\delta M_Z^2}
= s^2 + c^2\left( {\delta M_Z^2\over M_Z^2} - {\delta M_W^2\over M_W^2}\right)
\,,\nonumber
\eea
which then yield
\bea
{G_F\over \sqrt 2} & = & {e^2\over 8s^2M_W^2}\left[ 1+2{\delta e\over e}
-{c^2\over s^2}\left( {\delta M_Z^2\over M_Z^2}-{\delta M_W^2\over M_W^2}
\right) +{\Sigma^W(0)-\delta M_W^2\over M_W^2}\right.\nonumber\\
& & \quad \left. +({\rm vertex + box~graphs})\right] \,,\\
& = & {e^2\over 8s^2M_W^2}(1+\Delta r) \,.
\eea
Here $2\delta e/e=\Delta\alpha$, which is given above in the OS scheme, 
and $\delta M_i^2=\Re\Sigma^i(M_i^2)$ are the usual mass counterterms.
The vertex and box corrections are found to be small as are the light
fermion contributions.  We then can write
\be
\Delta r = \Delta\alpha-{c^2\over s^2}\Delta\rho+\Delta r_{\rm rem}\,.
\ee
The $\Delta\rho$ term arises from the $(t,b)$ contribution to the
$W$ and $Z$ self-energies and can be written to one-loop as the
familiar relation \cite{vel} (neglecting terms of order $m_b^2/m_t^2$)
\be
\Delta\rho={3\alpha\over 16\pi s^2c^2}{m_t^2\over M_Z^2}\approx
{3G_Fm_t^2\over 8\sqrt 2\pi^2}\,.
\ee
$\Delta r_{\rm rem}$ contains the remaining contributions, which are
generally not enhanced,
including those due to the Higgs boson.  The $\Delta\alpha$ and $\Delta\rho$ 
terms give the largest contributions to $\Delta r$; for $m_t\sim 
175$ GeV and a light Higgs they take on the values $\approx 0.0582$ and 
$\approx -0.0322$, respectively, while the $\Delta r_{\rm rem}$ pieces are 
$\lsim 0.005-0.010$.  Note that for this value of $m_t$ the two dominant
terms tend to cancel.  This makes it very important to understand $\Delta
\alpha$ as well as possible, and to calculate $\Delta\rho$ and $\Delta
r_{\rm rem}$ to higher-order in order to obtain a firm a set of predictions for
$M_W$ and $\sin^2\theta_w$.

\nn

As discussed previously $\Delta\alpha$ is now relatively well-known in 
comparison to just a few years ago with $\delta(\Delta\alpha)_{1998}\lsim 
0.2\delta(\Delta\alpha)_{1995}$.  Indeed most terms in $\Delta r$ have now been
computed beyond the level of one-loop, although a complete two-loop calculation 
has not yet been performed.  For example, in the case of $\Delta\rho$ two-
and three-loop QCD corrections have been calculated, as have the two-loop Higgs 
exchange terms.  The full $G_F^2m_t^4$ and $G_F^2m_t^2M_Z^2$ two-loop
contributions to $\Delta r$ are also computed, as are the two-loop Higgs 
dependent non-leading $m_t$ terms.\cite{dgs,hol2}

These sophisticated calculations lead to precise predictions for
$M_W$ and to uncertainties in $\sin^2\theta_w$ of only
$\sim 3\times 10^{-5}$ for a fixed set of input parameters.  In both cases 
these uncertainties are far smaller than those associated with the error on 
the top-quark mass measurement.
Given the uncertainties in $G_F$ and $\alpha(M_Z)$ 
discussed above, together with the determinations from LEP of
$M_Z=91186.7\pm 2.1$ MeV and $\alpha_s(M_Z)=0.119\pm 0.005$, the
uncertainty on $M_W$ for $m_t=173.8\pm 5.0$ GeV is 32 MeV from
$\delta(m_t)$ and 5 MeV from all other sources for a fixed value of $m_H$.

\begin{figure}[htbp]
\centerline{
\psfig{figure=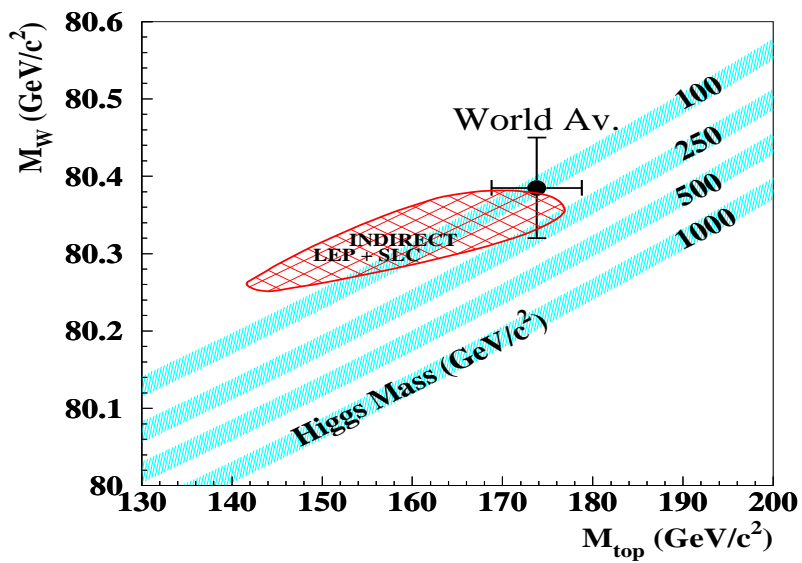,height=10cm,width=12cm,angle=0}}
\vspace*{-1.5cm}
\caption{The comparison of the indirect determinations (from LEP and SLD)
and the world average of direct measurements of $M_W$ and $m_t$.  Also displayed
is the SM relationship for the masses as a function of the Higgs mass.
From~\protect\cite{karlen}.}
\label{mtmw}
\end{figure}

We now examine how well these predictions match the experimental results.
The Tevatron value of $m_t=173.8\pm 5.0$ GeV is shown in Fig. \ref{mtmw}
along with the combined $W$-boson mass as directly measured by CDF/D0/UA2
and LEPII, \ie, $M_W=80.390\pm 0.064$ GeV.  Also displayed in the figure
is the prediction for $M_W$ in the SM as a function of $m_t$ for various
values of the Higgs-boson mass; the widths of the bands for
different Higgs masses shows an exaggerated $\pm 2\sigma$ uncertainty in
the SM prediction.  Given the measured value of $m_t$, it is
clear that the combined measurements of the $W$ mass prefer somewhat
lighter Higgs boson masses; this result is common with other indirect
determinations of the Higgs mass as will be discussed below.  

\nn

\subsection{Radiative Corrections II: Precision Measurements on the $Z$-Pole}

\begin{figure}[htbp]
\centerline{
\psfig{figure=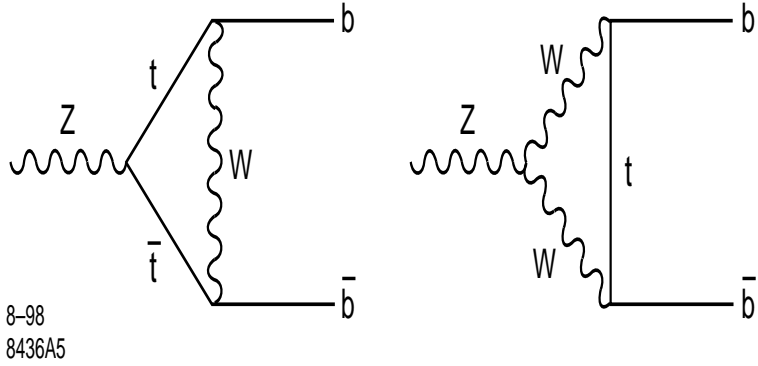,height=5cm,width=10cm,angle=0}}
\vspace*{0.5cm}
\caption{Feymann diagrams depicting heavy $m_t$ one-loop corrections to the
$Z\to b\bar b$ vertex.}
\label{zbbar}
\end{figure}

At tree-level the process $\epem\to f\bar f$ occurs through $\gamma$ and $Z$
exchange.  At one-loop (and higher) order one needs to include, as in the
case of $\mu$-decay, vertex corrections, external wave function renormalization,
gauge boson self-energies, bremsstrahlung, and box diagrams; the last of
which are very small numerically in the region of the $Z$-pole.  In this
region the $\gamma$ exchange amplitude can be approximated by a factorized
form
\be
A_{\gamma}=4\pi\alpha(s){Q_eQ_f\over s}[(1+F^e_V)\gamma_\mu
-F_A^e\gamma_\mu\gamma_5]\times [(1+F_V^f)\gamma^\mu-F_A^f\gamma^\mu\gamma_5]
\,,
\ee
with $\alpha(s)\simeq\alpha(M_Z)$ and $F_{V,A}^{e,f}$(s) representing
form factors which vanish as $s\to 0$ and are numerically small, $\sim
10^{-3}$ at $s=M_Z^2$.  Similarly one has for the $Z$ piece (note the
$s$-dependent width terms in the propagator)
\bea
A_Z& = & {\sqrt 2 G_FM_Z^2\sqrt{\rho_e\rho_f}\over (s-M_Z^2)+i\mbox{${s\over 
M_Z^2}$}M_Z\Gamma_Z}[(T_{3L}^e-2Q^e\kappa_es_w^2)\gamma_\mu-T_{3L}^e
\gamma_\mu\gamma_5]\\
& & \quad\quad\quad\quad\quad\quad\times [(T_{3L}^f-2Q^f\kappa_fs_w^2)
\gamma_\mu-T_{3L}^f\gamma_\mu\gamma_5]\nonumber\,,
\eea
such that it appears that the $Z$ has effective couplings to any fermion
of
\be
(\sqrt 2 G_FM_Z^2)^{1/2}[g_V^f\gamma_\mu-g_A^f\gamma_\mu\gamma_5]\,,
\ee
with
\be
g_V^f=\sqrt{\rho_f}(T_{3L}^f-2s_f^2Q_f)\,,\quad\quad\quad\quad g_A^f=
\sqrt{\rho_f}T_{3L}^f\,,
\ee
where the weak mixing angle is rescaled, $s_f^2\equiv \kappa_f s_w^2$,
due to $Z-\gamma$ mixing.
The quantities $\rho_f$ and $\kappa_f$ absorb all vertex and self-energy
corrections, have both universal and flavor dependent pieces, and are
generally complex.  The imaginary part of these corrections,
which arises from absorptive parts, is generally neglected.  This
is known as the Effective Born Approximation, which is excellent numerically.  
For most fermions the universal terms dominate, with the
exception being the case of the $b\bar bZ$ coupling, where
potentially large $m_t$ enhanced graphs such as those depicted in Fig.
\ref{zbbar} contribute.  Compared to the $d\bar dZ$ coupling, these effects 
can be approximated at one-loop as
\be
\sqrt{\rho_b}=\sqrt{\rho_d}(1-2x_t)\,,\quad\quad\quad s_b^2={s_d^2\over
1-2 x_t}\,,\
\ee
with $x_t\equiv G_Fm_t^2/8\sqrt 2\pi^2$.  This leads to a strong $m_t$
dependence in the $Z\to b\bar b$ width, which is not seen in the decays
to other fermion pairs.  This special correction is completely computed at
one-loop as are the leading two-loop $O(G_F^2m_t^4)$ and
$O(\alpha G_Fm_t^2)$ terms.  The correction factors $\rho_f\,,
\kappa_f$ for the other flavors are also known beyond leading order.
Note that although $\kappa_f$ and $s_w^2$ are scheme dependent, the
combination $s_f^2=\kappa_fs_w^2$ is scheme independent.  $\Delta\kappa_f$
is  especially small $(\kappa_f=1+\Delta\kappa_f)$ in the 
$\overline {\rm MS}$ scheme, \eg, $s_\ell^2=\sin^2\theta_
{\overline{\rm MS}}+0.00029$.

\nn

Near the $Z$-pole, the cross section can be written as $\sigma\sim|
A_\gamma+A_Z|^2$, so that
\bea
\sigma(s) & = & {12\pi\Gamma_e\Gamma_f\over 
|s-M_Z^2+i\mbox{${s\Gamma_Z\over M_Z}$}
|^2}\left[ {s\over M_Z^2}+R_f{s-M_Z^2\over M_Z^2}+I_f{\Gamma_Z\over M_Z}
+\cdots\right] \nonumber\\
& & \quad +{4\pi\alpha^2(s)\over 3s}Q_f^2N_c(1+\delta_{QED}
+\delta_{QCD}) \,,
\eea
with $R_f\,, I_f$ representing the $\gamma-Z$ interference terms and the
last terms being due to pure photon exchange.  The pure $Z$ resonance
term is usually written as
\be
\sigma_{res}=\sigma_0{s\Gamma_Z^2\over (s-M_Z^2)+{s^2\Gamma_Z^2\over M_Z^2}}\,,
\quad\quad \sigma_0={12\pi\over M_Z^2}{\Gamma_e\Gamma_f\over \Gamma_Z^2}\,.
\ee

Due to the small size of the box graphs and the ability to at least
approximately factorize the $\epem\to f\bar f$ amplitude, initial and
final state QED (and QCD) corrections can be handled separately.  While
final state corrections are best performed using perturbation theory,
initial state radiation (ISR) (real and virtual) is best accounted for by
the convolution approach,
\be
\sigma_{obs}(s)=\int dz~R(z)\sigma(s(1-z))\,,
\ee
with $R(z)$ being a radiator function.  $R(z)$ is known through
${\cal O}(\alpha^2)$ and all LL and NLL terms have been resummed.  ISR
reduces the value of the peak cross section by $\approx 25\%$ and shifts
the peak position by $\approx 90$ MeV.  The effects of ISR must be
deconvoluted from the data before it can be compared with the SM predictions.

There are several classes of observables on the $Z$-pole:

\vspace{0.5cm}
\noindent $(i)$ $Z$ partial widths

\nopagebreak
  The partial widths of the $Z$-boson can be written as (defining
$\beta\equiv 1-4m_f^2/M_Z^2$)
\be
\Gamma_f=N_c{G_F M_Z^3\over 6\sqrt 2\pi}\beta_f \left[{3-\beta_f^2\over 2}
|g_V^f|^2+\beta_f^2|g_A^f|^2\right](1+\delta_{QED}+\delta_{QCD}) \,,
\ee
with the corrections $\delta_{QED}$ being known through $O(\alpha^2)$ and 
$\delta_{QCD}$ through $O(\alpha^3_s,\alpha\alpha_S)$.  Commonly
defined ratios of widths are $R_{had}\equiv\Gamma_{had}/\Gamma_\ell\,,
R_{b,c}\equiv\Gamma_{b,c}/\Gamma_{had}$.

\vspace{0.5cm}
\noindent $(ii)$ Forward-Backward Asymmetries

\nopagebreak
  The fermion forward-backward asymmetries are defined by
\be
A^f_{FB}\equiv {\sigma_F^f-\sigma_B^f\over\sigma_F^f+\sigma_B^f}\,,
\ee
with
\be
\sigma^f_{F,B}=\int_{\cos\theta^{>}_{<}0} {d\sigma^f\over d\cos\theta}
d\cos\theta\,.
\ee
Here, the pure $\gamma$ terms are subtracted and final state QED and QCD
corrections are normally deconvoluted from all asymmetries, so that
\be
A^f_{F,B}={3\over 4}A_eA_f\,,
\ee
with
\be
A_f={2\beta_fg_V^fg_A^f\over \mbox{${3-\beta_f^2\over 2}$}|g_V^f|^2
+\beta_f^2|g_A^f|^2}\,.
\ee

\vspace{0.5cm}
\noindent {(iii) Left-Right Asymmetry}

\nopagebreak
The left-right polarization asymmetry, $A_{LR}$, requires a polarized electron
beam and is only measured by SLD.  It is given by
\be
A_{LR}={\sigma_L^f-\sigma_R^f\over\sigma_L^f+\sigma_R^f}=PA_e\,,
\ee
where $\sigma^f_{L,R}$ are the cross sections for left- or right-handed
polarized electron beams with $P$ being the degree of beam polarization.
$A_{LR}$ probes the leptonic $Z$ couplings directly and is independent
of the final state $f$.

\vspace{0.5cm}
\noindent {(iv) Polarized Forward-Backward Asymmetry}

\nopagebreak
  The polarized forward-backward asymmetry, $A_{FB}^{pol}(f)$, also
requires polarized initial beams and hence is also only measured by SLD.
It is defined by
\be
A_{FB}^{LR}(f)=A_{FB}^{pol}(f)={(\sigma_L^f-\sigma_R^f)_F-
(\sigma_L^f-\sigma_R^f)_B\over (\sigma_L^f+\sigma_R^f)_F
+(\sigma_L^f+\sigma_R^f)_B}={3\over 4}P A_f\,,
\ee
and thus directly probes the final state fermion couplings to the $Z$-boson.

\vspace{0.5cm}
\noindent {(v) Final State $\tau$ Polarization}

\nopagebreak
In the decay $Z\to\tau^+\tau^-$ the polarization of the final state
taus can be measured as a function of their production angle using the
tau decay modes $\mu\bar\nu_\mu\nu_{\tau}\,, e\bar\nu_e\nu_\tau\,, 
\pi\nu_\tau\,, \rho\nu_\tau$, and $a_1\nu_{\tau}$.  Assuming that taus
decay with the Lorentz structure of the SM, this inventive observable can be 
written as
\be
P_\tau(\cos\theta)=-{A_\tau(1+\cos^2\theta)+2A_e\cos\theta\over
1+\cos^2\theta+2A_eA_\tau\cos\theta}\,.
\ee
This clearly provides a measurement of both $A_{e,\tau}$.

\vspace{0.5cm}
All of these observables are well understood in the SM with uncertainties
due to uncalculated higher order corrections being below the level of 
$0.01\%$.  This is comparable to that due to the corresponding errors
in the input
parameters.  Note that ISR corrections to the various 
asymmetries can also be performed by the convolution technique discussed
above by using various radiating functions.

\begin{figure}[htbp]
\centerline{
\psfig{figure=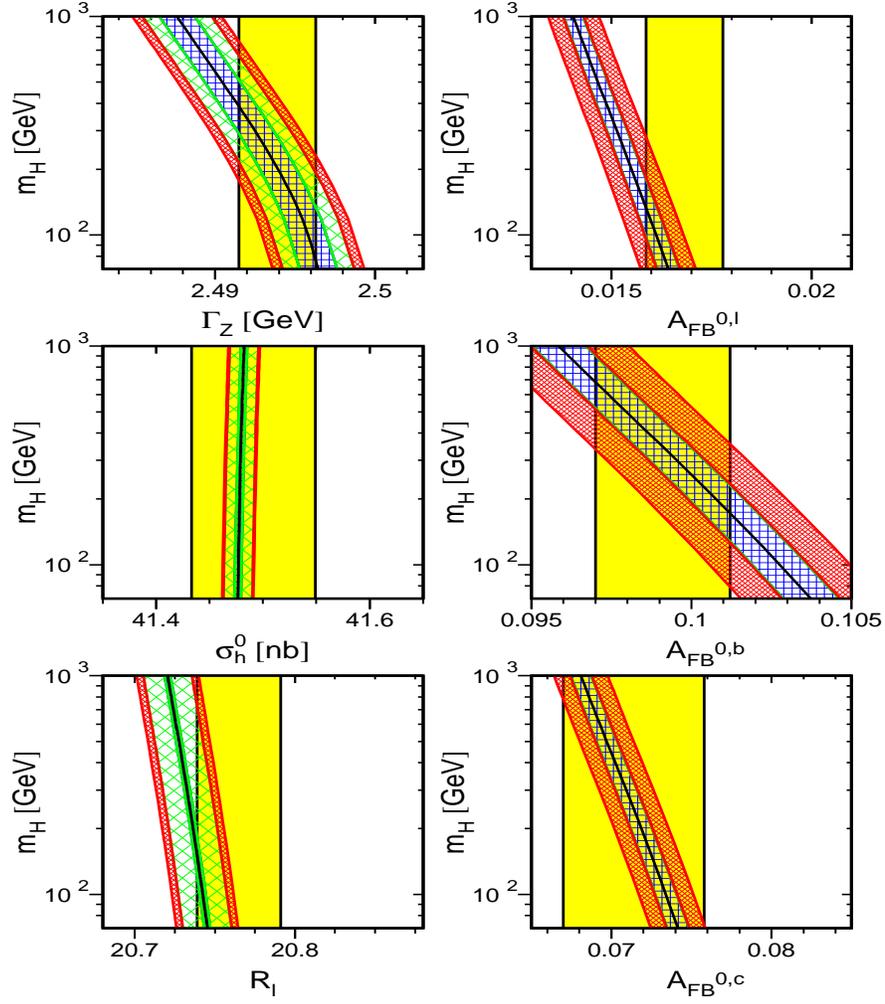,height=14cm,width=12cm,angle=0}}
\vspace*{0.5cm}
\caption{Comparison of LEPI measurements (represented by the vertical band) 
with the SM expectations as a 
function of the Higgs mass.  The width of the SM band is due to uncertainties
in $\alpha(M_Z^2)\,, \alpha_s(M_Z)$, and $m_t$.
From~\protect\cite{karlen}.}
\label{data1}
\end{figure}

\begin{figure}[htbp]
\centerline{
\psfig{figure=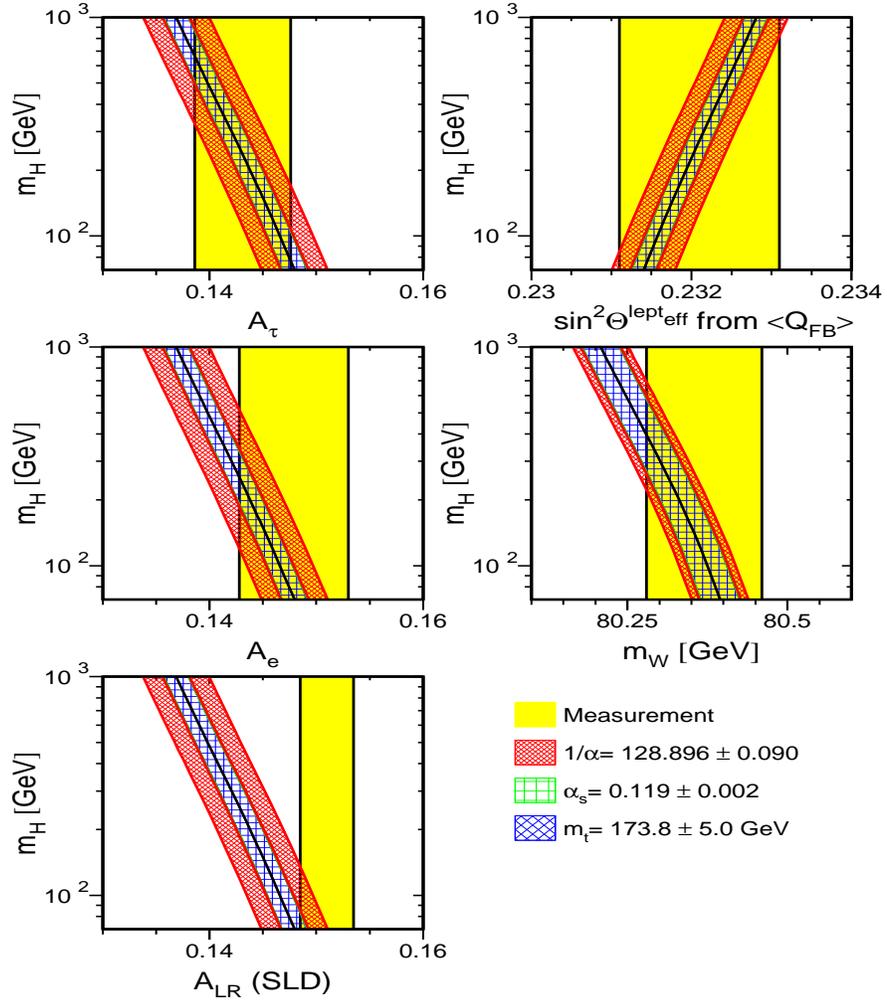,height=14cm,width=12cm,angle=0}}
\vspace*{0.5cm}
\caption{Comparison of LEPI and SLD measurements with the SM expectations as a 
function of the Higgs mass.  The width of the SM band is due to uncertainties
in $\alpha(M_Z^2)\,, \alpha_s(M_Z)$, and $m_t$.
From~\protect\cite{karlen}.}
\label{data2}
\end{figure}

\begin{figure}[htbp]
\centerline{
\psfig{figure=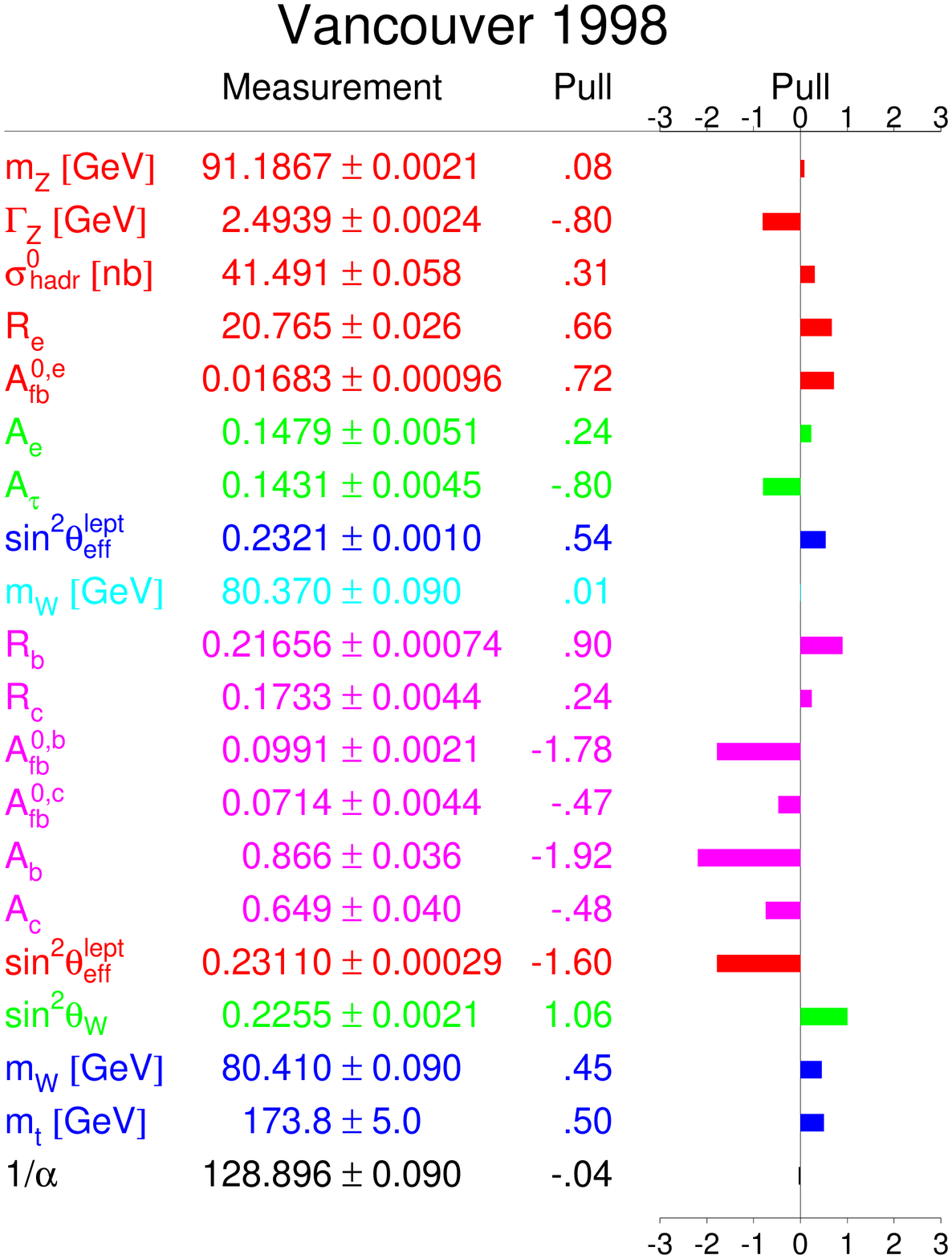,height=14cm,width=12cm,angle=0}}
\vspace*{0.5cm}
\caption{Compilation of the world's electroweak data as of summer 1998,
and the deviation (in number of $\sigma$) of each measurement from the SM fit.
From~\protect\cite{karlen}.}
\label{pull}
\end{figure}

Next we examine how well the SM predictions agree with the enormous amount
of $Z$-pole data collected at LEPI and SLC/SLD.  Figs. \ref{data1} and 
\ref{data2} display the comparison between the SM expectations and the data.  
As can be seen, there is generally
very good agreement between the two if all the values of the input parameters 
(including $m_t$ and $m_H$) are allowed to vary within their allowed ranges.  
The SM best fit to the data is obtained\cite{karlen} by allowing the values of 
the input parameters to vary and minimizing the overall
$\chi^2$ distribution.  Figure \ref{pull} shows the `pull' for each observable,
\ie, the number of standard deviations that each measurement is away from the 
SM fit.  Note the high level of precision of the measurements of the various
observables, with many being at the $0.1\%$ level.  This provides a quantum
level test of the SM predictions, and the agreement between theory and data
is amazingly good.  We also note that there is generally good agreement 
amongst the various experiments for
each of the $Z$-pole observables.  The LEPI and SLD $Z$-pole
data set can be used by itself to predict the correlated values of $M_W$ and
$m_t$; this allowed region is presented in Fig. \ref{mtmw}.  Here, the overlap 
with the direct measurements for both quantities is good and the 
preference for light Higgs masses is again observed.

\nn

\nn

\nn

\begin{figure}[htbp]
\centerline{
\psfig{figure=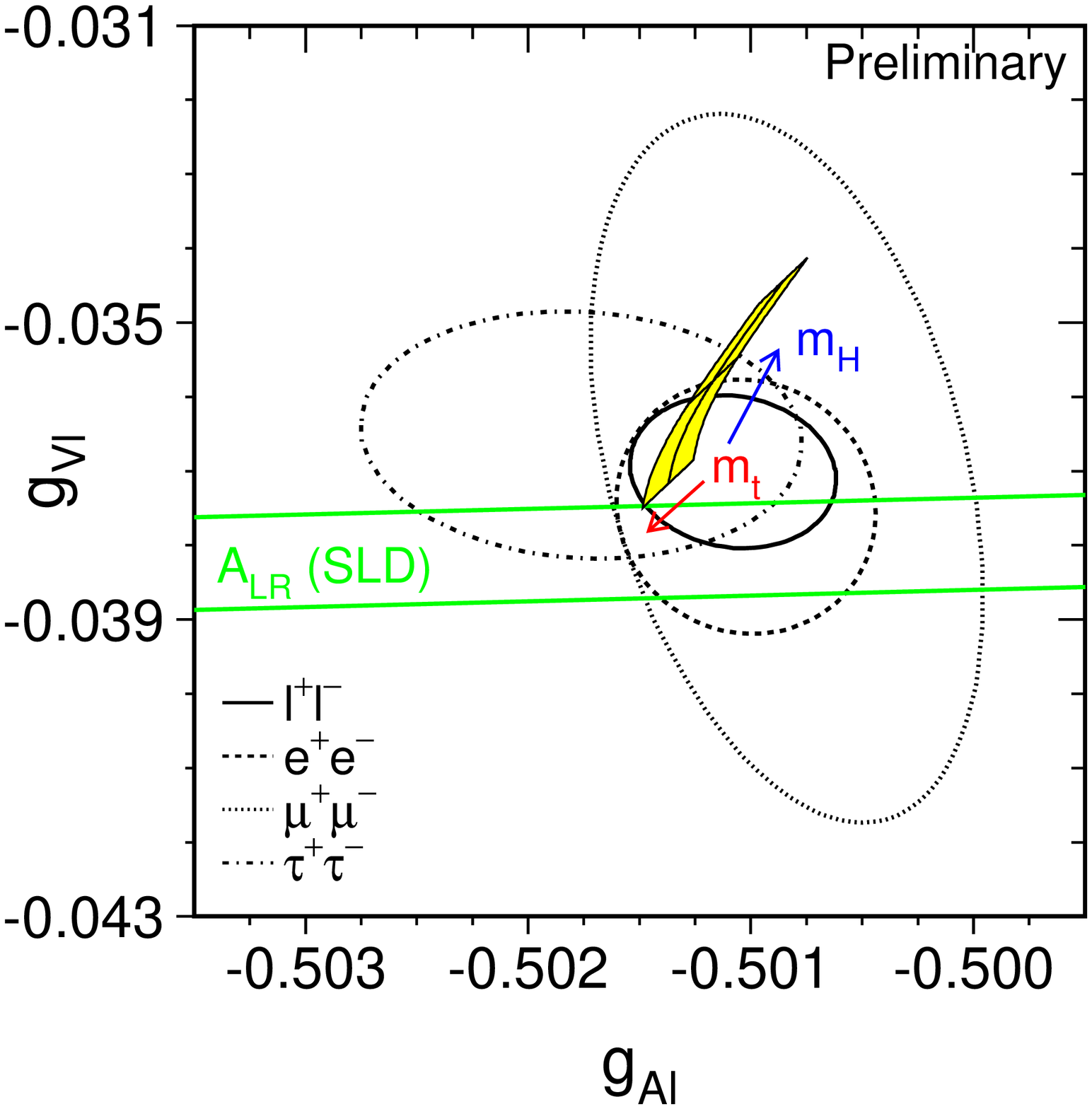,height=7cm,width=8cm,angle=0}}
\vspace{-1.0cm}
\centerline{
\psfig{figure=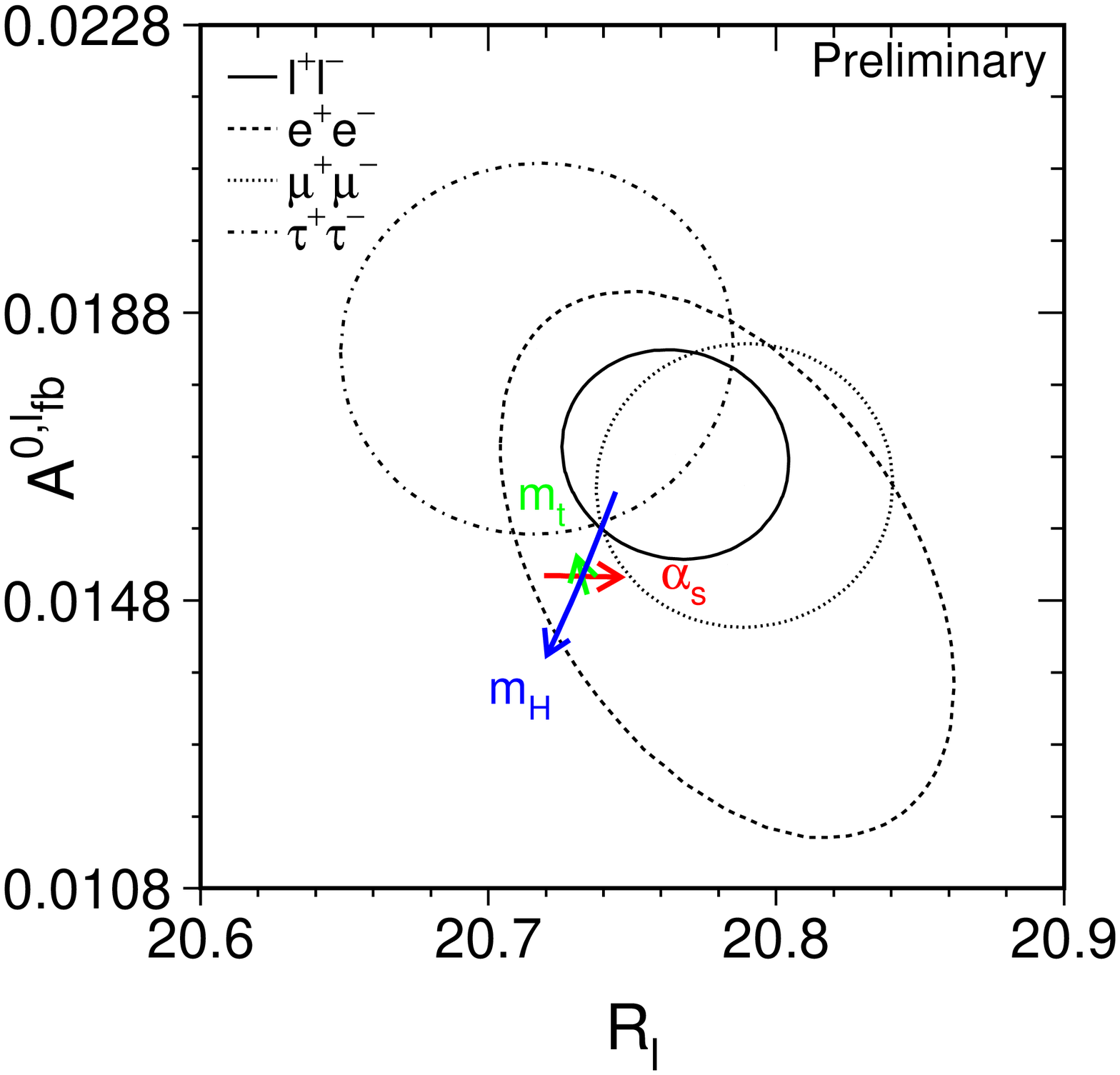,height=7cm,width=8cm,angle=0}}
\vspace*{-0.5cm}
\caption{(a) $1\sigma$ contours in the $g_{V\ell}-g_{A\ell}$ plane from LEPI
measurements and from $A_{LR}$ from SLD.
(b) $1\sigma$ contours in the $A_{FB}^{0,\ell}-R_\ell$ plane.  In both
figures, the solid contour represents the fit assuming lepton universality.
The lines with arrows (or banana shaped region) represent the variation in the 
SM prediction when the 
parameters are varied in the intervals $m_t=174.1\pm 5.4\gev\,,
m_H=300^{+700}_{-240}\gev$, and $\alpha_s(M_Z)=0.118\pm 0.003$.
From~\protect\cite{karlen}.}
\label{leptons}
\end{figure}

The results of the electroweak fits to the leptonic data are presented in
Fig. \ref{leptons}.  Here we see that the fits for the leptonic couplings 
$g_V$ and $g_A$ are in good agreement with the SM predictions.  There is
also good agreement with the assumption of universality.

\nn

One interesting derived result 
is that of the number of light neutrinos, which is obtained from the
invisible width of the $Z$, \ie, $\Gamma_{inv}=\Gamma_Z-\Gamma_{had}-
\Gamma_{e,\mu\tau}$.
$\Gamma_Z$ is determined from a fit to the lineshape, and
$\Gamma_{had}/\Gamma_Z$ and $\Gamma_{e,\mu\tau}/\Gamma_Z$ are
obtained from the cross section at the $Z$ peak for each of these final
states.  The data yield $\Gamma_{inv}=500.1\pm 1.9$ MeV.  
The number of light neutrino generations is then given by $N_\nu=(\Gamma_{inv}/
\Gamma_\ell)/(\Gamma_\nu/\Gamma_\ell
)_{SM}=2.994\pm 0.011$, which is extremely close to the value of three!
This error will be reduced even further when the new luminosity error analysis
presently underway is completed.  

\begin{figure}[htbp]
\centerline{
\psfig{figure=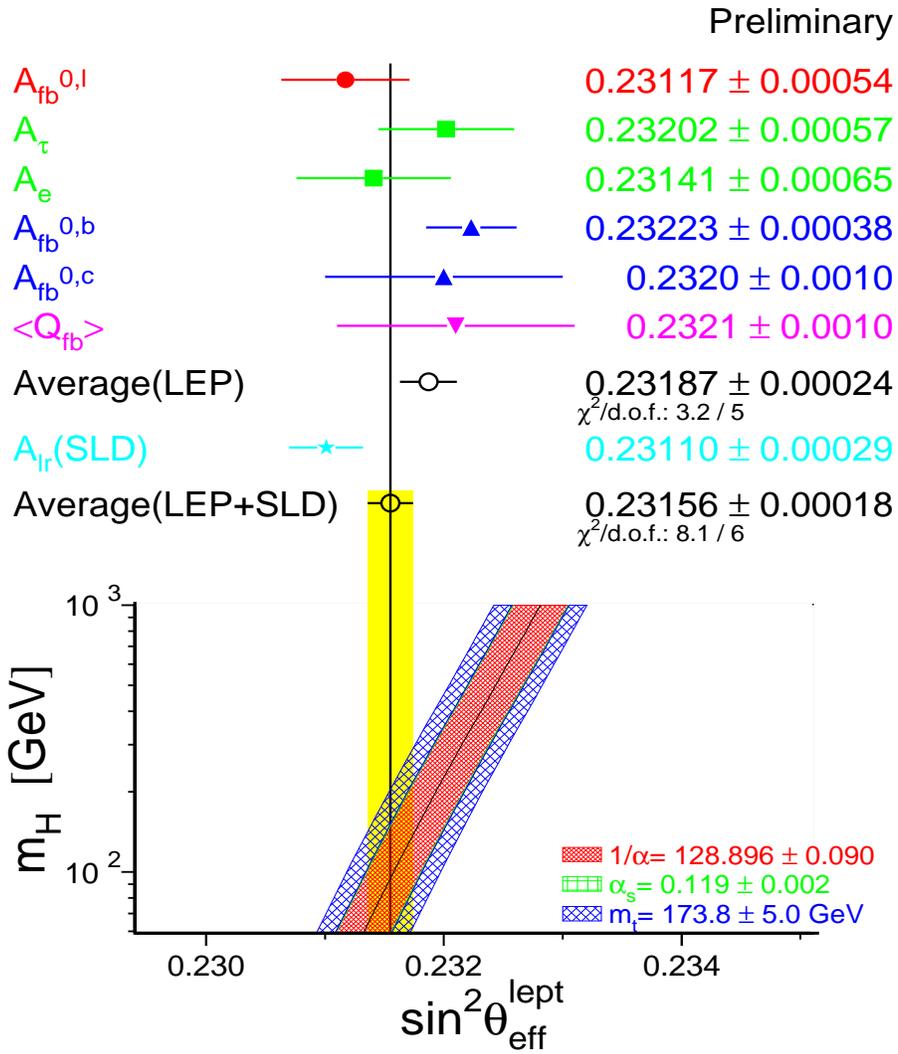,height=14cm,width=12cm,angle=0}}
\vspace*{0.5cm}
\caption{Compilation of various determinations of 
$\sin^2\theta_w~^{\rm lept}_{\rm eff}$.
Also shown is the SM prediction as a function of the Higgs mass.
From~\protect\cite{karlen}.}
\label{sinw}
\end{figure}

\begin{figure}[htbp]
\centerline{
\psfig{figure=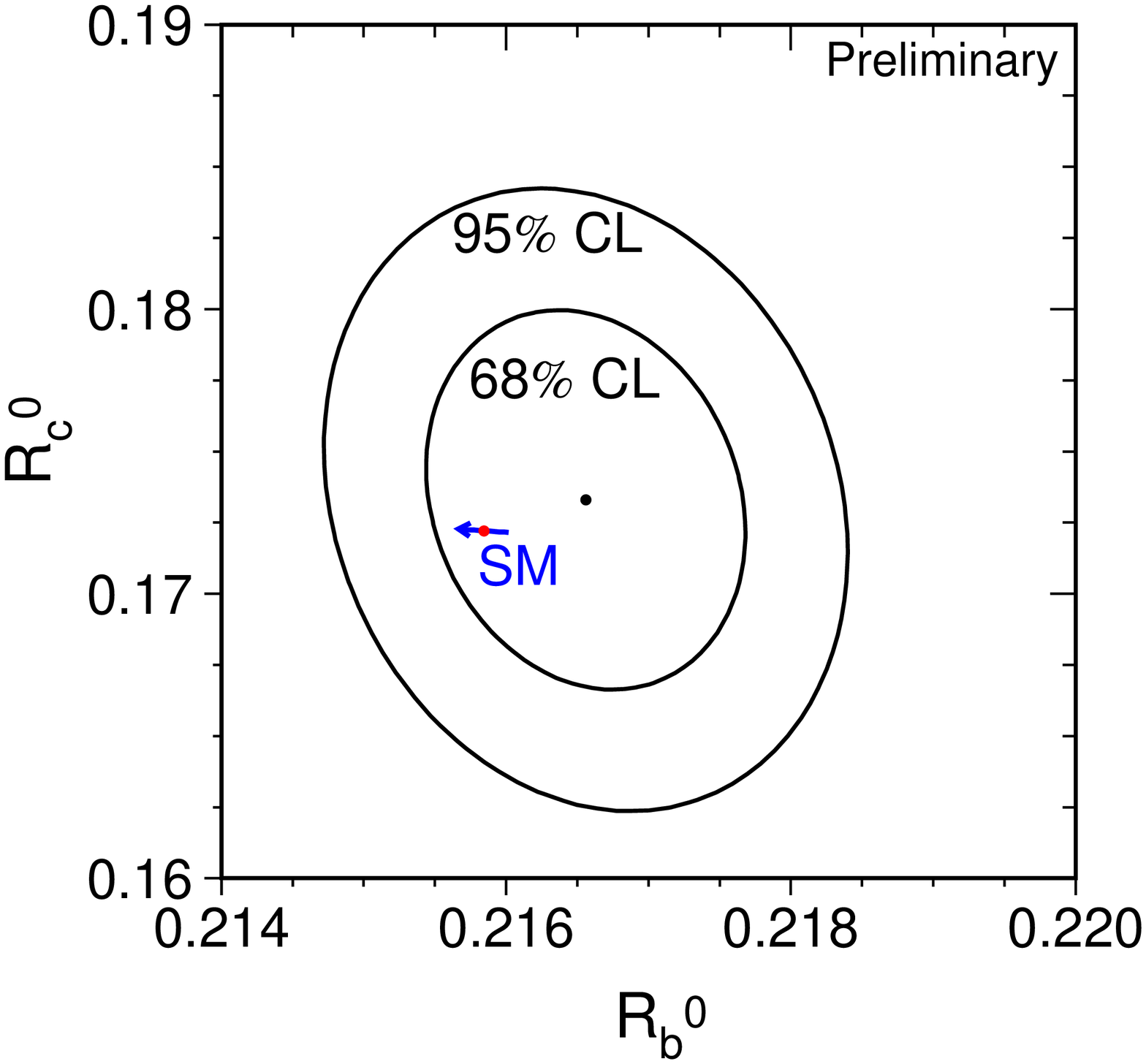,height=7cm,width=8cm,angle=0}}
\vspace{-1.0cm}
\centerline{
\psfig{figure=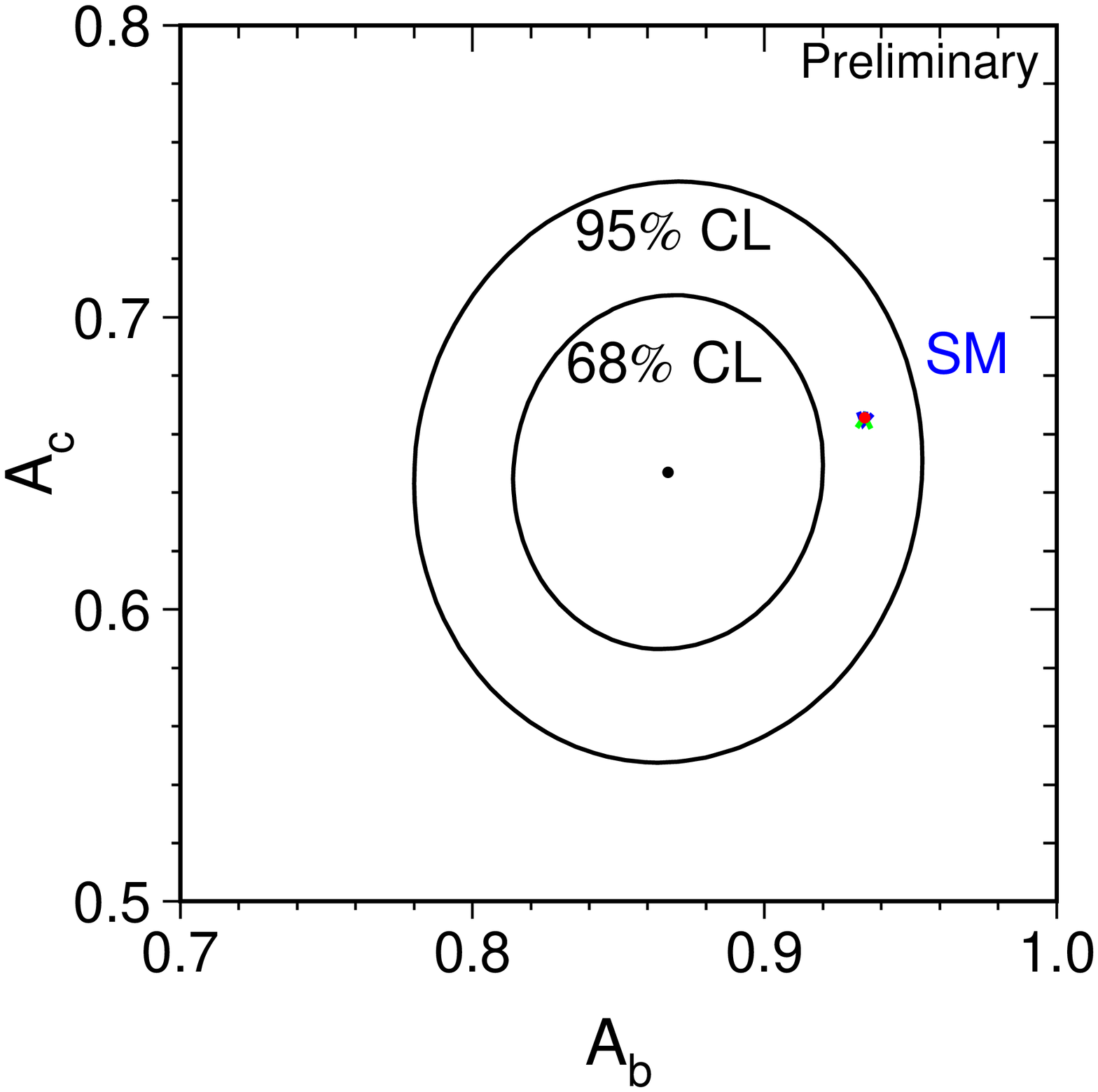,height=7cm,width=8cm,angle=0}}
\vspace*{-0.5cm}
\caption{Contours in the (a) $R_b-R_c$ from LEPI and SLD data and 
(b) $A_b-A_c$ planes from 
and SLD data alone.  The SM prediction for $m_t=174.1\pm 5.4$ GeV is also shown.
From~\protect\cite{karlen}.}
\label{bvsc}
\end{figure}

\begin{figure}[t]
\centerline{
\psfig{figure=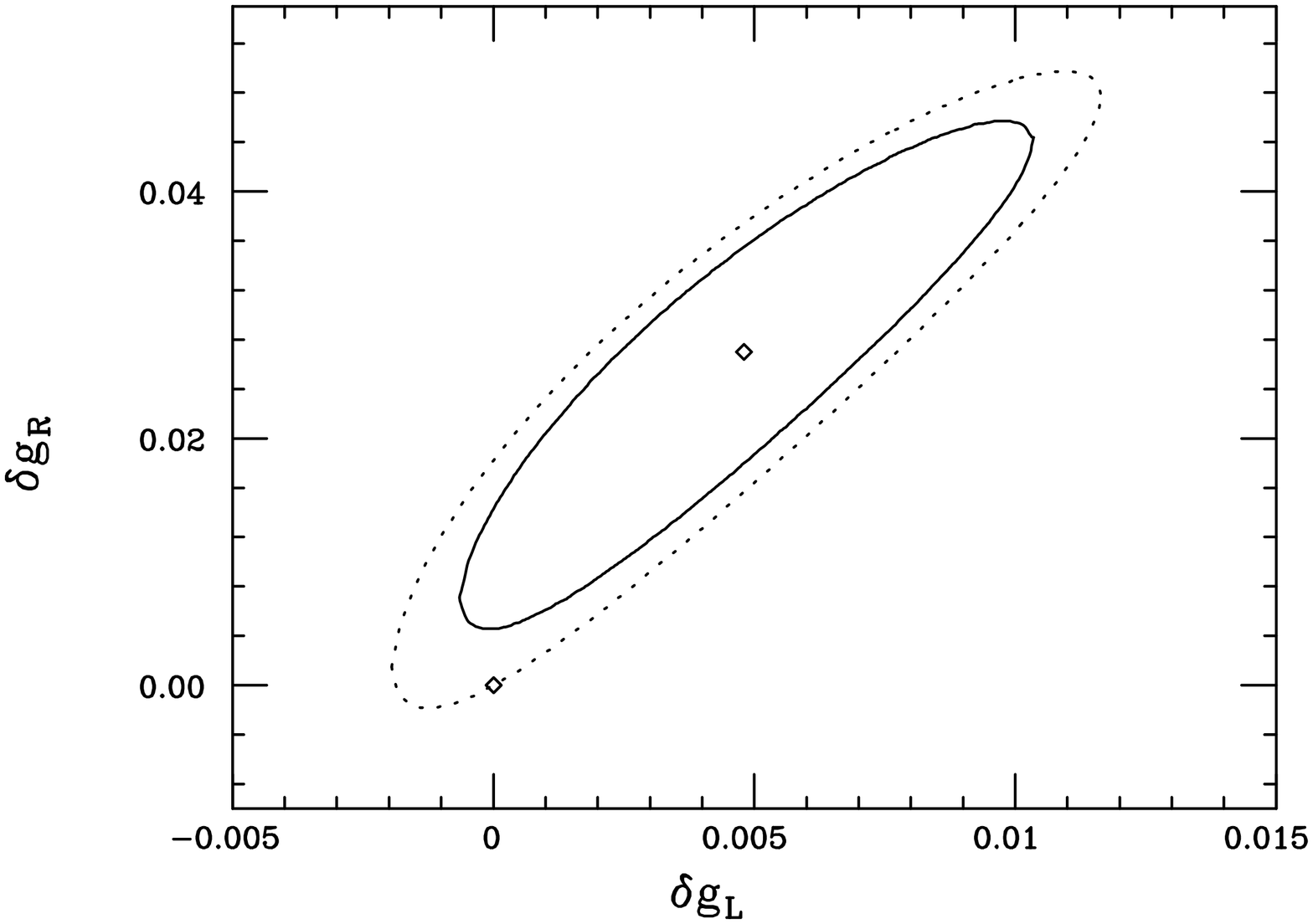,height=9cm,width=12cm,angle=-90}}
\vspace*{-0.5cm}
\caption{$95\%$ (solid curve) and $99\%$ C.L. (dashed) 
fit to the parameters $\delta g^b_{L,R}$ 
using the full LEPI/SLD $Z\to b\bar b$ data set.  The diamond at the center
(edge) represents the best fit (SM prediction).}
\label{glgrb}
\end{figure}

There are a few observables that are worth special attention and should be
watched as data analysis progresses further.  The first is 
$x_\ell=\sin^2\theta^{\rm lept}_{eff}\equiv (1-g_V^\ell/g_A^ell)/4$ which  
is defined in the effective renormalization scheme.
Figure \ref{sinw} compares the values of $x_\ell$ obtained
from 8 different observables, with the single most precise determination being
from $A_{LR}$ as measured by SLD.  The second most precise
determination is from the LEPI measurement of $A_{FB}^b$;  this yields
$x_\ell=0.23223\pm 0.00038$, which is somewhat high
compared to that obtained from the purely leptonic measurements.  These
leptonic observables, $A_{LR}\,, A_{FB}^{pol}(\ell)\,, A_{FB}^\ell\,, A_\tau$,
and $A_\ell$ yield the combined value of $x_\ell=0.23129\pm 0.00022$.  Hence 
the average value of $x_\ell$ extracted from the leptonic measurements disagrees
with that from $A_{FB}^b$ by roughly $3\sigma$.  However, when $x_\ell$
is extracted from $A_{FB}^{b,(c)}$,
the SM values of $A_{b,(c)}$ are assumed and used as input (recall that
$A_{FB}^f={3\over 4}A_eA_f$).  Meanwhile, note that both $A_{FB}^b$ and $A_b$ 
are roughly $2\sigma$ low in comparison with the best fit SM expectations as 
shown in Fig. \ref{pull}.  Perhaps there should be some suspicion 
that there is some new physics in the $Zb\bar b$ vertex and the following
analysis should be employed: $(i)$
Use the combined leptonic measurements of LEPI and SLD to determine
$x_\ell$ or $A_\ell$. $(ii)$ Determine the value of $A_b$ for LEPI using the
relation given above, so that $A_b^{LEP}$ and $A_b^{SLD}$  can be combined 
directly and compared with the SM expectation.  This
procedure implies that $A_b^{ave}=0.882\pm 0.019$, which is $3\sigma$
below the SM prediction of $0.925$.  
The results of the fits to the heavy quark data is
summarized in Fig. \ref{bvsc}.  In order to explicitly display the discrepancy
with the SM, Fig. \ref{glgrb} shows the results of fitting the values of
$\delta g^b_{L,R}$ to the $Z\to b\bar b$ data set, where these quantities are
defined as
$g^b_{L,R}=g^b_{L,R}|_{SM}+\delta g^b_{L,R}$.
As can be seen from the figure, the SM prediction (which uses $m_t=173.8\gev\,,
m_H=300\gev\,, \alpha_s(M_Z)=0.119$) lies on the $99\%$ C.L. exclusion
contour.  The best fit to these quantities yields $g^b_{R}=0.027$ and
$g^b_{L}=0.0048$.  This represents a $30-40\%$ shift in the right-handed
b-quark coupling!  One should not be overly concerned with these $3\sigma$
discrepancies at present, but they should be watched carefully in the future.

\nn

\nn

\nn

\subsection{Radiative Corrections III: Low-Energy Precision Measurements}

\subsubsection{Deep Inelastic Neutrino Scattering}

Data on cross sections from $^(\bar\nu^)N$ charged and neutral current
(CC, NC) cross sections at modest values of $Q^2$ have been used for over
20 years \cite{pdg98} to obtain information on the $u$ and $d$ quark couplings 
to the $Z$-boson.  (Here, $N$ represents an isoscalar nuclear target.)
Such measurements are easily influenced by systematic effects as well as
uncertainties in the parton density functions and are thus conventionally
quoted in terms of cross section ratios, 
\be
R^{\nu,\bar\nu}=\sigma^{\nu,\bar\nu}_{NC}/\sigma^{\nu,\bar\nu}_{CC}\,,
\ee
where many potentially dangerous effects cancel.  However, 
possibly large uncertainties remain in these ratios due to sea quark
contributions.  

The most recent measurements of this type was performed by the NuTeV
Collaboration \cite{nutev} who have made use of the 
Paschos-Wolfenstein \cite{pw} relations, which we now describe.
These relations greatly  reduce the uncertainties
due to both sea quark contributions and the excitation of charm.  Neglecting
the contributions of the initial state $b$ and $t$ quarks, the integrated
charged current  $^(\bar\nu^)N$ cross sections can be written in LO as
\bea
\sigma^\nu_{CC} & \sim & (u+d)[1-s^2_\theta(f-1)]+2s[1+c^2_\theta(f-1)]
+{1\over 3}(\bar u+\bar d+2\bar c)\,,\nonumber\\
\sigma^{\bar\nu}_{CC} & \sim & {1\over 3}(u+d+2c)+2\bar s[1+c^2_\theta
(f-1)]+(\bar u+\bar d)[1-s^2_\theta(f-1)]\,,
\eea
where the $q/\bar q$ symbol represents the integrated contribution
of that particular parton density function to the cross section, 
$s_\theta=\sin\theta_c\,, c_\theta= \cos\theta_c$ (where $\theta_c$ is the 
Cabbibo angle), and $f$ represents a penalty factor for the 
excitation of charm off the light
quark sea, which we will discuss below.  Since the parton densities are
integrated (over $x$ and $y$), it is assumed that $s=\bar s$ and $c=\bar c$.  
In this case, 
\be
\sigma^\nu_{CC}-\sigma^{\bar\nu}_{CC}\sim (u+d-\bar u-\bar d)
[{2\over 3}-s^2_\theta(f-1)]\,,
\ee
and all sea quark terms cancel.

Similarly, the corresponding difference for the NC
processes can be calculated, giving (the sea quark terms cancel here as well)
\be
\sigma^\nu_{NC}-\sigma^{\bar\nu}_{NC}\sim {2\over 3}(u+d-\bar u-\bar d)
[g_L^2(u)+g_L^2(d)-g_R^2(u)-g_R^2(d)]\,,
\ee
so that the ratio can be defined 
\be
R^-\equiv {\sigma^\nu_{NC}-\sigma^{\bar\nu}_{NC}\over
\sigma^\nu_{CC}-\sigma^{\bar\nu}_{CC} }={\mbox{${1\over 2}$}-\sin^2\theta_w
\over 1+\mbox{${3\over 2}$}s^2_\theta(f-1)}\,,
\ee
where the SM expressions for $g_{L,R}(u,d)$ have been inserted.
This ratio, in the limit $f\to 1$, is the Paschos-Wolfenstein relation.
In LO, one sees that a measurement of $R^-$ will determine 
$\sin^2\theta_w$ if $f$ is known.  Experimentally, $f$ is determined by
measuring the charm production cross section in CC reactions via the dimuon
final state and then parameterizing it in terms of the slow rescaling
formalism.  

Of course, both the QCD and electroweak radiative corrections
need to be performed.  The electroweak corrections have both universal
and non-universal contributions, \eg,
\be
g_L(u)={1\over 2}-{2\over 3}\sin^2\theta_w \quad \to\quad
\rho_{\nu N}[{1\over 2}-{2\over 3}\kappa_{\nu N}\sin^2\theta_w]+\lambda_{Lu}\,,
\ee
with similar redefinitions for the remaining couplings. These corrections are
known in both the OS and $\overline{\rm MS}$ schemes.  NuTeV chooses the
OS scheme and extracts the preliminary value of $\sin^2\theta_w|_{\rm OS}$ at 
one-loop (taking $m_t=175$ GeV and $m_H=150$ GeV) 
\be
\sin^2\theta_w|{\rm OS}=0.2253\pm 0.0019\pm 0.0010\,,
\ee
which relates to
\be
M_W=80.26\pm 0.11 \gev\,.
\ee
This is in good agreement with the direct $M_W$ measurements performed at
LEPII and the Tevatron.

\subsubsection{Atomic Parity Violation}

Atomic parity violation occurs through the parity violating terms in the
$Z$-boson exchange between the electrons and the nucleus of an atom.
The effective Hamiltonian shows that there are two possible contributions due
to the $g_V$ and $g_A$ couplings; $g_V^eg_A^{\rm nucl}$ and 
$g_A^eg_V^{\rm nucl}$.  A
short analysis shows that atomic parity violating effects are coherent so
that the $g_A^eg_V^{\rm nucl}$ contribution
would be enhanced in heavy nuclei.  In addition, the claimed
uncertainties in the atomic wave function for Cesium and Thallium, for which the
most precise data is available, is of order $\sim 1\%$. The size of atomic
parity violation is described by what is called the weak charge,
\be
Q_w\sim -2[g_A^eg_V^uN_u+g_A^eg_V^dN_d]\,,
\label{qw}
\ee
where $N_{u,d}$ represents the number of $u$- and $d$-quarks in the relevant
nucleus and $g_{V,A}^{e,u,d}$ are the relevant $Zf\bar f$ couplings.  In
leading order one finds directly
\be
Q_w=Z(1-4\sin^2\theta_w)-N\,.
\ee
Including the weak corrections entails making the replacements
\bea
g_A^eg_V^u & \to & \rho_{eq}\left[ -{1\over 2}+{4\over 3}\kappa_{eq}\sin^2
\theta_w\right] +\lambda'_u\,,\nonumber\\
g_A^eg_V^d & \to & \rho_{eq}\left[ {1\over 2}-{2\over 3}\kappa_{eq}\sin^2
\theta_w\right] +\lambda'_d\,,
\eea
with the corrections being known in both the OS and
$\overline{\rm MS}$ schemes.  The most precise data,\cite{pdg98} in
comparison to the SM predictions are listed in Table \ref{qwdata}, where the
error on the SM prediction arises from varying $m_t$ and $m_H$.  We see that
there is excellent agreement at the $1\sigma$ level.

\begin{table}
\centering
\begin{tabular}{|c|c|c|} \hline\hline
& Data & SM Prediction \\ \hline
$Q_w(Cs)$ & $-72.41\pm 0.84$ & $-73.12\pm 0.06$ \\
$Q_w(Th)$ & $-114.8\pm 3.6$ & $-116.7\pm 0.1$ \\ \hline\hline
\end{tabular}
\caption{Comparison of measurements of the atomic weak charge in Cesium and
Thallium with the SM predictions.}
\label{qwdata}
\end{table}

\subsection{Radiative Corrections IV: Higgs Mass Bounds}

Assuming the SM describes all of the present data, the $Z$-pole results
and the low-energy data,
combined with direct measurements of $M_W\,, m_t$, and $\alpha_s$,
can be used to constrain the Higgs mass since it is the
only remaining free parameter.  Presently, direct searches at LEP have
bounded the Higgs mass from below, with the current preliminary 
limit \cite{dt} being  $m_H>93.6$ GeV.

\begin{figure}[htbp]
\centerline{
\psfig{figure=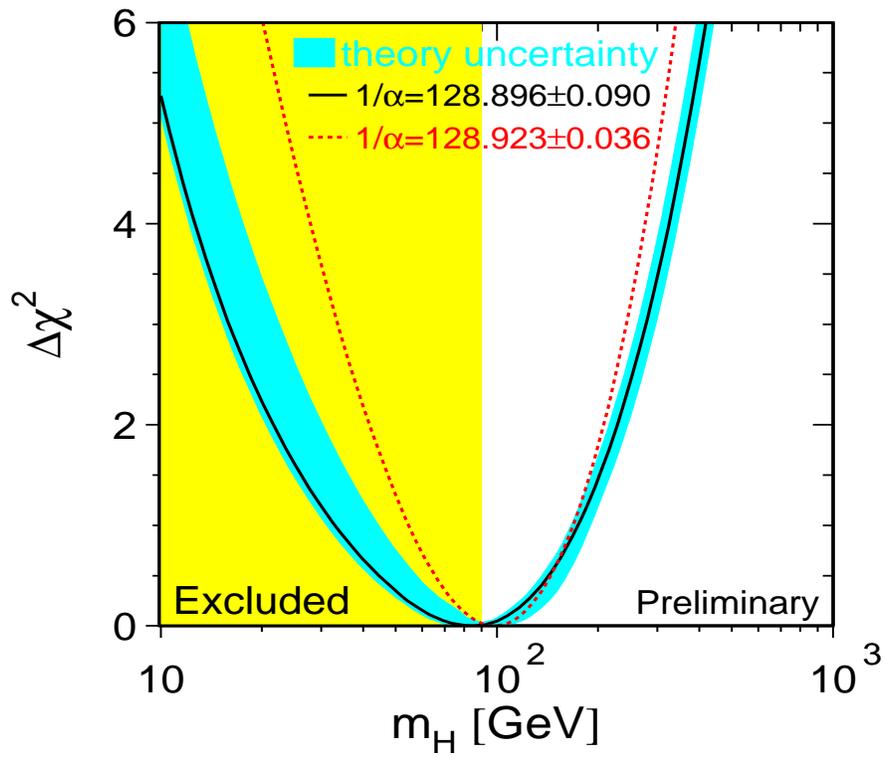,height=10cm,width=12cm,angle=0}}
\vspace*{0.5cm}
\caption{The $\Delta\chi^2$ fit using all data as a function of the Higgs
mass.  The vertical band displays the $95\%$ C.L. exclusion limit on $m_H$
from the direct search.  From~\protect\cite{karlen}.}
\label{higgsbounds}
\end{figure}

Figure \ref{higgsbounds}
displays the $m_H$ dependent $\Delta\chi^2$ distribution for all of
the above data.  The best fit is obtained for 
\be
m_H=84^{+91}_{-51}\gev\,,
\ee
with a $95\%$ C.L. upper limit of 280 GeV.  Much of the weight toward lower
values is driven by the $Z$-pole leptonic asymmetries and the determinations
of $M_W$.  If one includes the experimental lower bound from direct searches
in the fit, the one-sided $95\%$ C.L. upper limit on $m_H$ is not much
affected.  Note that these bounds would change if new physics were present.
This result has generally led to the speculation that the 
discovery of the Higgs may not be too far in the distant future.

\nn

\subsection{$S\,, T\,, U$ Formalism}

A convenient parameterization which describes potential new physics
contributions to electroweak radiative corrections is given by the
$S\,, T\,, U$ formalism of Peskin and Takeuchi.\cite{pestak}  These parameters
are defined such that they vanish for
a reference point in the SM (\ie, a specific value for the top-quark and Higgs 
masses) and deviations from zero would then signal the existence of new physics.
We note that an alternative set of parameters, $\epsilon_i$, also exist
\cite{guido} which not require a reference point for the SM.

These parameters are only sensitive to new physics which contributes to the
so-called oblique corrections.  As discussed in
the previous sections, there are three classes of radiative corrections to
4 fermion processes: vacuum polarization corrections, vertex corrections, and
box corrections.  The vacuum polarization corrections are usually referred
to as oblique corrections, as they only affect the propagation and mixings
of the gauge bosons and do not change the form of the interaction itself.
They are thus independent of the final state fermions
and affect all processes with electroweak gauge boson exchange universally.
Whereas the direct corrections, \ie, the vertex and box corrections, are
clearly dependent on the final state fermions and are specific to each
process.  For new physics to affect the direct corrections, it must couple
directly to the light external fermions.  Such couplings are expected to
be suppressed in most cases, with one exception being the $Zb\bar b$ vertex.

The assumptions inherent in the $S\,, T\,, U$ formalism about the nature of the 
new physics are:
\begin{enumerate}
\item{The electroweak gauge group is given by $SU(2)_L\times U(1)_Y$, and
thus there are no additional electroweak gauge bosons beyond the $\gamma\,
Z$ and $W$.}
\item{New physics couplings to light fermions are suppressed, and hence only
oblique corrections need to be considered.}
\item{The new physics scale is large compared to the electroweak scale.}
\end{enumerate}
The first two assumptions indicate that 
only four vacuum polarization functions need to be considered, namely
the self-energies of the $\gamma\,, Z$, and $W$, and $\gamma-Z$ mixing.
Defining the notation
\be
\int d^4xe^{iq\cdot x}\langle J_X^\mu(x)J_Y^\nu(0)\rangle =
ig^{\mu\nu}\AQ{XY}+(q^\mu q^\nu {\rm term}) \,,
\ee
where $J_X$ is the current that couples to gauge boson $X$, these four
vacuum polarization functions can be denoted as $\AQ{\gamma\gamma}(q^2)\,
\AQ{ZZ}(q^2)\, \AQ{WW}(q^2)$\, and $\AQ{\gamma Z}(q^2)$.  Note that the
$q^\mu q^\nu$ terms of the vacuum polarization tensors are neglected here
as they only correct the longitudinal components of the gauge boson
propagators and are thus suppressed by $m_f^2/M_{W/Z}^2$.

The third assumption allows for the expansion of the new physics contributions 
to the self-energies
in powers of $q^2/M_{new}$ about $q^2=0$, where $M_{new}$ represents the
heavy scale of the new interactions.  Keeping only constant and linear terms
in $q^2$ thus yields
\bea
\AQ{\gamma\gamma} 
& = & q^2 \AP0{\gamma\gamma} + \cdots \nonumber\\
\AQ{Z\gamma} 
& = & q^2 \AP0{Z\gamma} + \cdots \\
\AQ{ZZ} 
& = & \A0{ZZ} + q^2 \AP0{ZZ} + \cdots \nonumber\\
\AQ{WW} 
& = & \A0{WW} + q^2 \AP0{WW} + \cdots \nonumber
\label{LINEAR}
\eea
for the part of the vacuum polarization functions which arise solely from 
the new interactions.  This approximation thus allows us to express the new
contributions in terms of six parameters.  Three of these may be absorbed
into the renormalization of the input parameters $\alpha\, G_F$, and $M_Z$,
leaving three that are measurable.  One choice for these parameters is given 
by \cite{pestak}
\bea
\alpha S & = & 4s_w^2 c_w^2
               \left[ \AP0{ZZ}
                          -\frac{c_w^2-s_w^2}{s_wc_w}\AP0{Z\gamma}
                          -\AP0{\gamma\gamma}
               \right]\,,  \nonumber \\
\alpha T & = & \frac{\A0{WW}}{\mw^2} - \frac{\A0{ZZ}}{\mz^2}\,, \\
\alpha U & = & 4s_w^2
               \left[ \AP0{WW} - c_w^2\AP0{ZZ}
                         - 2s_wc_w\AP0{Z\gamma} - s_w^2\AP0{\gamma\gamma} 
               \right]\,. \nonumber
\eea
This definition ensures that the parameters $T$ and $U$ vanish if the
new physics observes custodial isospin symmetry.  In fact $T$ represents
the shift of the $\rho$ parameter due to new physics,
\be
\rho = 1 + \delta\rho_{\rm SM} + \alpha T\,. 
\ee

\begin{figure}[htbp]
\centerline{
\psfig{figure=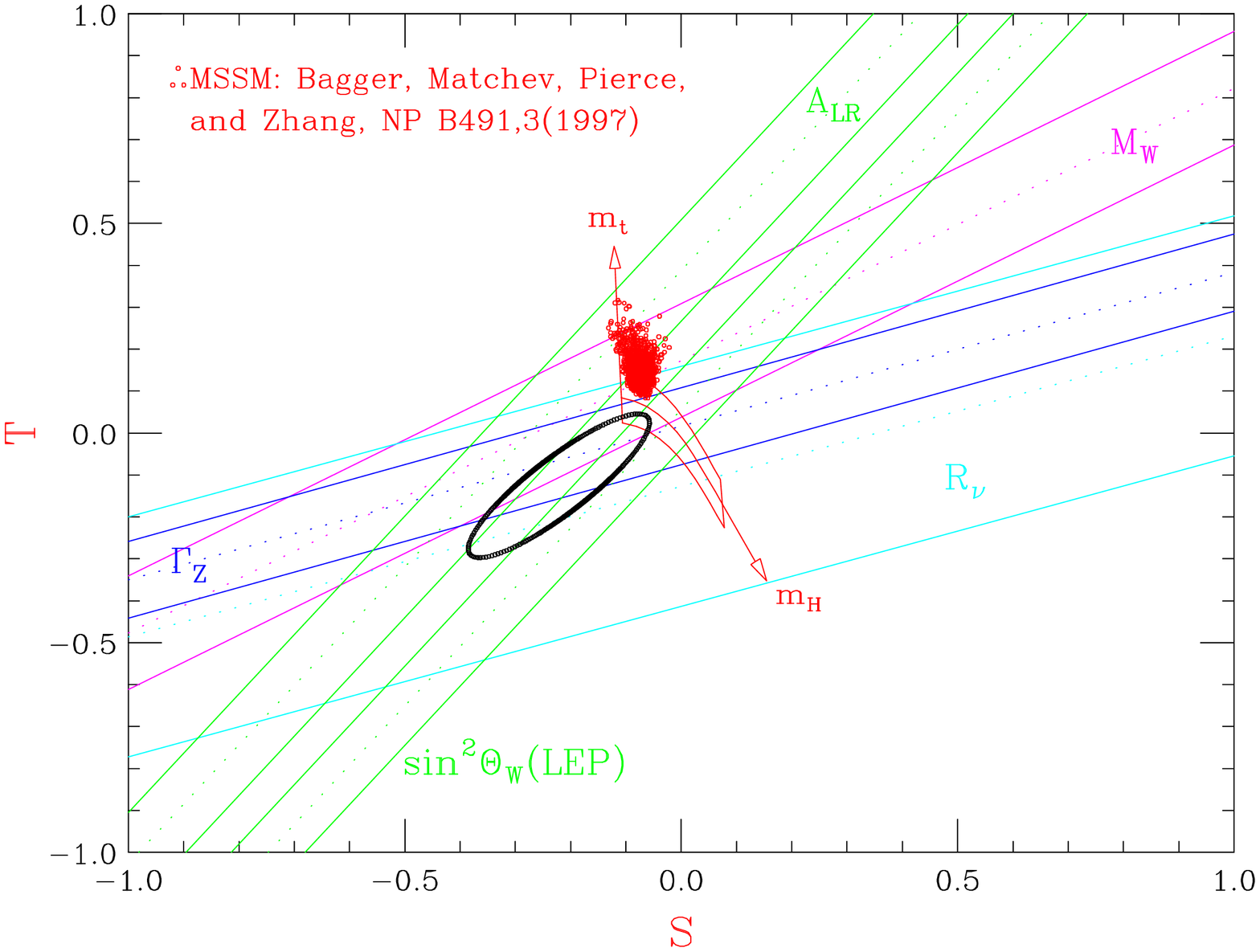,height=12cm,width=14cm,angle=90}}
\vspace*{-1cm}
\caption{The $68\%$ C.L. regions in the $S-T$ plane determined by various
measurements as labeled.  Also shown are the results from a global fit to the 
world's electroweak data which yields 
the $68\%$ elliptical confidence region.  The area predicted
by the SM as described in the text is shown as the banana shaped region,
while the MSSM expectations are denoted by the series of small circular
points.}
\label{stu}
\end{figure}

The electroweak observables can be expressed in terms of the parameters
$S\,, T\,, U$ in a straightforward fashion \cite{stuobserve} given the
input values for $\alpha\,, G_F$, and $M_Z$ and a reference value for
$m_H$.  These relations are
\bea
M_Z^2 & = & M^2_{Z0} {1-\alpha T \over 1-G_FM^2_{Z0}S/2\sqrt 2\pi}\,, 
\nonumber\\
M_W^2 & = & M^2_{W0} {1 \over 1-G_FM^2_{W0}(S+U)/2\sqrt 2\pi}\,,
\nonumber\\
\Gamma_Z & = & {1\over 1-\alpha T}{M_Z^3\Gamma_{Z0}\over M^3_{Z0}}\,,\\
A_i & = & {1\over 1-\alpha T}A_{i0} \,,\nonumber
\eea
where $A_i$ represents a neutral current amplitude and  $M_{Z0}\,, M_{W0}\,, 
\Gamma_{Z0}\,, A_{i0}$ are the Standard Model expressions in the 
$\overline{\rm MS}$ scheme.
Note that the $Z$-pole measurements do not depend on $U$, and
that this parameter only affects the $W$-boson mass.
A global fit \cite{pdg98} to the precision electroweak data presented in
Fig. \ref{pull} yields the determination
\bea
S & = & -0.16 \pm 0.14 \quad (-0.10)\,, \nonumber \\
T & = & -0.21 \pm 0.16 \quad (+0.10)\,, \\
U & = & 0.25 \pm 0.24 \quad (+0.01)\,, \nonumber
\label{stufit}
\eea
taking $m_t=175\pm 5$ GeV.  The central values assume $m_H=M_Z$ while the
numbers in parentheses describe the change for $m_H=300$ GeV.  These parameters
are all consistent with their Standard Model values of zero within roughly
$1\sigma$.  The corresponding $68\%$ C.L. allowed region \cite{baird} 
(with updated data from the 1998 summer conferences) in the 
$S-T$ plane is displayed in Fig. \ref{stu}.  Also shown are the $68\%$ C.L.
bands in this plane from the measurements of $A_{LR}$, $\sin^2\theta_w^{\rm 
eff}$ from LEP, $\Gamma_Z$, $M_W$ (assuming $U=0$), and the ratio
of charged to neutral current cross sections.  The Standard Model prediction is
represented by the banana shaped region, where the right-hand edge of this
area corresponds to $m_H=88 $ GeV and $m_t=173.9\pm 5.2$ GeV.  Increasing the
Higgs mass to 1 TeV sweeps out the bounded area.  The predictions of the
Minimal Supersymmetric Standard Model (MSSM) from Bagger \etal,\cite{damien}
are also shown as a series of small circular points, where each point
corresponds to a separate choice for the 5 MSSM parameters.  We see that
again the global fit is consistent with the SM, and that in the absence of
the LEP $\sin^2\theta_w^{\rm eff}$ measurements the remaining measured bands
would overlap in the region of negative $S$.

\vspace*{-0.5cm}
\nn

As an example of how new physics can affect these parameters, let us
consider the case of a new color-singlet heavy fermion $SU(2)_L$ doublet,
$(N\,, E)$, with $m_{N,E}\gg M_Z$.  The contribution of this doublet to
$S\,, T\,, U$ is given by
\bea   
S & = & \frac{1}{6\pi}\left[ 1 - Y \ln\left(\frac{m_N^2}{m_E^2}
                                      \right)
                      \right]\,, \nonumber\\
T & = & \frac{1}{16\pi s^2 c^2 \mz^2}
        \left[ m_N^2 + m_E^2 - \frac{2 m_N^2 m_E^2}{m_N^2 - m_E^2}
                               \ln\left(\frac{m_N^2}{m_E^2}
                                  \right)
        \right]\,,  \\
U & = & \frac{1}{6\pi}
        \left[ -\frac{5 m_N^2 - 22 m_N^2 m_E^2 + 5 m_E^2}
                     { 3( m_N^2 - m_E^2 )^2 }
        \right.  \nonumber\\
  &   & \phantom{\frac{1}{6\pi}}
        \left.
               +\frac{m_N^6 - 3 m_N^4 m_E^2 - 3 m_N^2 m_E^4 + m_E^6}
                     { (m_N^2 - m_E^2)^3 }
                \ln\left( \frac{m_N^2}{m_E^2}
                   \right)
        \right]\,, \nonumber
\eea
where $Y$ represents the hypercharge of the doublet.
The expression for $T$ is just the familiar contribution of a fermion $SU(2)_L$ 
doublet to the $\rho$ parameter and restricts the size of the mass splitting
within the doublet.  The experimental bounds on $T$, or $\rho$, 
tightly constrain the mass splitting to be very small within any new fermion 
doublet, \ie, $\Delta m\equiv |m_N-m_E|\ll m_{N,E}$.  In fact, any new
doublets must be nearly degenerate!  In this case the above expressions
simplify to
\bea
S & \approx & \frac{1}{6\pi} \approx 0.05\,, \nonumber \\
T & \approx & \frac{1}{12\pi s^2 c^2}
              \left[\frac{(\Delta m)^2}{\mz^2}\right]\,, \\
U & \approx & \frac{2}{15\pi}
              \left[\frac{(\Delta m)^2}{m_N^2}\right]\,. \nonumber
\eea
This conflicts with the global fit value for $S$ at the $1.5-2\, \sigma$
level.  A generalization of the contribution to $S$ from a 
multiplet of heavy degenerate chiral fermions is given by
\be
S=C\sum_i [T_{3L}^i-T_{3R}^i]^2/3\pi \,,
\ee
where $T_{3L,R}$ is the third component of left-,right-handed weak isospin
of fermion $i$ and $C$ is the number of colors.
This is a serious problem for theories which contain a large number
of extra fermion doublets, such as technicolor models.  For example, in
technicolor models with QCD-like dynamics, a full technigeneration 
yields \cite{mitch} $S\sim 1.62$, which is clearly excluded!  However,
models of walking technicolor can avoid these difficulties and can yield
smaller or even negative values of $S$.\cite{walktc}  Consistency with
the values of $S$ and $T$ has now become a standard viability test
in constructing theories beyond the Standard Model.

If the third assumption above is relaxed and the scale of the new physics
is near the electroweak scale, then the linear approximation in Eq. (120)
no longer applies and  more parameters need to be introduced.
In this case, Burgess \etal \cite{stuvwx} have shown that it is sufficient
to introduce three additional parameters, bringing the total number to six.
The definitions of $S$ and $U$ become slightly modified, $T$ is unchanged,
and the additional parameters are denoted as $V\,, W\,, X$.  The revised
set of parameters are defined by
\bea
\alpha S
& = & 4s_w^2 c_w^2\left[ \BZ{ZZ}
                    -\frac{c_w^2-s_w^2}{s_wc_w}\B0{Z\gamma}
                    -\B0{\gamma\gamma}
              \right]\,, \nonumber\\
\alpha T 
& = & \frac{\A0{WW}}{\mw^2} - \frac{\A0{ZZ}}{\mz^2}\,, \nonumber \\
\alpha U
& = & 4s_w^2\left[ \BW{WW} - c_w^2\BZ{ZZ} \right. \nonumber\\
& & \quad\quad \left. - 2s_wc_w\B0{Z\gamma} - s_w^2\B0{\gamma\gamma}
          \right] \,, \nonumber\\
\alpha V & = & \APZ{ZZ} - \BZ{ZZ} \,, \\
\alpha W & = & \APW{WW} - \BW{WW} \,, \nonumber\\
\alpha X & = & -s_wc_w\left[ \frac{\AZ{Z\gamma}}{\mz^2} - \B0{Z\gamma}
                        \right]   \,,\nonumber
\label{STUVWX}
\eea
Clearly in the limit $M_{new}\to\infty$, $S$ and $U$ coincide with their
original definitions and $V\,, W\,, X$ vanish.

\begin{figure}[t]
\centerline{
\psfig{figure=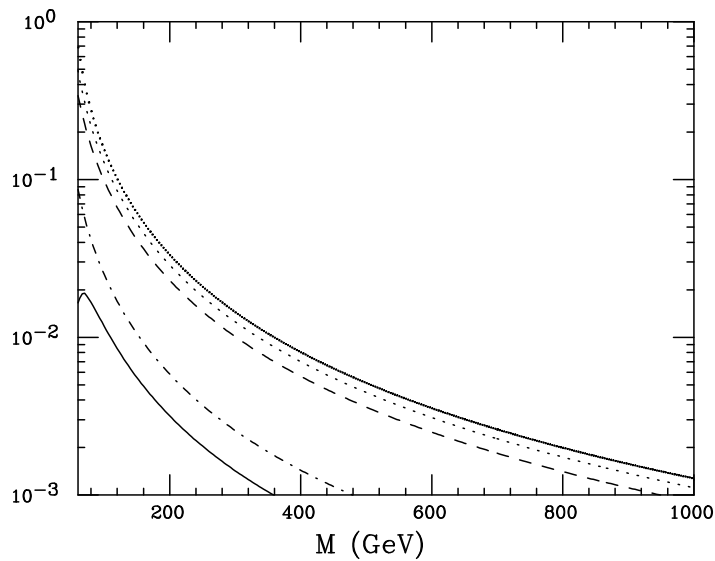,height=9cm,width=12cm,angle=-90}}
\vspace*{-1cm}
\caption{\small Contribution of 3 generations of degenerate $E_6$ exotic
fermions of mass $M$ and their SUSY partners to the oblique parameters.
From top to bottom the curves correspond to the parameter $-V, -W, -S,
X,$ and $-U$, respectively.}
\label{exotics}
\end{figure}

It is interesting to note that some types of new physics may be quite
close to the electroweak scale and yet make little contribution to the
oblique parameters.  As an example we consider the low-energy sector of
string-inspired SUSY $E_6$ theories wherein the particle spectrum of the
MSSM is augmented by three generations of vector-like fermions and their
supersymmetric partners.  If, for simplicity, the small mixing between 
these states and the SM fields is neglected and they are taken to be 
degenerate, then the new contributions to $T$ vanish automatically due
to the vector-like nature of the exotic fields.  The corresponding
contributions to the other oblique parameters are presented in Fig.
\ref{exotics} and are seen to be small ($\lsim 0.1$) for $m\gsim 150$
GeV.  This example demonstrates that new physics may be lurking nearby
without manifesting itself in the oblique corrections.

\vspace*{-0.5cm}
\nn

\subsection{Gauge Boson Pair Production}

One element of the SM remains to be directly tested with significant precision,
namely the non-Abelian self-couplings of the weak gauge bosons.  Deviations
from the SM gauge theory predictions for the Yang-Mills self-interactions
would clearly signal the existence of new physics, possibly arising from, \eg, 
loop corrections of new particles, substructure in the gauge boson
sector, or additional interactions involving new gauge bosons.  In addition, 
precise measurements of the $WWV$ three-point function
(where $V=\gamma$ or $Z$) can provide information on the nature of electroweak 
symmetry breaking.

\begin{figure}[htbp]
\centerline{
\psfig{figure=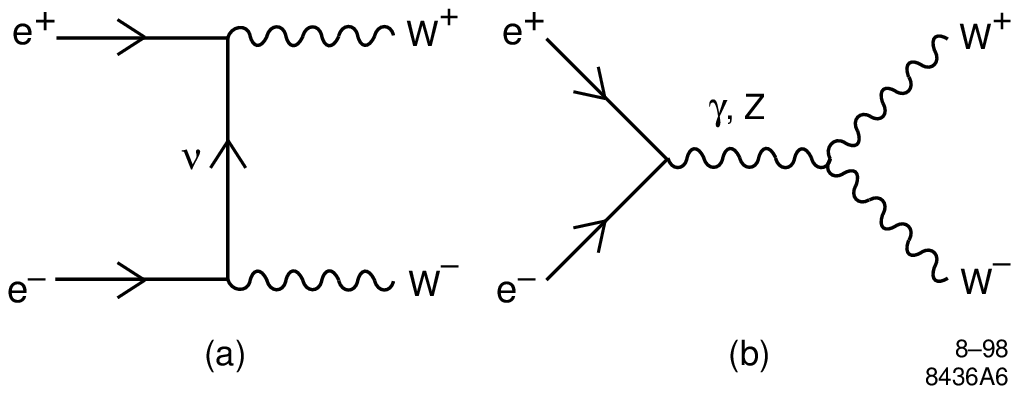,height=5cm,width=10cm,angle=0}}
\vspace*{0.5cm}
\caption{Feymann diagrams mediating the reaction $\epem\to W^+W^-$.}
\label{eeww}
\end{figure}

\begin{figure}[t]
\centerline{
\psfig{figure=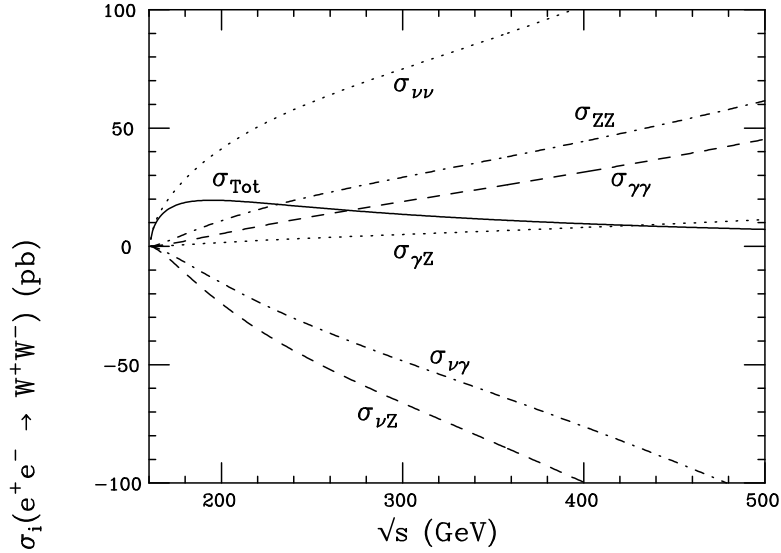,height=9cm,width=12cm,angle=-90}}
\vspace*{-1cm}
\caption{The energy dependence of each separate contribution as labeled, as 
well as the total cross section for $W$ pair production in \epem\ collisions.}
\label{wpairs}
\end{figure}

\begin{figure}[htbp]
\centerline{
\psfig{figure=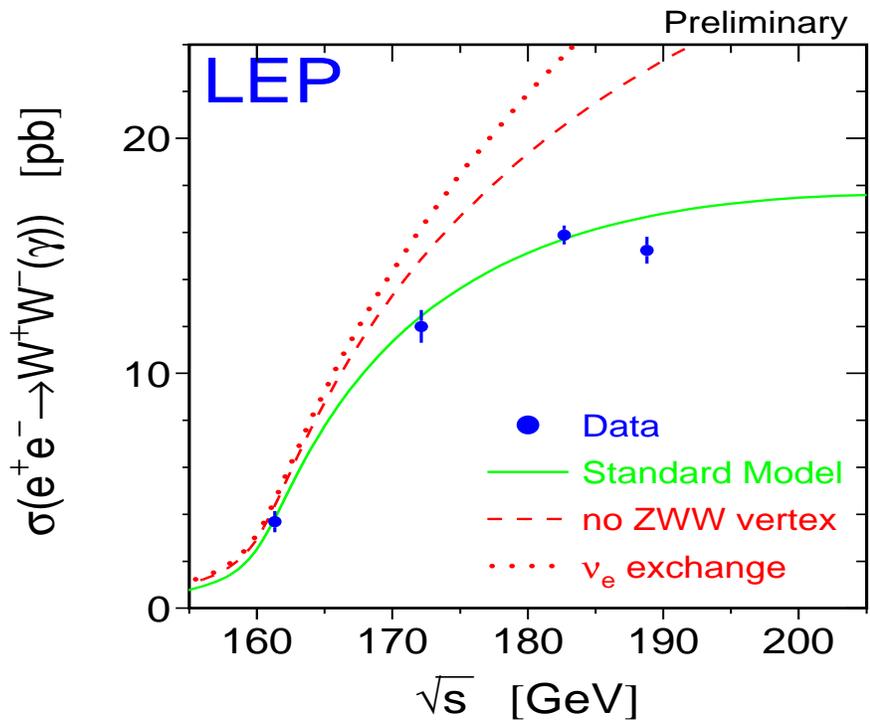,height=11cm,width=12cm,angle=0}}
\vspace*{-0.5cm}
\caption{Energy dependence of the cross section for $\epem\to W^+W^-$ as
measured at LEP II.  From Ref.~\protect\cite{karlen}.}
\label{lepww}
\end{figure}

The pair production of gauge bosons in \epem\ or $q\bar q$ annihilation
efficiently probes the structure of the trilinear couplings.  In particular,
the energy dependence of the cross section for the reaction $\epem\to W^+W^-$
is critically dependent on the gauge cancellations.  For example, the
contribution of the neutrino exchange diagram depicted in Fig. \ref{eeww}(a)
grows very rapidly with energy,
\be
\sigma_{\nu\nu}\simeq {\pi\alpha^2 s\over 96x_w^2M_W^4}\,,
\ee
and violates the optical theorem.  This presented a serious problem for the
Fermi theory of weak interactions, which was resolved only when 
gauge theories were introduced for the electroweak interactions.  Including the
SM s-channel gauge boson exchange diagrams of Fig. \ref{eeww}(b) yields
the behavior
\be
\sigma_{\rm total}\simeq {\pi\alpha^2\over 2x_w^2s}\ln {s\over M_W^2} \,
\ee
for the total cross section.
These gauge cancellations are explicitly illustrated in Fig. \ref{wpairs},
where the energy dependence of each contribution as well as for the total cross
section for $\epem\to W^+W^-$ is shown.  This energy dependence has just now 
been measured at LEPII over a limited energy range near and just above
threshold, as shown in 
Fig. \ref{lepww} from Ref. \cite{karlen}.  These measurements are preliminary,
but are in complete agreement with the SM, \ie, the slightly low data point at 
$\sqrt s=189$ GeV is no cause for concern at present.  A higher energy \epem\ 
collider is necessary in order to get a better lever arm
in determining the high $\sqrt s$ behavior displayed in Fig. \ref{wpairs}.

\nn

\nn

\nn

The cross section for $\epem\to W^+W^-$ is quite large, given roughly by
20 units of $R$ where $R$ is defined in Section 2.1,  and presents a serious
source of background for new physics signatures in high energy \epem\ 
collisions.  However, if polarized beams
are available, this cross section can be substantially reduced by adjusting
the initial electron beam polarization to be mainly right-handed.
In addition, the angular dependence of this reaction is peaked in the
forward direction due to the t-channel pole, with the $W^+$ being produced 
preferentially along the $e^+$ direction.  This peaking sharpens with
increasing center of mass energy and may also be used to differentiate $W$ pair
production from new physics.  

Potential deviations from the SM form of the $WWV$ trilinear couplings
are parameterized in terms of the  most general Lorentz invariant
effective Lagrangian \cite{hhpz}
\bea
{\cal L}^{eff}_{WWV} & = & g_{WWV}\Bigl[ ig_1^V(W^\dagger_{\mu\nu}W^\mu V^\nu
-W^\dagger_\mu V_\nu W^{\mu\nu})+i\kappa_V W^\dagger_\mu W_\nu V^{\mu\nu}
\nonumber\\
& & \quad + {i\lambda_V\over M_W^2}W^\dagger_{\lambda\mu}W^\mu_\nu 
V^{\nu\lambda} - g_4^V W^\dagger_\mu W_\nu(\partial^\mu V^\nu +\partial^\nu
V^\mu) \\
& & \quad  +g_5^V\epsilon^{\mu\nu\rho\sigma}(W^\dagger_\mu
\stackrel{\leftrightarrow}{\partial}_\rho W_\nu)V_\sigma
+\tilde\kappa_V W^\dagger_\mu W_\nu \tilde V^{\mu\nu}+{i\tilde\lambda_V\over 
M^2_W}W^\dagger_{\lambda\mu}W^\mu_\nu\tilde V^{\nu\lambda}\Bigr] \,,\nonumber
\label{wwveff}
\eea
where $W_{\mu\nu}\equiv\partial_\mu W_\nu-\partial_\nu W_\mu$,
$V_{\mu\nu}\equiv\partial_\mu V_\nu-\partial_\nu V_\mu$, $(A
\stackrel{\leftrightarrow}{\partial}_\mu B)\equiv A
(\partial_\mu B)-(\partial_\mu
A)B$, and $\tilde V_{\mu\nu}\equiv {1\over 2}\epsilon_{\mu\nu\rho\sigma}
V^{\rho\sigma}$.  The overall normalization is defined such that
$g_{WW\gamma}=-e$ and $g_{WWZ}=-e\cot\theta_w$.  The coupling $g_5^V$ violates
$C$ and $P$ separately, while $g_4^V\,, \tilde\kappa_V$, and $\tilde\lambda_V$ 
are $CP$-violating.  At tree-level within the SM 
the couplings take on the values $g_1^V=\kappa_V=1$ and $\lambda_V=g_5^V=
g_4^V=\tilde\kappa_V=\tilde\lambda_V=0$.  For on-shell photons, electromagnetic
gauge invariance fixes $g_1^\gamma=1$ and $g_5^V=0$.  Convention dictates that
anomalous values of the $CP$-conserving couplings are denoted as
$\Delta g_1^Z\equiv g_1^Z -1$, $\Delta\kappa_V\equiv\kappa_V-1$, 
$\lambda_V$, and $g_5^Z$.  The $C$ and $P$-conserving terms in the effective
$WW\gamma$ interaction correspond to the lowest-order terms of an
electromagnetic multipole expansion via
\bea
Q_W & = & eg_1^\gamma\,,\nonumber\\
\mu_W & = & {e\over 2M_W}(g_1^\gamma+\kappa_\gamma+\lambda_\gamma)\,,\\
q_W & = & -{e\over M_W^2}(\kappa_\gamma-\lambda_\gamma) \,,\nonumber
\eea
where $Q_W$ represents the charge, $\mu_W$ the magnetic dipole moment, and $q_W$
the electric quadrupole moment of the $W$-boson.  Whereas the two 
CP-violating couplings are related to the electric dipole moment, $d_W$, and
magnetic quadrupole moment, $\tilde Q_W$, of the $W$ by
\bea
d_W & = & {e\over 2M_W}(\tilde\kappa_\gamma+\tilde\lambda_\gamma)\,,\\
\tilde Q_W & = & -{e\over M_W^2}(\tilde\kappa_\gamma-\tilde\lambda_\gamma)\,.
\nonumber
\eea
It is important to keep in mind that these parameterized couplings are
form factors and hence are functions of $q^2$, \eg, $\lambda(q^2)=\lambda
(1+q^2/\Lambda^2)^{-n}$, where $q$ is the momentum transfer, $\Lambda$ is
the form factor scale, and $n=2$ for $WWV$ couplings.  Other parameterizations 
of these triple gauge boson couplings can be found in Ref. \cite{hisz}.

In any model with new physics that couples to the $W$, anomalous trilinear
couplings will be induced.  However, theoretical arguments suggest that the 
expected values of the induced couplings are small.  
Several general analyses have been
performed \cite{anomsize} where the contributions of new physics to these
couplings have been parameterized by both linear and non-linear effective 
Lagrangians.  In either case, the resulting anomalous couplings are found
to be suppressed by factors of $(M_W^2/\Lambda^2)$, where $\Lambda$ is the 
scale of the new interactions.  For new physics at the TeV scale or above,
this leads to typical values of the anomalous couplings of $10^{-2}$ or less.
In the SM, loop contributions generate these anomalous 
couplings at the level of $\sim 10^{-3}$,\cite{anomwsm}  and supersymmetric 
contributions yield \cite{anomwsusy} similarly small values.

Corresponding trilinear $ZZ\gamma$ and $Z\gamma\gamma$ couplings are not 
present in the SM, but may also arise from non-standard interactions.
In this case, the most general Lorentz invariant Lagrangian involving one
on-shell $Z$-boson and one on-shell photon is \cite{hhpz} (where $V=Z$ or
$\gamma$ is not necessarily on on-shell)
\bea
{\cal L}_{ZV\gamma} & = & -ie\left[ \left( h_1^VF^{\mu\nu}+h_3^V
\tilde F^{\mu\nu}\right) Z_\mu { (\Box + M_V^2)\over M_Z^2}V_\nu \right. 
\nonumber\\
& & \quad\quad \left. + \left( h_2^VF^{\mu\nu}+h_4^V
\tilde F^{\mu\nu}\right) Z^\alpha { (\Box + M_V^2)\over M_Z^2}\partial_\alpha
\partial_\mu V_\nu \right] \,,
\label{zvgeff}
\eea
where $F^{\mu\nu}$ is the photon field strength tensor and clearly,
$h_i^V=0$ in the SM.  $h^V_{1,2}$ are $CP$-violating.  Here, the
$(\Box-M_V^2)/M_Z^2$ factor is implied by Bose symmetry and indicates
that these couplings arise from higher dimensional operators
than in the $WWV$ case.  It is hence expected that they should be quite
suppressed and take on very small values in any new physics scenario.
These couplings
are related to the magnetic and electric dipole and quadrupole transition
moments of the $Z$ by
\bea
\mu_Z & = & -{e\over\sqrt 2 M_Z}{E^2_\gamma\over M_Z^2}(h_1^Z-h_2^Z) 
\,,\nonumber\\
d_Z & = & -{e\over\sqrt 2 M_Z}{E^2_\gamma\over M_Z^2}(h_3^Z-h_4^Z) 
\,,\nonumber\\
Q_Z & = & -{2\sqrt{10} e\over M_Z^2}h_1^Z\,,\\
\tilde Q_Z & = & -{2\sqrt{10} e\over M_Z^2}h_3^Z\,,\nonumber
\eea
where $E_\gamma$ represents the photon energy.  Due to the higher dimension
operators in this case, $n=3\,,4$ for $h^V_3\,, h^V_4$, respectively, in the
$q^2$ dependent form factor for these couplings.

\begin{figure}[htbp]
\centerline{
\psfig{figure=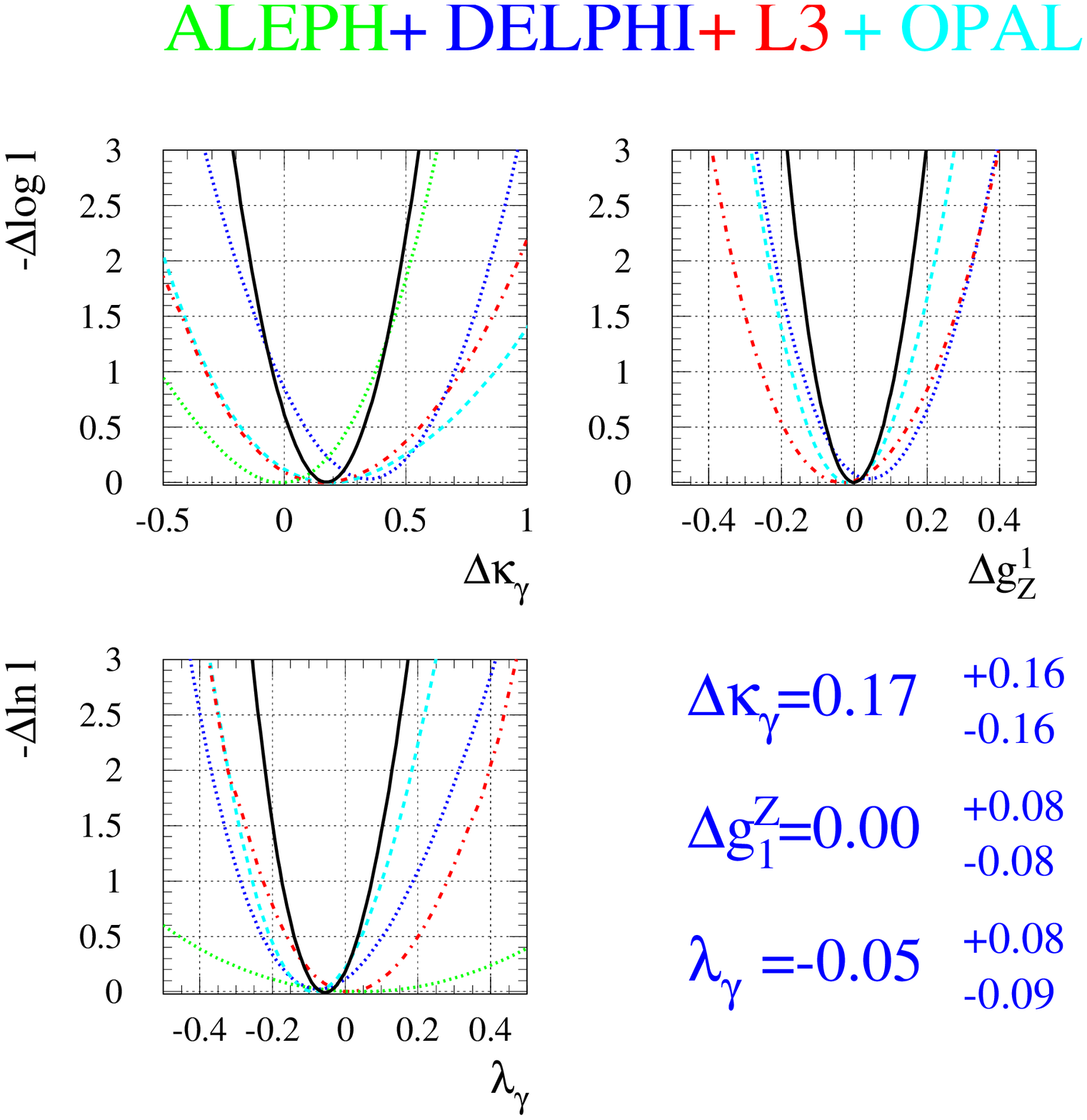,height=11cm,width=12cm,angle=0}}
\vspace*{-0.5cm}
\caption{Log likelihood distributions and allowed ranges for the anomalous
trilinear coupling parameters $\Delta\kappa_\gamma\,, \Delta g_1^Z$, and
$\lambda_\gamma$ as determined at LEPII.  From Ref.~\protect\cite{htp}.}
\label{wwvlepii}
\end{figure}

We now examine the current and prospective bounds placed on the anomalous 
trilinear couplings by present and future experiments.  In principle, all
14 free parameters in the Lagrangian of Eq. (131) are
simultaneously present in $W$ pair production at LEPII.  It would be an
impossible and meaningless exercise to consider all these parameters at
once and thus physical insights are employed to suggest which couplings
are the most likely to have observable effects at LEPII.  The data is then
analyzed with the following assumptions: (i) Only the CP-conserving couplings
are considered as bounds on the neutron electric dipole moment constrain
combinations of the CP-violating interactions.  (ii) Only the dimension 6
operators are considered as the higher dimension terms are more likely to
be suppressed. (iii) The gauge boson self-interactions should resemble, at
least approximately, those of the SM.  Hence SU(2)$_L\times$U(1)$_Y$ gauge
invariance is imposed and operators which would have produced large effects
at LEPI/SLC are excluded.  These assumptions reduce the set of free
parameters to just three, $\Delta g_1^Z\,, \Delta\kappa_\gamma$, and
$\lambda_\gamma$, with the constraints $\Delta\kappa_Z=\Delta g_1^Z-\Delta
\kappa_\gamma\tan^2\theta_w$ and $\lambda_Z=\lambda_\gamma$.  The 
preliminary combined results \cite{htp} from LEPII (including single $W$
production as well) are presented in Fig. \ref{wwvlepii}, where only one
parameter is taken to be non-zero at a time.  We see that all three
couplings are consistent with zero at this level of sensitivity, which is
$1-2$ orders of magnitude above the theoretical expectations.  We also note
that the momentum dependent form factors for these couplings are not
numerically relevant at LEPII.  The CP-conserving $ZV\gamma$ interactions are 
also constrained by LEPII via the reactions $\epem\to q\bar q\gamma,\nu\bar\nu
\gamma$.  The current bounds \cite{htp} are $|h^\gamma_3|<0.34$ and
$|h^\gamma_4|<0.55$, assuming a form factor scale of $\Lambda=1$ TeV.
We note that the $WW\gamma$ and $WWZ$ couplings may be probed separately via
the processes $\epem\to\nu\bar\nu\gamma$ and $\epem\to\nu\bar\nu Z$,
respectively, if the possibility of anomalous $ZV\gamma$ interactions are
neglected.

\begin{figure}[htbp]
\centerline{
\psfig{figure=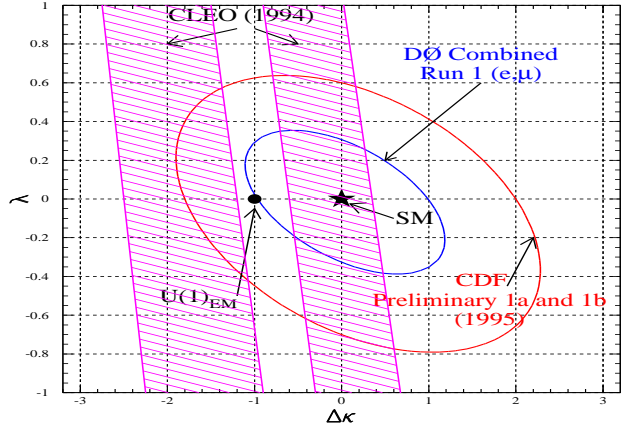,height=9cm,width=10cm,angle=0}}
\vspace*{-0.5cm}
\centerline{
\psfig{figure=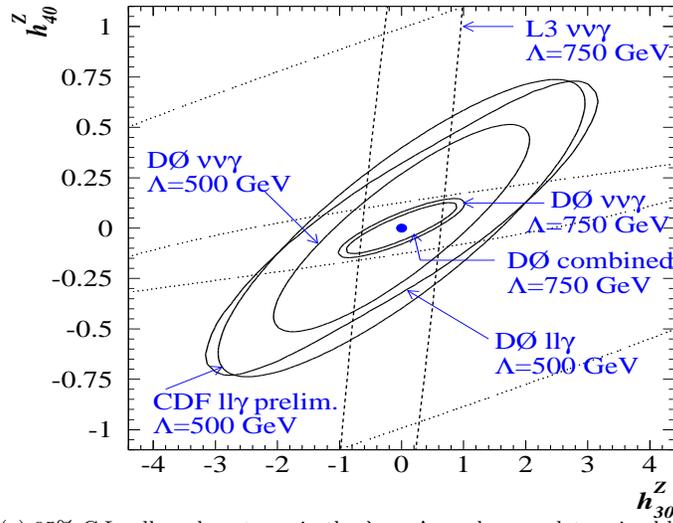,height=7cm,width=9cm,angle=0}}
\vspace*{-0.5cm}
\caption{(a) $95\%$ C.L. allowed contours in the $\lambda_\gamma-
\Delta\kappa_\gamma$ plane as determined by CDF and D0 from $W\gamma$
production (taking $\Lambda=1.5$ TeV).  $95\%$ C.L. allowed bands as
determined by CLEO from $B\to X_s\gamma$.  (b) $95\%$ C.L. allowed contours
in the $h_3^Z-h_4^Z$ plane for the various final state channels; the values
of $\Lambda$ and the experiments are as labeled.  From 
Ref.~\protect\cite{diboson}.}
\label{wwvtev}
\end{figure}

The trilinear gauge couplings are also constrained by diboson production
at the Tevatron.  The effect of these interactions is to increase the
diboson production cross section and to enhance the $p_T$ spectrum of the
gauge bosons at large values of $p_T$.  At hadron colliders it is important
to include the form factor dependence of the couplings in order to preserve
unitarity, taking $q^2=\hat s$ where $\sqrt{\hat s}$ is the subprocess center
of mass energy.  The $WW\gamma$ self-interactions are cleanly measured in
$W\gamma$ production, which has the advantage of being independent of the 
$WWZ$ vertex.  The $95\%$ C.L. bounds obtained \cite{diboson} by D0 from
such events, $-0.93<\Delta\kappa_\gamma<0.94\, (\lambda_\gamma=0)$ and
$-0.31<\lambda_\gamma<0.29\, (\Delta\kappa_\gamma=0)$ taking $\Lambda=1.5$ TeV,
are not as stringent as those obtained at LEPII, but are free of the
assumptions discussed above.  The results from the 2-parameter fit are
displayed in Fig. \ref{wwvtev}(a).  Also shown in the figure for comparison
are the constraints from the observation of the rare inclusive decay
$B\to X_s\gamma$ by CLEO.\cite{cleo}  We see that the case of $\lambda_\gamma
=\kappa_\gamma=0$ is excluded at the $95\%$ C.L.  The inclusion of the $WW$
and $WZ$ production channels with the $W\gamma$ events increases the
statistical sensitivity, but necessarily introduces a dependence on the
$WWV$ vertex.  Employing the same set of assumptions as used at LEPII yields
\cite{diboson} the $95\%$ C.L. bounds $-0.33<\Delta\kappa_\gamma<0.46$ and
$|\lambda_\gamma|<0.21$ from D0.  Combining these results with those from
LEPII gives
\bea
-0.15 & < & \Delta\kappa_\gamma<0.41\,,\nonumber\\
-0.16 & < & \lambda_\gamma<0.10 \,,
\eea
at $95\%$ C.L.  We note that the Tevatron analysis is not very sensitive to
$\Delta g_1^Z$.  Similarly, the $ZV\gamma$ interactions are constrained via
$Z\gamma$ production at the Tevatron.  Combining the $\epem\gamma/\mu^+\mu^-
\gamma$ and $\nu\bar\nu\gamma$ channels, D0 finds \cite{diboson}
\be
|h_3^{Z,\gamma}|<0.37\,, \quad\quad |h_4^{Z,\gamma}|<0.05\,, 
\ee
at $95\%$ C.L., taking $\Lambda=750$ GeV.  The $95\%$ C.L. allowed contours
in the $h_3^{Z,\gamma}-h_4^{Z,\gamma}$ plane are presented in 
Fig. \ref{wwvtev}(b) for the various final states with $\Lambda=500$ and 750
GeV.  The increased sensitivity of the $\nu\bar\nu\gamma$ channel is
illustrated in this figure.

\begin{figure}[htbp]
\centerline{
\psfig{figure=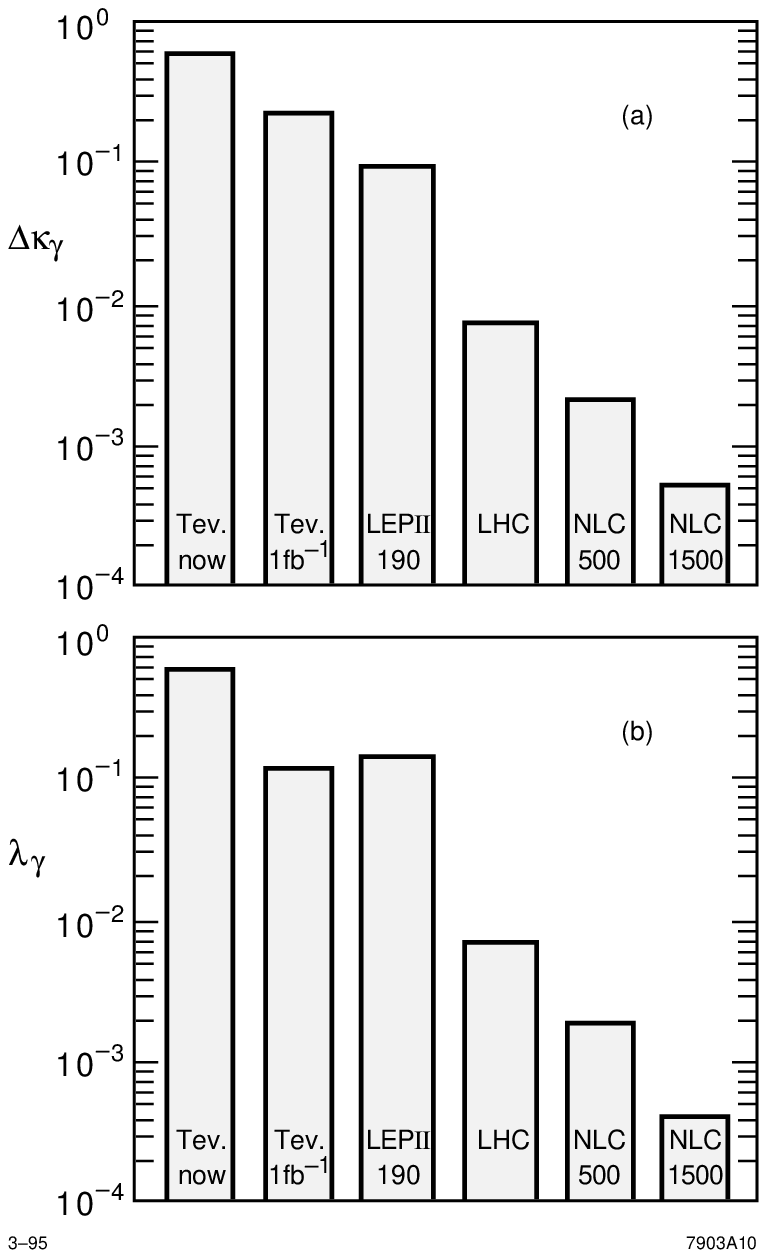,height=16cm,width=11cm,angle=0}}
\vspace*{0.5cm}
\caption{Comparison of $95\%$ C.L. constraints on $\Delta\kappa_\gamma$
and $\lambda_\gamma$ at present and future accelerators.  From 
\protect\cite{nlcbook}.}
\label{wwvfut}
\end{figure}

A comparison from Ref. \cite{nlcbook} of the anticipated level of 
sensitivity that can be reached on
$\Delta\kappa_\gamma$ and $\lambda_\gamma$ at $95\%$ C.L. from a variety
of processes at the Tevatron, LEPII, LHC, and NLC is presented in Fig.
\ref {wwvfut}.  We see from the figure that the region where one expects
the effects of new physics to appear, as discussed above, will only start to 
be probed at the LHC, and the predicted level of the SM loop
corrections can only be tested at a higher energy linear collider.

\nn

\nn

\vspace*{-0.5cm}
\nn

\section{Summary}

We have reviewed the basic components of the SM and the experiments which
have probed them.  However, due to time constraints there are many features
of the SM which we have omitted, such as heavy quark decays, the GIM mechanism,
and rare and forbidden processes.  Such processes, as well as the precision
measurements discussed here, provide powerful constraints
on potential scenarios of new physics.\cite{htt}

In summary we see that all elements of the SM enjoy an outstanding agreement
with an overwhelming set of experimental data, even at the quantum level.
In fact searches for new physics is tightly constrained by the success of the 
SM.  However, in light of the many questions that remain unanswered by the
theoretical framework of the SM, we hope that unexpected discoveries reveal
themselves in the near future!

\section*{Acknowledgments}
I thank P. Burrows and T. Rizzo for their assistance in preparing these
lectures and J. Bagger, the TASI staff, and the students
for providing a stimulating atmosphere for the school.

%
\def\IJMP #1 #2 #3 {Int. J. Mod. Phys.  {\bf#1},\ #2 (#3)}
\def\MPL #1 #2 #3 {Mod. Phys. Lett.  {\bf#1},\ #2 (#3)}
\def\NPB #1 #2 #3 {Nucl. Phys. {\bf#1},\ #2 (#3)}
\def\PLBold #1 #2 #3 {Phys. Lett. {\bf#1},\ #2 (#3)}
\def\PLB #1 #2 #3 {Phys. Lett.  {\bf#1},\ #2 (#3)}
\def\PR #1 #2 #3 {Phys. Rep. {\bf#1},\ #2 (#3)}
\def\PRD #1 #2 #3 {Phys. Rev.  {\bf#1},\ #2 (#3)}
\def\PRL #1 #2 #3 {Phys. Rev. Lett. {\bf#1},\ #2 (#3)}
\def\PTT #1 #2 #3 {Prog. Theor. Phys. {\bf#1},\ #2 (#3)}
\def\RMP #1 #2 #3 {Rev. Mod. Phys. {\bf#1},\ #2 (#3)}
\def\ZPC #1 #2 #3 {Z. Phys.  {\bf#1},\ #2 (#3)}

\end{document}